# Evidence for BSM spin 0 and spin 2 resonances at LHC

# Possible Interpretations

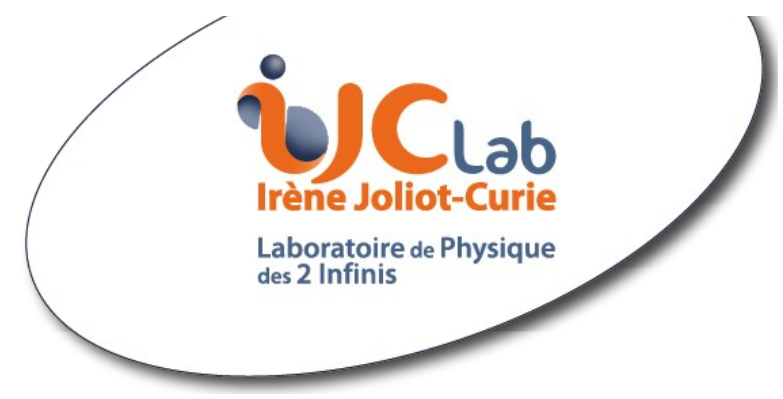

June 22, 2026

Alain Le Yaouanc[1], François Richard[2]

Université Paris-Saclay, CNRS/IN2P3, IJCLab, 91405 Orsay, France

***Abstract**: Nine statistically significant decay channels are observed in LHC data around a mass of **650 GeV**. We interpret three of them as coming from a **20 GeV wide resonance** observed in e+e-, 2 photons and ZZ into four leptons which could be a **J=2 Kaluza Klein graviton resonance** called **T700 (T for tensor with J=2)**. This hypothesis is reinforced by noting that this signal is produced by **VBF** and not by **ggF** since it disappears in ZZ when treated as a scalar. Given that the six other excesses have poor mass resolution, one cannot exclude the presence of an additional wide scalar resonance called **H650.** Assuming a Randall Sundrum RS model, we conclude that LHC observes the **predicted sequence T380, T700 and T1000**, with prospects to discover heavier states by reconstructing hadronic states. At variance with the RS model, T700 decouples from gluon and photon pairs, suggesting a **composite model interpretation** in which **preons are charge neutral and colourless.** T700 does significantly couple to e+e- which has implications for future e+e- colliders. **Perturbative unitarity requirements** predict **$T^{++}$->$W^+W^+$ and $T^+$->ZW resonances,** again indicated by LHC data. With **BR(e+e-)~0.25%**, measured by ATLAS and CMS, this scenario offers excellent prospects for abundantly (**Gigafactory**) producing a sequence of **narrow resonances at future e+e- colliders.** For heavy scalars, the situation is less clear. Following ATLAS and CMS, we expect that the **top loop contribution** to the gluon-gluon fusion mechanism **ggF** could produce a **deficit rather than an excess** in the mass distribution of top pairs, which prevents a standard estimate of the **statistical significance** for heavy resonances. It seems that the pseudo-scalar and scalar resonances **A490 and H650**, indicated by other channels, create observable deviations in the **$t\bar{t}$** analyses presented by ATLAS and CMS. The **scalar resonances** seem to form the **triple Higgs doublet structure predicted by Weinberg**. The present note summarises these arguments and collects available indications in view of electing a **future collider.***

## Talk given at IRN Terascale meeting at IJCLab Orsay, April 20-22nd 2026

1 Alain Le Yaouanc <alain.le-yaouanc@ijclab.in2p3.fr>

2 François Richard <francois.richard@ijclab.in2p3.fr>

# Introduction

There are many ways in which a resonance can be missed at LHC. We will illustrate this by various examples and try to interpret consistently the many indications produced so far.

Our best candidate for a BSM resonance is described in Appendix I and in [1]. LHC observes **9 significant excesses of events around 650 GeV** which could come from two nearby resonances. There is a resonance at 700 GeV with a total width of ~20±10 GeV precisely measured in **$ZZ \to \ell^+\ell^- + \ell'^+\ell'^-$, γγ and e+e-**. This resonance is unlikely to be a heavy scalar expected in models with extra scalar doublets, given the presence of detectable **γγ and e+e-** modes as shown in figure 1 and 2.ll_i

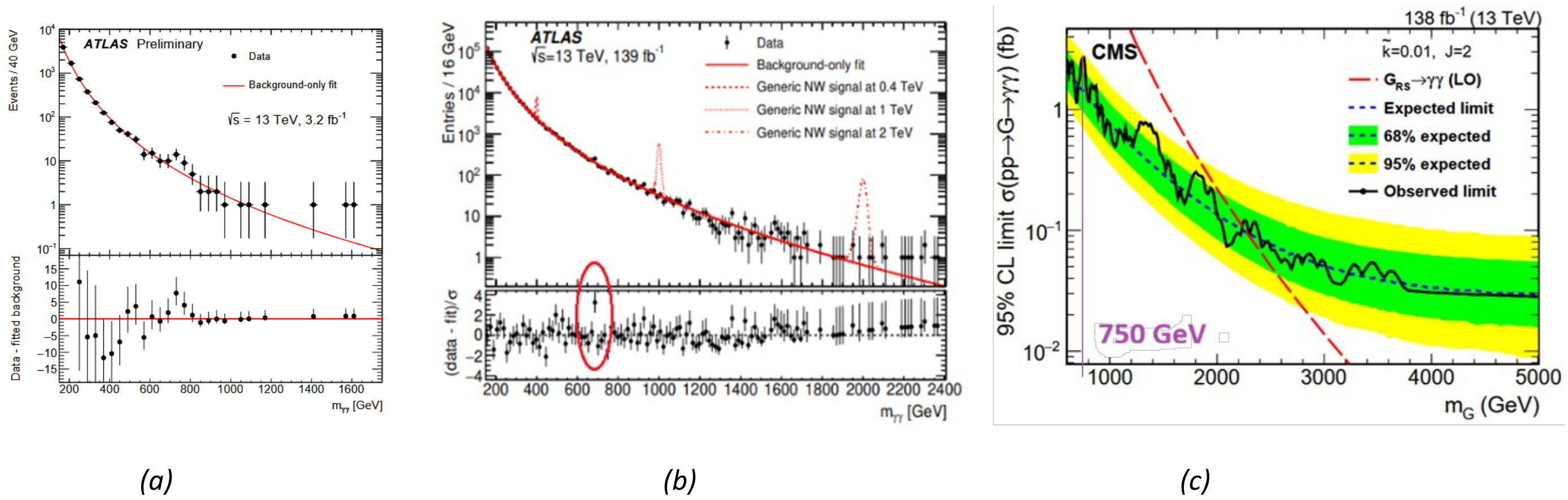

*(a)* *(b)* *(c)*

*Figure 1: (a) γγ mass distribution initially observed by ATLAS[3] indicating a **wide peak** at a mass of 750 GeV.*
*(b) Mass distribution observed by ATLAS [55] with the full RUN2 sample indicating a **narrow peak** at 700 GeV.*
*(c) Mass distribution from CMS [55] indicating a mild excess at 750 GeV. The red dashed prediction assumes full coupling of gg to KK gravitons.*

The signal observed in the two photon mode at **700 GeV** recalls the indication observed by ATLAS at **750 GeV with a cross section of ~5fb**. The displacement in mass and the larger width could be due to an initial **imperfect energy calibration**. The final cross section seen in (b) (c), about 3-5 times lower than in (a), suggests a **positive statistical fluctuation** in the initial sample**.** After collecting 4 times more statistics and in absence of a reinforcement of the signal, this effect was declared a statistical fluctuation.

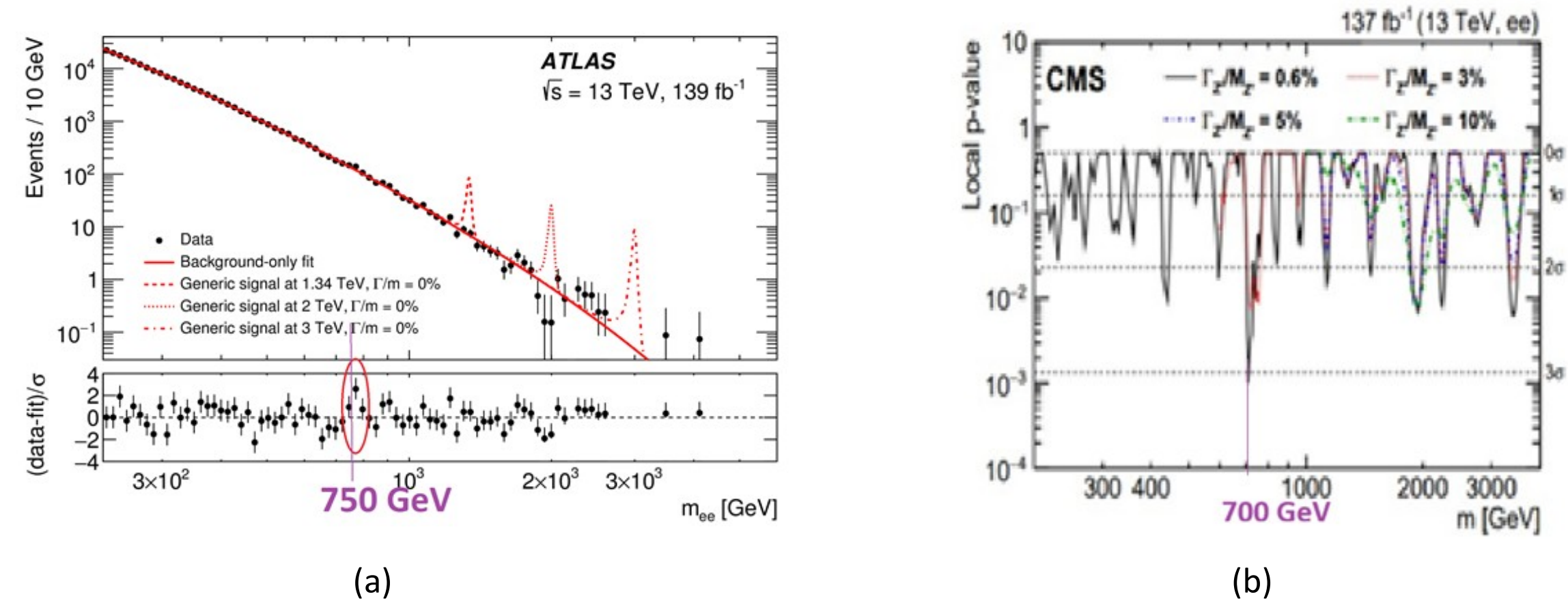

(a) (b)

*Figure 2: (a) e+e- mass distribution observed in ATLAS indicating a small excess around 750 GeV. (b) P-value distribution observed by CMS showing a narrow excess in the e+e- distribution at 700 GeV.[4]*

---

3 ATLAS indication for a two photon indication at 750 GeV ATLAS-CONF-2015-081

4 See 1903.06248 for ATLAS and 2103.02708 for CMS.

With the full sample of RUN2, ATLAS observes a narrower signal at a displaced mass which suggests that it was perhaps premature to draw a negative conclusion. CMS shows a weaker indication at 750 GeV. This revision is strongly reinforced par the occurrence of 8 independent indications in the same region of mass.
For **lepton pairs** searches, we only retain e+e- instead of µ+µ- for a superior mass resolution. The best evidence comes from CMS and is consistence with a 700 GeV mass as shown in figure 2b, with a non-negligible indication of ATLAS consistent with the same mass. In Appendix V we discuss possible interpretations of this evidence which is not incompatible with the standard RS model.
For what concerns the ZZ mode, the situation is less clear. In the **ggF+VBF** mode[5], with genuine cuts, ATLAS [70] observes in figure 3a a wide excess around 650 GeV after subtracting the standard background. Assuming a scalar resonance, an anti Drell Yan selection was applied which did not result in a reinforcement of the indication as shown in figure 3a.

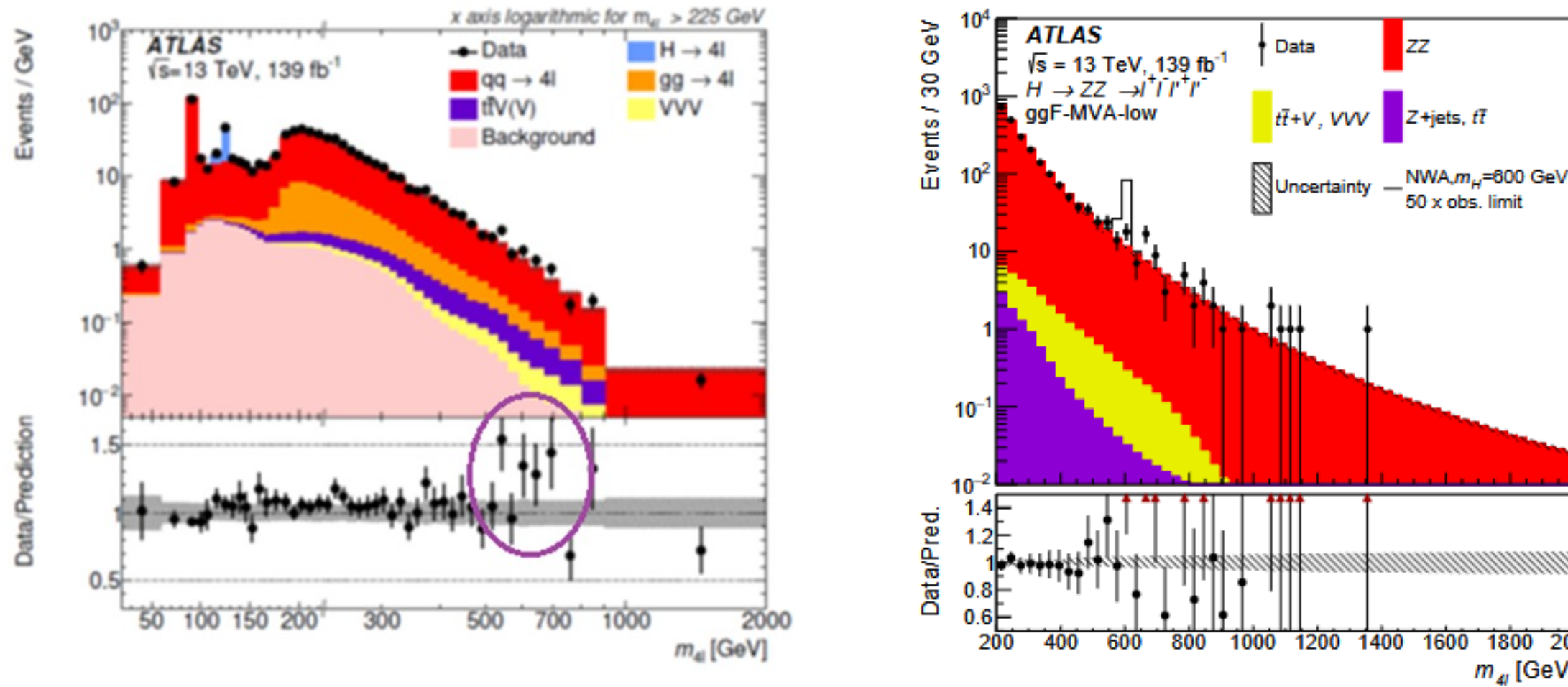

*Figure 3: The ZZ mass distribution observed in 4 leptons by ATLAS indicating a wide structure around 650 GeV. On the right-hand side one applies a selection against DY assuming a scalar resonance.*

Taking into account interference effects, [56] claims that this resonance also sits near 700 GeV and has a width of order 20±5 GeV. Admittedly the analysis performed in [56] assumes that the resonance is a scalar and is therefore not relevant to our interpretation, but indicates the potential effect of interferences. Another evidence against a scalar interpretation is illustrated by figure 4 for the VBF case. When applying an anti Drell Yan selection, the ZZ signal disappears, as seen in figure 4b. Figure 4c shows that a J=2 state has a distribution similar to the DY background, hence the observed disappearance. Note that for the ggF process this selection should not matter, implying that **the dominant contribution to figure 3 comes from VBF. The same happens in the CMS analysis [49].**

(a) (b) (c)

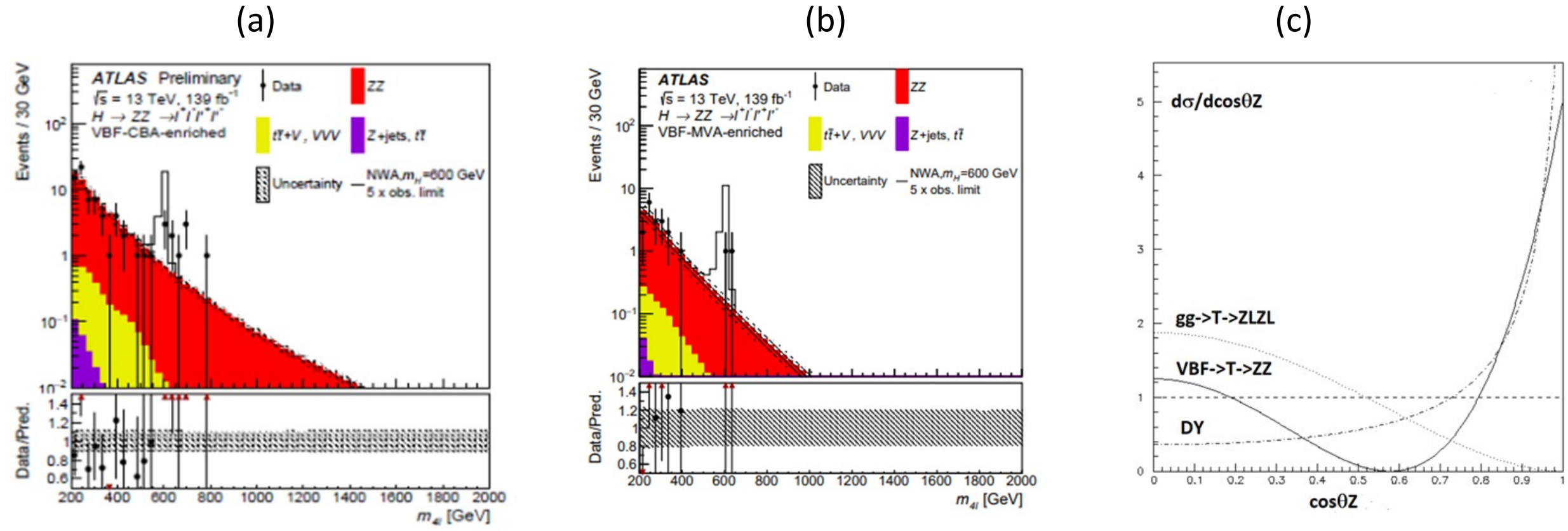

*Figure 4:(a) shows the excess observed in VBF while (b) shows an almost disappearance of this excess applying a selection against the Drell Yan background. (c) shows a forward peaking of VBF→T700→ZZ analogous to the DY distribution.*

5 Noting that these plots are abusively labelled ggF while they contain both ggF and VBF.

Figure 4a shows **9 candidates** consistent with the 700 resonance. Only **two of them survive to the anti DY selection** in conformity with a J=2 angular distribution**.**

Note in passing that the VBF selection practised in 4a does improve the ratio s/b with respect to figure 3 but not the statistical significance with respect to figure 3. This is due to the VBF selection which has a low efficiency, about 5 to 10 times lower than for the VBF+ggF selection.

The excess observed in figure 3 could therefore be entirely due to VBF.

A similar phenomenon is observed for the T380->ZZ candidate which is discussed in section I.1.

These features have led us to assume that this resonance could be a **KK graviton,** a state predicted in the Randall Sundrum (RS) model (section I and Appendix V). The likely presence of a **nearby wide scalar** resonance is also suggested by the observed large number of channels implying a total width inconsistent with the measured width of T700. Most of these states are reconstructed with poor mass resolution and could therefore receive contributions from a wide scalar called hereafter **H650.** RS predicts the masses and the widths of a sequence of KK gravitons and, assuming that **T700** is the **second rank graviton recurrence**, we will show that **T380 and T1000** indications fulfil the mass predictions of RS (see section I.2 and I.3).

**T700** is also observed [62] in **WW->**$\ell^+\ell^-\nu\nu$**,** with a poor mass resolution. Reference [62] claims that it could come entirely from the VBF process with a cross section of **160±50 fb.**

The coupling of T700 to ZZ and WW could imply, through the **perturbative unitary sum rules SR** (see [51] and Appendix IV), the existence of charged resonances $T^+$ and $T^{++}$ which are also indicated by LHC data (section I.4). This feature suggests that T380 has an isospin I=2.

The presence of T1000 observed in h125h125 has some measurable consequences on the **Higgs self coupling measurements** which we have tried to estimate (section I.3).

For heavy resonances decaying into top pairs (section II) and produced by ggF, interference with the QCD SM process leads to a **deficit rather than a bump**, **as assumed in naïve searches.** This results in observable deviations (see sections II.1 and II.3) but, so far**, standard methods are unable to quantitatively account for the significance of these effects.** This process would therefore constitute an other source of confusion for resonance searches. Note that for ggF→ZZ, the QCD interfering background is much smaller than the DY background which does not interfere, hence a reduced risk due to the loop effect.

The various scalar candidates can be interpreted within the **three Higgs doublet model** proposed by Weinberg (section I.5) or within NMSSM (Appendix II).

Four top channel has been measured by ATLAS and CMS. A slight excess is observed which we attempt to interpret within the RS model (section II.5). Section III will deal with the consequences of our findings for future colliders. Section IV deals with possible consequences for DM searches.

# I. The KK graviton scenario

Table 1 summarises the main figures concerning the KK graviton resonances. For the 700 GeV resonance, the **ggF->ZZ** angular distribution goes like $(z^2-1)^2$ (see appendix III) which would pass the scalar cuts applied by ATLAS and CMS. With the scalar cuts practised by ATLAS and CMS, the signal goes away meaning that the observed excess mainly comes from the VBF contribution and is due to a J=2 particle. The DY background dominates over the ggF background and should not interfere with the VBF→ZZ signal. In the RS model, one predicts a sequence of KK graviton resonances $G_{iKK}$ which have **quantified masses** [16]. These masses are proportional to the **zeros of the Bessel function** $J_1(x)$**:**

**$x^G_i$=3.83,7.02,10.17,13.32,16.5,19.6,22.8,25.9,29.05,32.2,35.3…**

| m $G_{KK}$ GeV | 380 | **700±20** | 1000 | 1322 | 1640 | 1945 | 2260 | 2590 | 2905 |
|---|---|---|---|---|---|---|---|---|---|
| $\Gamma$ $G_{KK}$ GeV | 3 | **20±7** | 60 | 130 | 260 | 440 | 680 | 1042 | 1440 |
| VBF→$G_{KK}$ fb | 360 | **100±50** | 38 | 15 | 7 | 3.4 | 1.7 | 0.9 | 0.4 |

*Table 1: expected parameters of the first KK graviton resonances*

We assume that **T700 is the second KK resonance**. The mass of the first resonance is given by (3.83/7.02)700=380 GeV in agreement with observation (see section I.1). The expected ggF cross section is huge at 380 GeV, therefore incompatible with observation. One therefore needs to assume that the **coupling to gluons of these resonances is drastically suppressed** and this should also apply to T700.

The predicted mass for the first recurrence is larger than the measurement found in the process A→hh+Z, presumably due to the poor energy resolution of the 4b states (see II.1) but agrees with T700→T380+h→$b\bar{b}$+$b\bar{b}$. One has **mi~100$x^G_i$** in GeV, while the total width itself goes like **$\Gamma$i~0.06($x^G_i$)$^3$** in GeV.

Assuming that VBF is dominant, the two main inputs to determine this cross section are the following:

- The observed **VBF cross section for T700 into WW** ~160±50 fb [62]
- The total width which can be deduced from the ZZ and $\gamma\gamma$ measurements

To reconcile the width measurements, one has to take into account interference effects in ther ZZ case as discussed in reference [56] which finds **$\Gamma_{tot}$=20±7 GeV,** which is consistent with the ATLAS measurement shown in figure 1b. In the RS one has hh/ZZ~1. In [16], one predicts $\Gamma_{tt}$~9$\Gamma_{ZZ}$=45 GeV, which is obviously **inconsistent** with these figures. A way out is therefore to assume that the top quark does not belong to the same brane as h/W/Z ( see Appendix V) and that its width is below 5 GeV. Neglecting the top pair contribution one concludes that **$\Gamma_{WW}$~10±3.5 GeV** , therefore **VBFww=50±25 fb** in tension with the result from CMS.

One has [16]:

$$\Gamma\left(G \to W_L^+ W_L^-\right) \approx \frac{(c\, x_n^G)^2\, m_n^G}{480\pi}$$

with $x_2$=7.02 and mG=700 GeV, hence c=**k/$\overline{\text{MPl}}$ =0.52±0.1**, with $\overline{\text{MPl}}$=2.4 × $10^{18}$ GeV, in agreement with expectations from [16] which predicts a value between 0.5 and 2 (see Appendix V for more details).

Given that ZZ/WW=0.5, one expects an **25 fb** VBF cross section for the ZZ mode, which, given figure 4a, implies A$\varepsilon$=40±20%.

The VBF cross section is simply deduced from VBF for Higgs:

**VBFGkk/VBFH=(2J+1)$\Gamma_{GWW}$ /$\Gamma_{HWW}$=0.43±0.15**

Knowing[6] VBFH one deduces the values of Table 1. Note that at 700 GeV, a SM Higgs would be ~10 times wider than a KK graviton resonance.

6 VBFH can be found up to 3 TeV in: https://twiki.cern.ch/twiki/bin/view/LHCPhysics/CERNYellowReportPageBSMAt13TeV#VBF_Process

For masses above 2 TeV the SM background starts to be smaller than the expected signal which justifies using **hadronic final states** since of the resonances become so wide that one does not benefit any more from the mass resolution of the leptonic modes. The wide tresonances tend to overlap abover 2.5 TeV such that large interference effect will start to become significant. **Figure 24 from Appendix I** shows that observable deviations start to be visible above 2.5 TeV. With more data and a full calculation of the expected patterns one may hope to confirm these indications.

LHC measures **BR(e+e-)~0.25±0.7%**. This value is consistent LEP/SLC deviations of AFBb and ALR (see Appendix V) provided that there are heavy KK Z bosons and that the right handed electron eR is close to the Higgs brane with ceR=0.45.

For what concerns **BR(γγ)~0.3±0.1%,** this value[7] seems well below the naïve expectation of **universal coupling of gravitons to the stress tensor** (Figure 30 in Appendix V). As explained in Appendix V, **RS predicts an enhancement factor of about 25** for $\gamma\gamma$ for the first graviton recurrence, therefore inconsistent with the data which show no indication at 380 GeV. It seems therefore that there is also **no direct coupling og KK gravitons to photons**. If the ATLAS and CMS indications are confirmed, this means that there are significant **loop contributions** of the KK fermions and bosons as suggested by figure 35 of [59] for the SM Higgs boson. In summary one has the following ratios:

**WW/ZZ/γγ/e+e-/tt=1/0.5/0.6%/0.5%/<0.5**

Note the **suppression of the top pair BR** with respect to an a priori expectation [16] which says that tt/ZZ=9. This requires a location of the top quark with ct>0.4 (see figure 33). The requirement that ggF<100 fb may require a stronger suppression but, unfortunately, the top loop calculation for KK graviton is not available.

To explain the absence of direct coupling to gluons and photons, one may invoke the **duality argument** 'à la Maldacena' (see Appendix V), assuming that gravitons are made of **colourless preons, without electric charge**, in similarity to the Higgs boson [78]. Intuitively, there is indeed no evident connection between the colour force and gravitation.

**Duality does not mean identity** and the two descriptions are **rather complementary than identical.** The 5D picture is in essence geometric, assuming that interactions do not take place when particles do not overlap in the extra dimension. A contrario, the fact that there is overlap does not necessarily mean that particles will interact. For instance, if the SM Higgs boson itself overlaps with the gluon distribution in 5D, this does not mean that it will **directly interact** with gluons. Therefore the 5D description does not pronounce on the properties related to the preonic content of composite particles in 4D.

For a neutral tensor resonance coupled to WW/ZZ, one naturally expects **charged resonances** with similar masses to satisfy **perturbative unitarity**[8]. This seems to be case, as shown in section I.4.

In **summary,** confronting the RS model prediction to experimental data, we conclude that T700 :

- Is a resonance observed in ZZ/WW/$\gamma\gamma$/e+e-
- Is a J=2 particle, produced in the VBF mode, therefore missed when applying angular selections against the DY background
- Has a VBF cross section in WW quantitatively compatible with the RS prediction
- Has a mass compatible with the two other KK recurrences, T380 and T1000, predicted by RS

7 Note that the acceptance for this mode is optimal since it is centrally produced (see Appendix III)

8 See [39] and Appendix IV for a thorough discussion of a KK graviton contribution to perturbative unitarity.

- Has a BR into top pairs more than times below the naïve RS prediction
- Has a BR into e+e- consistent with LEP/SLC measurements provided ceR=0.45
- Has a **negligible BR into gg**, in strong contradiction with the standard RS prediction
- Has a **BR into $\gamma\gamma$** which indicates that there is no direct coupling as for gg but only loop contributions

So far, the naïve picture was telling that with a huge cross section due to its gg coupling and with a large $\gamma\,\gamma$ BR, a KK graviton with mass below **4.5 TeV** was excluded by LHC (see figure 34). The main lesson of this analysis is that these statements are **invalidated** by observations. Table 1 tells us that gravitons with masses heavier than 2 TeV have a non negligible cross section in a mass region where the SM background vanishes. In Appendix I, figure 24, we see that in an analysis performed by CMS by selecting WW+ZZ hadronic final states observes indications of the presence of these heavy resonances.

## I.1 The T380 case

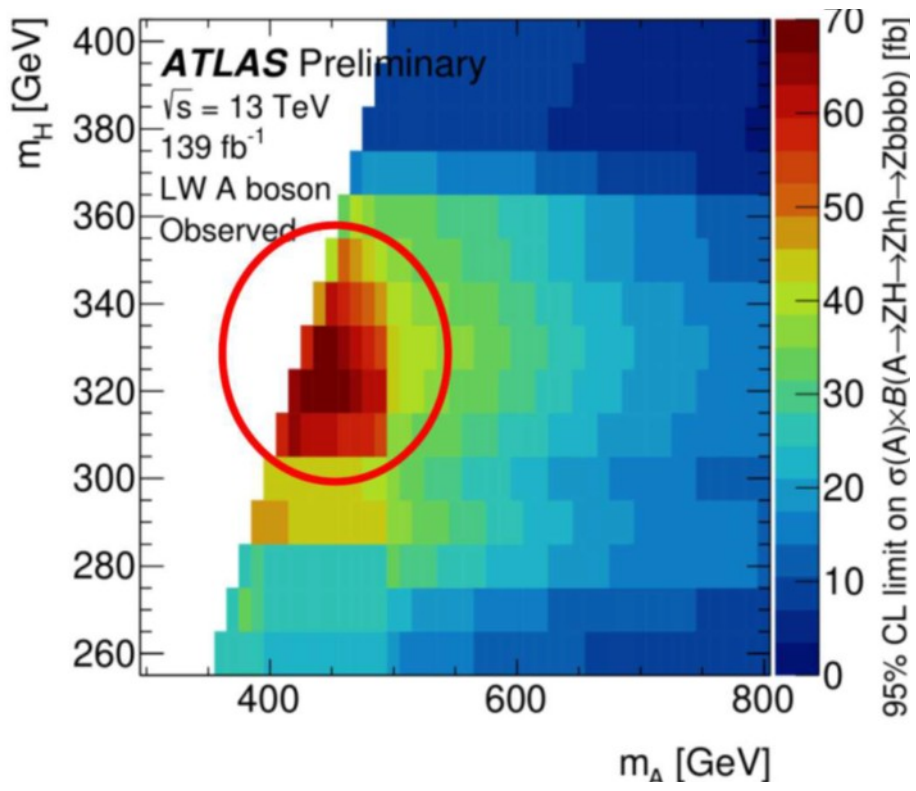

*Figure 5: Observed upper bounds at 95% CL on $\sigma$(A)xBR(A→ZH→Zhh→$\ell^+\ell^-b\bar{b}b\bar{b}$ in the (mA,mH) plane.*

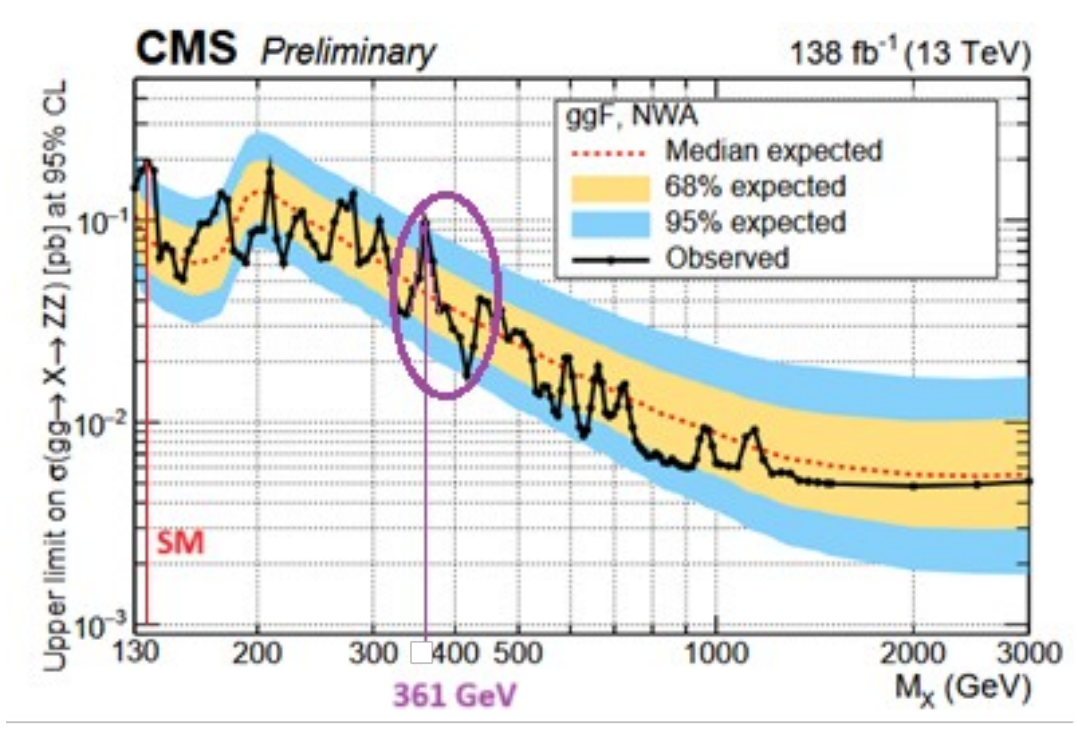

*Figure 6: upper limits on $\sigma$(gg→X→ZZ) at 95% CL from CMS.*

The mass and width of T380 are predicted by RS assuming that T700 is the second KK recurrence. This predictions agrees, within errors, with 5 indications:

- T700→ T380+h125 → $b\bar{b}+b\bar{b}$ [32]
- T380→ hh→ $\tau^+\tau^-+\gamma\gamma$ [30]

- T380→ ZZ shown in fig. 6 [49], with the observed mass, **361 GeV**, is slightly shifted probably due to interference effects. Note also the narrowness of this indication which agrees with the RS prediction (3 GeV). As expected the measured cross section ~100 fb falls below the VBF prediction deduced from T700, σ~240±70 fb, which is due to the selections practiced by CMS under the assumption of a **scalar resonance.** We also notice that T380 survives better to these selections than T700 which could be due to its narrowness[30].
- A490→ T380+Z→ hh+Z→ $b\bar{b}b\bar{b}$+Z [17] shown in figure 5.
- H+→T380+W→hh+W->$b\bar{b}b\bar{b}$+W shown in figure 9 [17]

Figure 5 shows a local excess of 3.8 s.d., displaced with respect to expectations which can be interpreted as due to poor mass reconstruction of the $b\bar{b}b\bar{b}$ final state.

**VBF production of $t\,\bar{t}$ resonances** has not been tried at LHC. They could in principle allow to observe T380 and T700 without the contamination from the toponium nor the isodoublet resonances which decouple from W/Z. Admittedly the expected cross sections are low since one expect a reduced BR of these resonances into top quarks

Finally one may ask **why T380 is not observed into two photons,** while T700 shows a signal. This is true for ATLAS while this information is not available in CMS where the mass spectrum starts at 500 Gev (see figure 1c). To answer quantitatively to this question one should be able to compute loop contributions to these two resonances. Unfortunately CMS It seems likely that with more statistics this signal should become visible with HL-LHC.

## I.2 Indication for the third KK graviton recurrence decaying into hh?

Figure 10 indicates that the detection of KK graviton resonances in the mode hh seems optimal for the third (1 TeV) and fourth (1.3 TeV) recurrences. As expected from figure 35b, there are two nearby deviations, one at 1.1 TeV (3.3 s.d.) , followed by a second one at 1.3 TeV (2 s.d.) [18]. This evidence, if real, is likely to improve shortly given the very active search for this channel. Fig.7b shows that $G_{KK3}$ is visible mainly in the $b\bar{b}\tau^+\tau^-$ mode. One expects for $G_{KK3}$→hh a cross section ~ 10 fb. This cross section is to be compared to the SM hh cross section, 31 fb for κλ~1, which is negligible above 800 GeV. Figure 10 shows that the SM contribution will be best measured for high masses, where it will be contaminated by the KK graviton contribution, implying that one will find **κλ** significantly above 1.

(a)

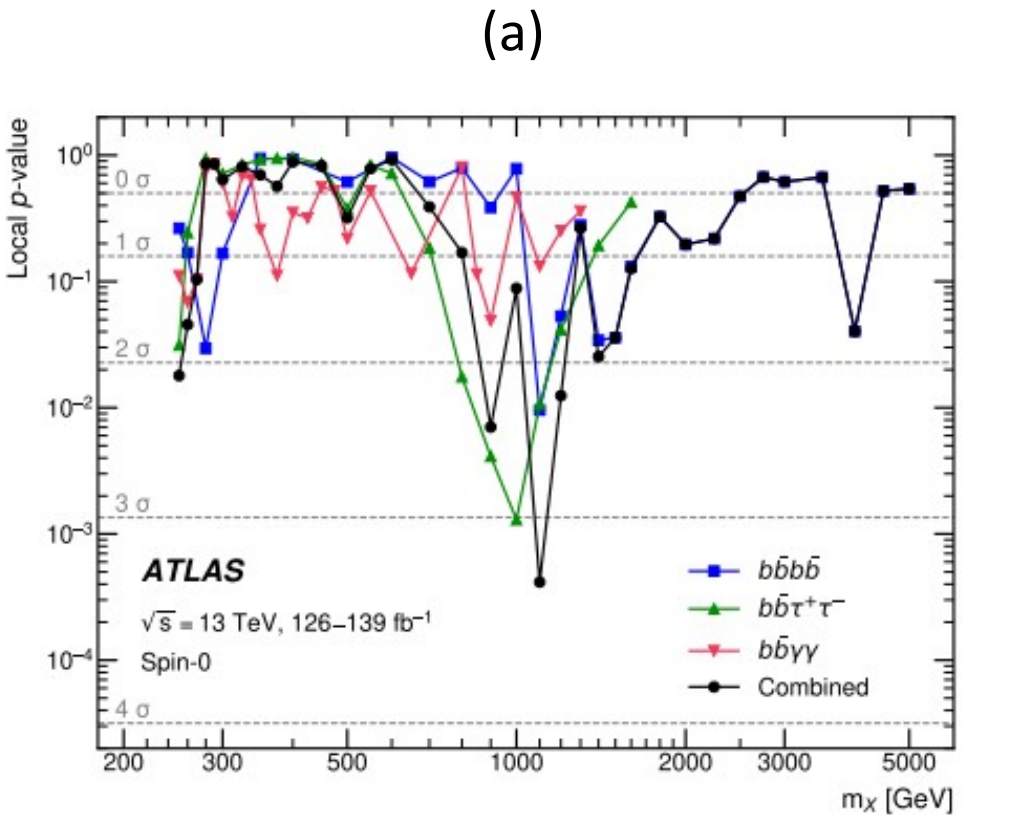

(b)

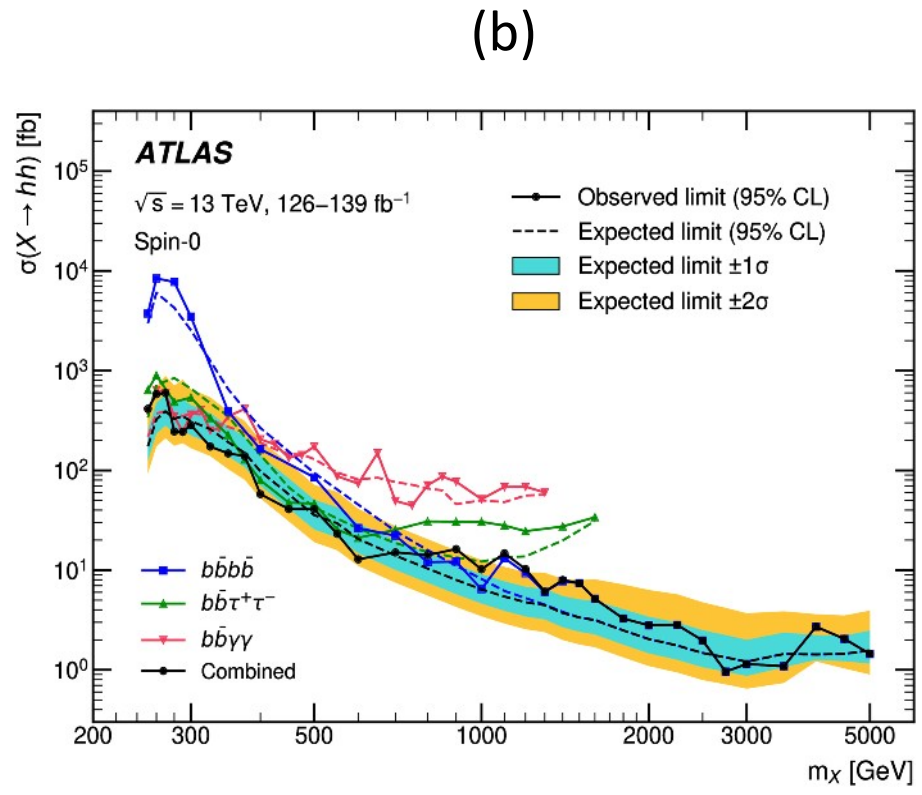

*Figure 7: Evidence for T1000→ hh. (a) indicates an excess at 1.1 TeV, mainly coming from the $b\bar{b}\tau^+\tau^-$ mode as indicated by (b).*

## I.3 Impact on Higgs self coupling measurements

Tagging by W has a better visibility for two reasons. One has $q\bar{q}$→T+W/T+Z=2, as for the SM. On top of that, one has BR(W->νℓ)/BR(Z->$\ell^+\ell^-$)=3, which also favours T+W.

Figure 10 indicates what can be expected from LHC measurements together with the HEFT predictions[9]. One sees that T380 seems to offer the best prospect. But in the **inclusive search** used at LHC to measure μhh, figure 10 tells us that **background is overwhelming** for this mass preventing observation, while this does not seem the case for T1000. So far, the measurements of **μhh** tend to be below 3, noting however that it is still insensitive to the contribution from T380.

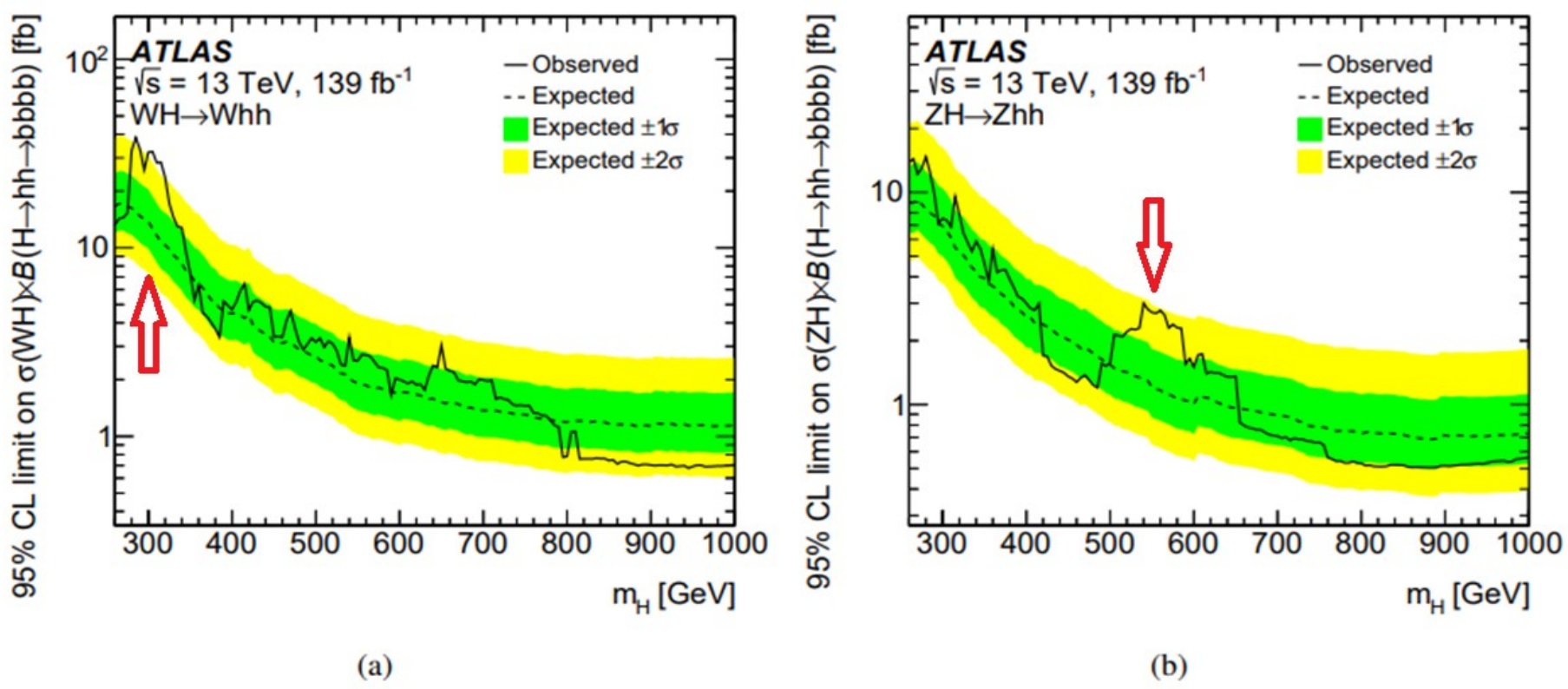

*Figure 9 : Observed (black solid curve) and expected (black dashed curve) 95 % CL upper limits on the prediction cross-sections at sqrt(s)=13 TeV of a heavy resonance H in the decay mode H→ hh → $b\bar{b}b\bar{b}$ in association with (a) a W boson and (b) a Z boson as a function of the resonance mass. The green (inner) and yellow (outer) bands represent ±1σ and ±2σ uncertainty in the expected limits.*

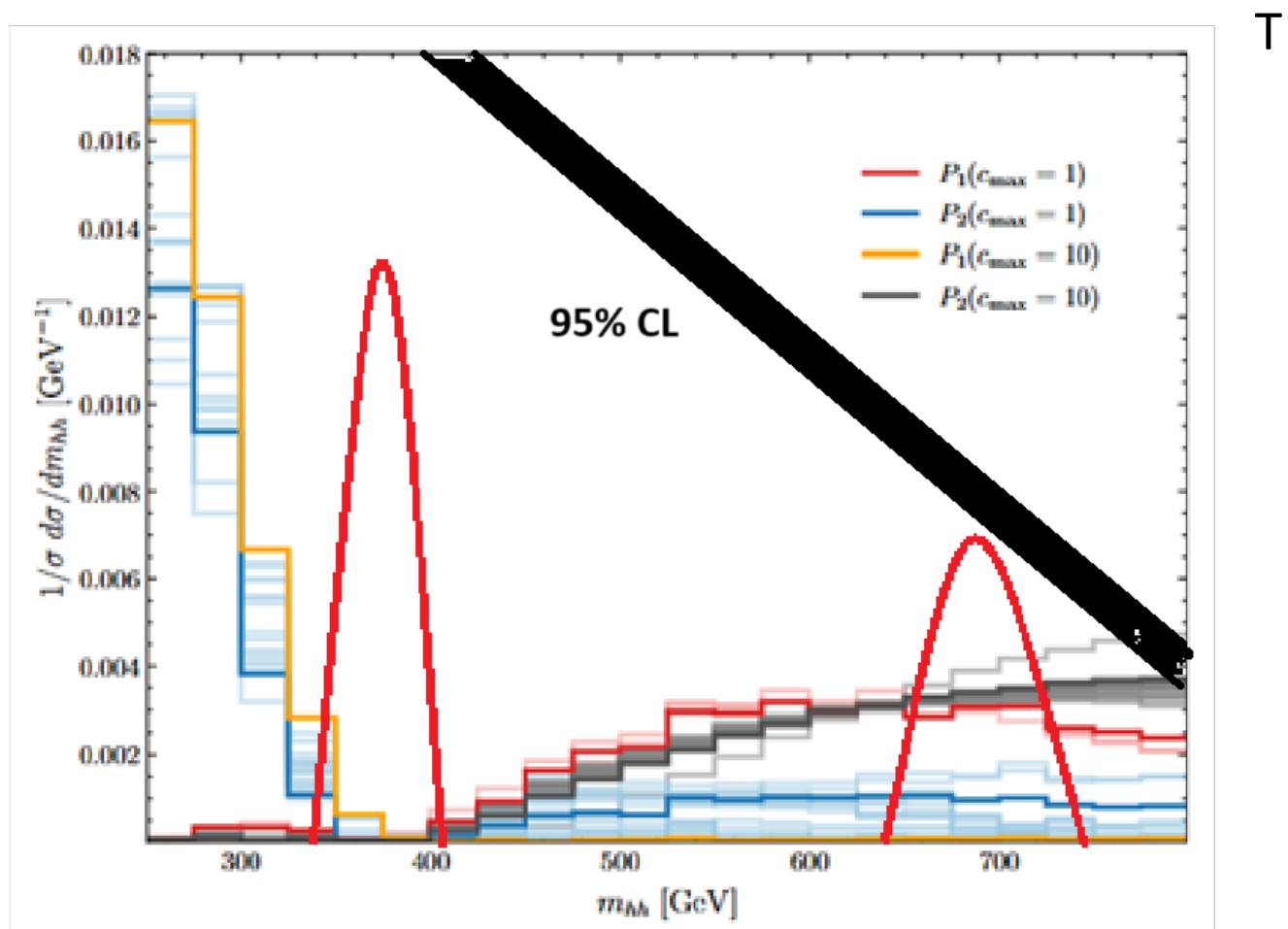

*Figure 10: Expected hh mass spectrum with various HEFT parametrisation of BSM physics with resonances above a TeV. The two red bumps correspond to expected signals due to T380 and T700. The black band corresponds to the present 95% C.L. upper limit given by ATLAS.*

9 See R. Gröber: An introduction to EFT https://indico.ijclab.in2p3.fr/event/12138/contributions/39315/

## I.4 The charged states

Figure 11 shows the results obtained by ATLAS and CMS for $W^+W^+$ and ZW resonances interpreted in terms of the Georgi Machacek model. The poor **mass resolution** with $W^+W^+$ does not allow a separation between the two first KK resonances, hence, presumably, the wide excess (black curve of 10a) observed by both experiments [19,20]. This means that the mass of this resonance could be overestimated and in fact could be closer to the masses of T380 and $T^+$375, implying a smaller isospin violation. There is also an ambiguity in evaluating the $T^{++}$cross section given that it could integrate the two neighbouring $G_{KK}$. The cross section of $T^{++}$450 could therefore be overestimated, hence $BR(W^+W^+)$ could be closer to 1 and the need for a $H^+H^+$ contribution be less pressing.

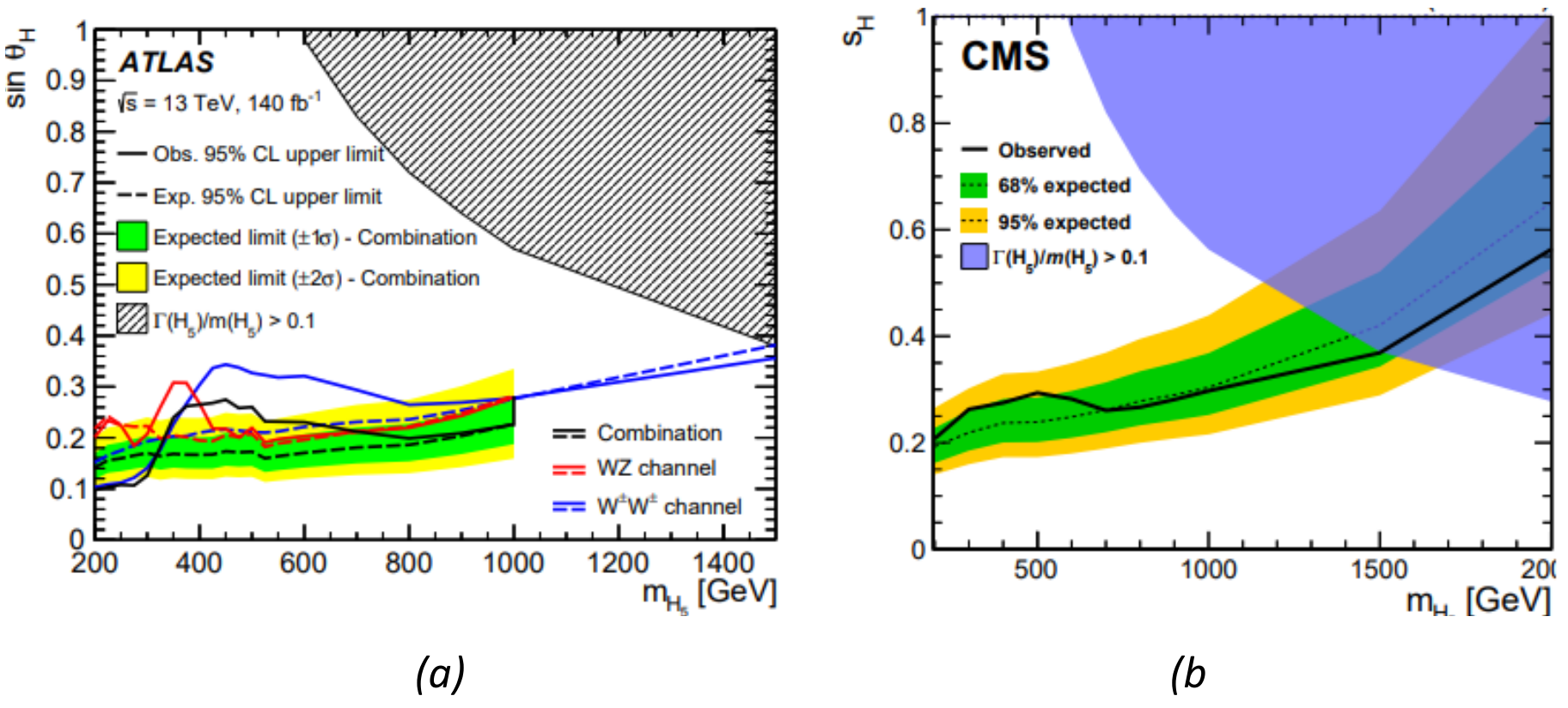

*(a)* *(b*

*Figure 11: Evidence for T+ and T++ in ATLAS and CMS.*

Note that these states were selected as scalar resonances, following the Georgi Machacek model, while they could be J=2 graviton recurrences according to the RS model, with anisotropic angular distributions, as for T700. Figure 12 indicates the present status of the WZ searches.

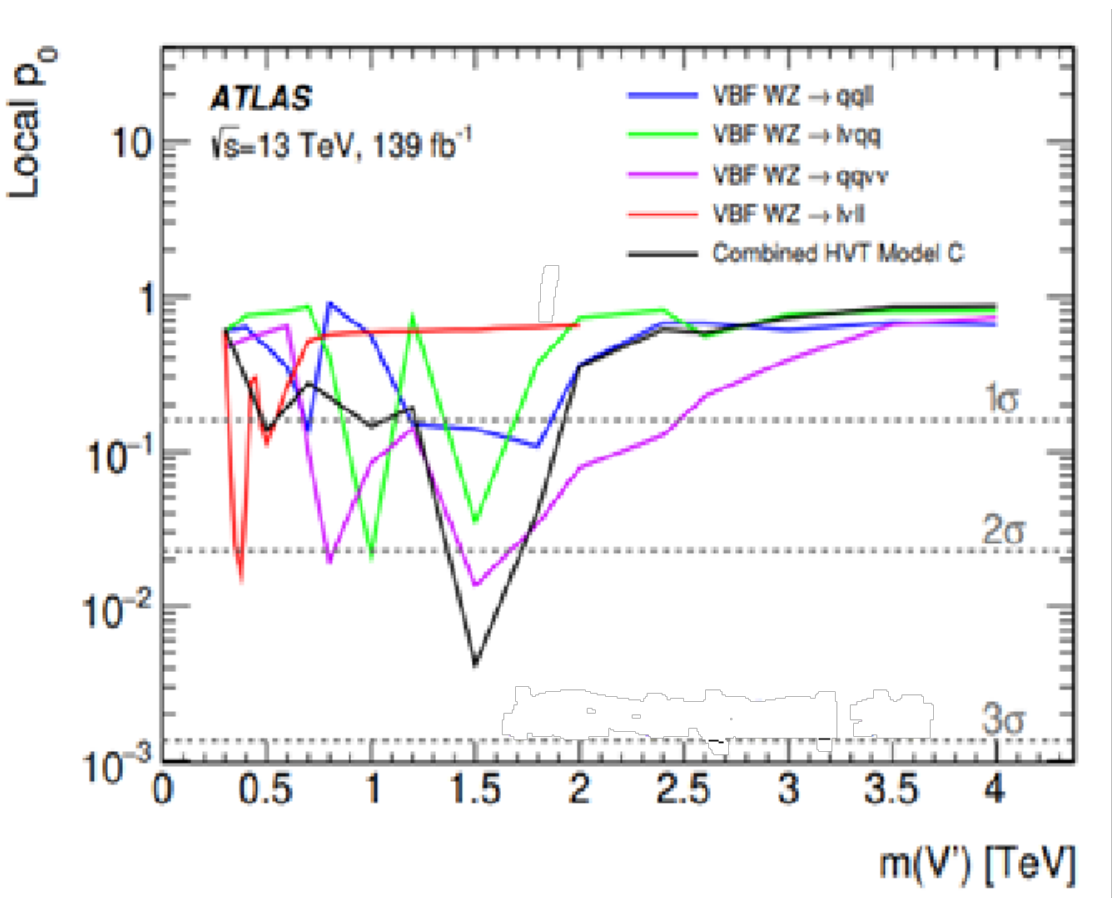

*Figure 12: Hint for an extra $T^+ \rightarrow WZ$ near 1.6 TeV.*

## I.5 Fake gravitons?

Composite models also predict spin 2 particles which could fake gravitons as pointed out in [22]. **Angular distributions** are primary tools to confirm/discard the graviton hypothesis. They can clearly eliminate the scalar interpretation but are a priori indistinguishable from those of a spin 2+ composite state.

As already pointed out, RS provides a precise prediction of the **mass spectrum of KK graviton recurrences** which is likely to be consolidated by HL-LHC data. The agreement with the RS predictions will constitute the best proof of this interpretation.

## I.6 Summary of our interpretations

One assumes a **3HD** [46] model due to Weinberg and **tensor KK isoquintuplets TQ1, TQ2, TQ3...**
This scenario has several merits :

- It integrates all BSM candidates
- It solves the T700→ZZ puzzle given that H650 does not couple to ZZ
- It includes $T^{++}$→$W^+W^+$ and $T^+$→$W^+Z$ which improves perturbative unitarity (Appendix IV)
- It predicts T700→γγ
- It predicts **charged KK graviton recurrences** at 700 GeV, not yet observed but less compulsory
- It predicts a graviton TQ3 at 1 TeV, still weakly indicated
- It predicts heavier gravitons indicated in hadronic modes (see Appendix I)

From TQ2=700 GeV, the Randall Sundrum model predicts TQ1=380 GeV, TQ3=1000 GeV and TQ4=1300 GeV. Concerning **A152,** so far observed [45] into 2 photons accompanied by b quarks, leptons or missing transverse energy, it can be interpreted as a cascade originating from the tensors. One should therefore select these topologies and identify them as coming from a cascade involving Z bosons. This should allow to reconstruct the parent(s).

| | | | | |
|---|---|---|---|---|
| | HD1 | **h125**→ SM modes | **WL** | **ZL** |
| **J=0 I=1/2** | HD2 | **h95**->2γ | t→**$H^+$130**+b→bc+b | **A152**→ 2γ+Z/W |
| | HD3 | **H650**→A490Z→ttZ<br>→ T380 h125<br>→ $t\bar{t}$ | **$H^+$700**→A490+$W^+$<br>→T380+$W^+$ | **A490**→T380+Z→hh+Z<br>→ $t\bar{t}$ |
| **J=2 I=2** | TQ1 | **$T^{++}$450**->$W^+W^+$ | **$T^+$375**→Z$W^+$ | **T380**→ZZ,hh,bb,(tt) |
| | TQ2 | **T700**→ZZ/WW/h125+h95/2γ/e+e- | | |
| | TQ3 | **T1000**→h125+h125 | | |
| | TQn | **VBF→T1322+T1640**→ ZZ<br>**T2930→**ZZ+WW | | |

It is unclear whether T700 (and heavier recurrences) has partners forming an I=2 representation.
In Appendix IV we discuss the issue. H650→A490Z→ttZ from ATLAS (which can also be understood as A650→ H490+Z) shows a large cross section interpreted as coming from the process $b\bar{b}$→A/H.
Quantitatively ATLAS measures a CL upper limit of 600 fb for an expectation of 260 fb. Such a large excess seems in tension with the ttZ measurements from ATLAS and CMS which agree with the SM.
The heavy Higgs H+ is observed in two instances (see Appendix I) but its mass is not precise.

## 1.6 Are there other KK resonances ?

An other issue is to understand **why LHC does not observe other kinds of KK particles**? In [71] a mechanism is described where bulk fields develop localised corrections to their kinetic terms which can modify the phenomenological prediction. **Large brane kinetic terms** allows to create a hierarchy between the diverse KK families.

A more radical way[10] out is to invoke an **analogue of the Higgs mechanism.** As the additional longitudinal degrees of freedom of Z and W come from the SM Higgs doublet, the additional degrees of freedom for the massive graviton KK modes (5 in the massive vs 2 in the massless case) come from one vector and one scalar KK mode. Experimentally, these “eaten” vectors and scalar degrees of freedom would not be visible as resonances. The same thing could also happen for the fermionic KK modes in a SUSY theory. They can become the missing degrees of freedom of the massive gravitino KK modes.

The recent claim by CMS [84] of an **8 TeV resonance decaying into four jet forming two sub-masses around 2 TeV** [84], allows some hope for an answer.  A possible interpretation is discussed in Appendix V in terms of an 8 TeV KK diquark decaying into a pair of KK quarks appears plausible. If confirmed, this would imply a tremendous hierarchy between the various KK particles.

## I.7 SUSY and/or extra-dimensions? The radion?

These findings are puzzling since they suggest that an **UV completion of the model** requires combining **SUSY,** explaining the scalar part, in particular the mass of h125 and **extra-dimensions**, with the sequence of graviton tensors. This combination has been also suggested in [23] where the authors resurrect a symmetry breaking mechanism of SUSY proposed in 1979 in by **Scherk and Schwarz** which relies on these ideas [48]. The driving idea of [23] is that the Higgs mass predicted at tree level by SUSY requires tremendous corrections (see Appendix II) which would result in an other type of fine tuning. The Scherk-Schwarz mechanism considerably alleviates this problem. The lighter the KK particles, the less FT. Reference [23] gives a simplified version with a universal extra-dimension while in the present case one assumes that the gravitons are the lightest KK particles.

The RS model predicts an additional scalar, the **radion.** Its mass is unknown. It could hide among the various scalar candidates or appear in the **four top channel** as discussed in II.3.

# II. Results from LHC on $t\bar{t}$ resonances

Sophisticated methods are presently used by both experiments to cope with the huge QCD background. In particular, they use angular selections given that the QCD background is **forward peaked**. They take into account interference effects. They however assume that there are only scalar resonances of the type A and H, not necessarily mass degenerate. According to our work, this is true for A490 but not for H650 where there could be a nearby tensor resonance T700 with a forward peaked VBF angular distribution. ATLAS allows gHtt to be negative while this does not seem the case for CMS. ATLAS also takes into account the loop effect which modifies the mass distribution both for A and H.

10 We are strongly indebted towards Severin Lüst for suggesting this mechanism.

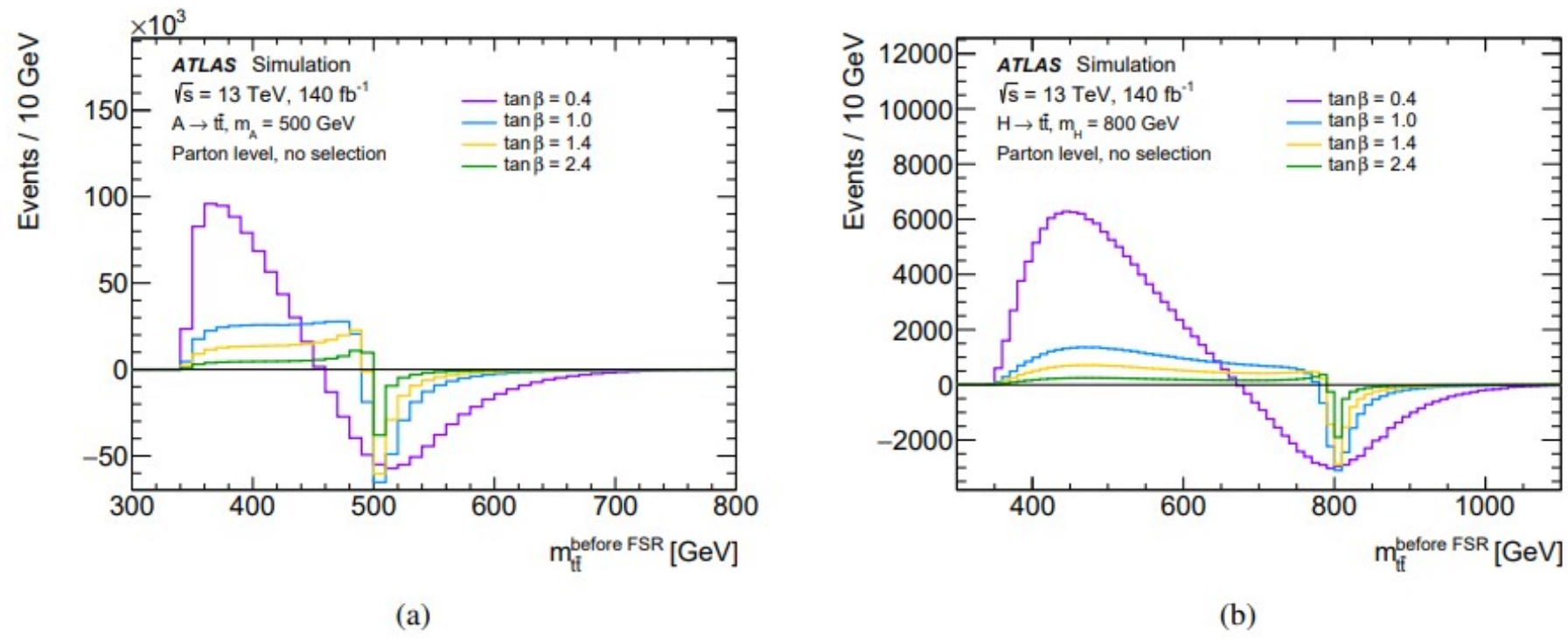

*Figure 13:Signal+interference distribution of the top quark pair at the parton level for A500 (a) and H800 (b) for various values of tan$\beta$ with type-I 2HDM in the alignment limit.*

Both experiments seem to be focussed on the ggF process, since in the 2HD model H is decoupled from ZZ/WW. Observing ZZ/WW final states or scalar resonances produced by VBF violates this principle which is the case for the $G_{KK}$ candidates.

CMS has produced **strong evidence (> 5 s.d.),** for a new resonance observed into t $\bar{t}$ near threshold [2]. From an angular analysis, [2] concludes that this resonance can be either a CP-odd scalar toponium **ηt343** or a CP-odd scalar A365, with a marked preference for the former interpretation, see [3-5]. A365 has a mass incompatible with the resonance A490 observed in the reaction[12]:

**A490→T380+Z→ h125h125+Z→ $b\bar{b}b\bar{b}+\ell^+\ell^-$**

The following diagram (a) illustrates the usual BSM mechanism for producing heavy scalars, with a top loop contribution, while (b) is the main SM background induced by gluon fusion ggF:

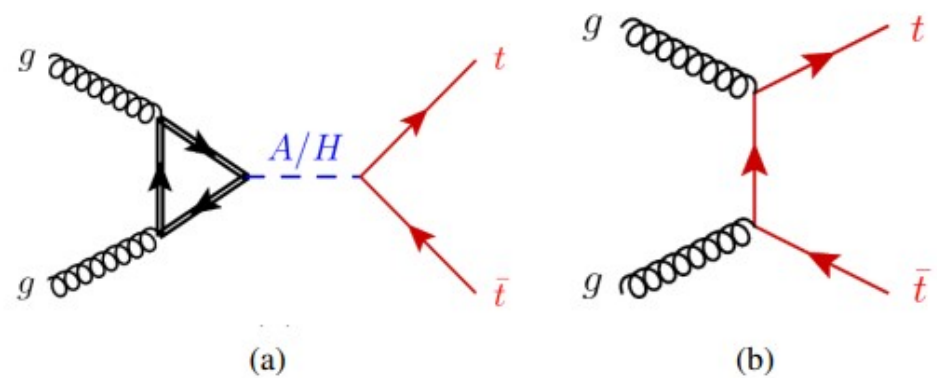

A **Kaluza Klein graviton resonance** can also be produced through a top quark loop or through direct coupling to the **energy-momentum tensor**:

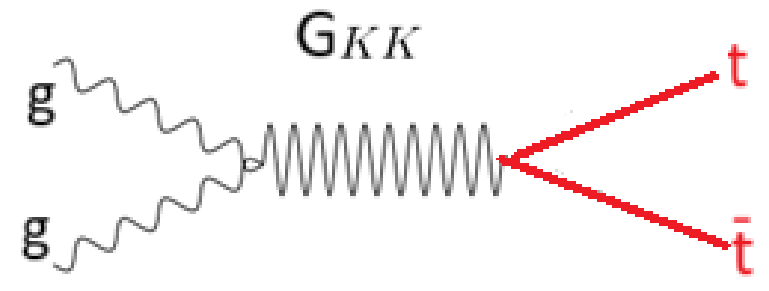

where $G_{KK}$ is a sequence of tensors, three of them being indicated by LHC data, T380 and T700 and T1000. Note that even if T380 seems decoupled from the two gluons, its coupling to top quarks would imply a significant coupling through the loop diagram and therefore a ggF contribution as for the scalar Higgs.
For ggF, one expects (Appendix II) large couplings of A490 and H650 to top pairs which will produce complex patterns. The reason for ignoring so far these searches is that the ggF process does not allow straightforward 'peak finding', which is the usual practice in our field.

12 See section II.1 for an explanation of the readjusted masses for A490 and T380 in the reaction A→TZ.

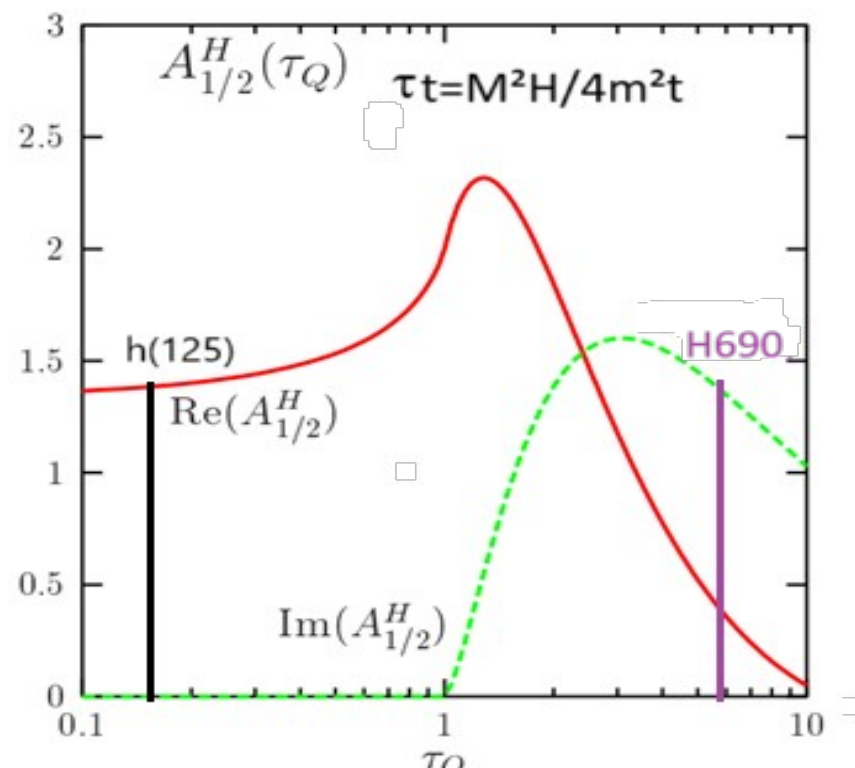

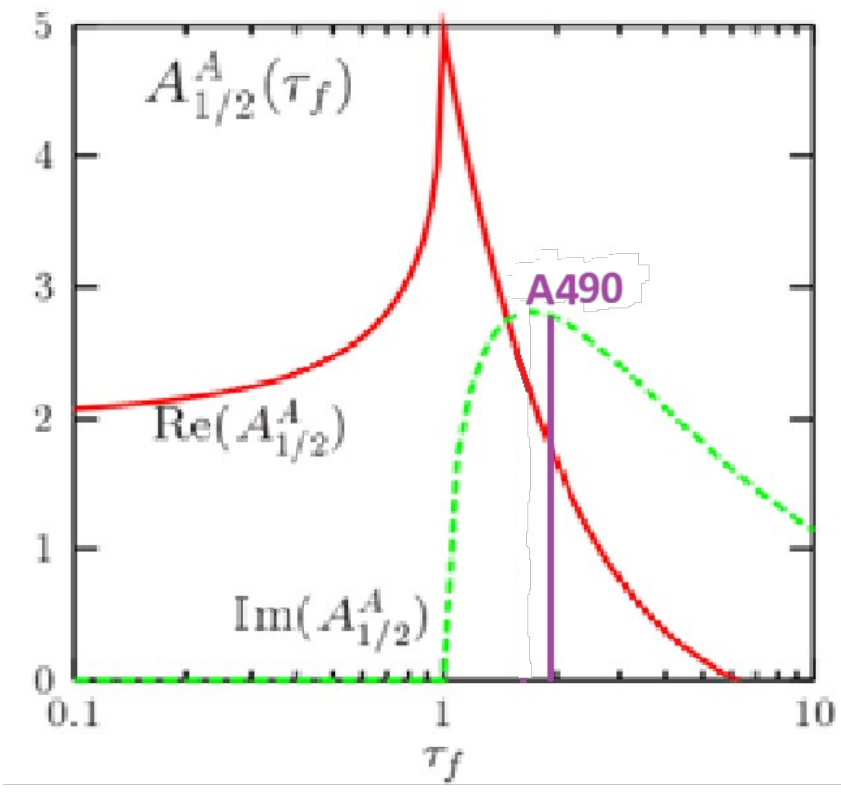

*Figure 14: Variation of real and imaginary parts of the loop amplitude vs. τ=M²H,A/4m²t for a CP-even and CP-odd resonance. The two resonances H700 and A490 are indicated.*

Figure 13 from [6] beautifully illustrates this aspect in the case of a heavy scalar resonance produced by ggF. For heavy resonances, the ggF process is not a genuine Breit Wigner amplitude since the top loop contribution to this process contains a real and an imaginary part, the later being dominant for a heavy Higgs as shown in figure 14.

This explains the negative contribution shown in figure 13. At resonance, the product of these two terms becomes real and positive, while the t-channel QCD amplitude itself is negative, hence a **'deficit' instead of a 'bump'** in the cross section. Similar features are reported in CMS [7]

These findings bear some primordial consequences for searches, teaching us how to interpret mass limits, in particular when they fall **below expectation.** Figure 13 is obtained by subtracting the SM background from the predicted observable cross section. Both A490 and H650 will be affected since, as indicated in figure 14 from [8], the dominance of the imaginary part sets up earlier in the A case. Near top threshold, as shown by figure 14, this effect disappears both for A and H. In particular, the production of a **CP-odd toponium** ηt343 appears to escape to this dramatic complexity. With a cross section of **~6 pb** and an expected width of 3 GeV (about twice the top quark width), this state is becoming a well established and unexpected discovery of LHC, which however seems to belongs to the SM sector [3-5]. A full proof of the **toponium interpretation** would require a precise measurement of the **width** and the mass, which is prevented by the experimental resolution.

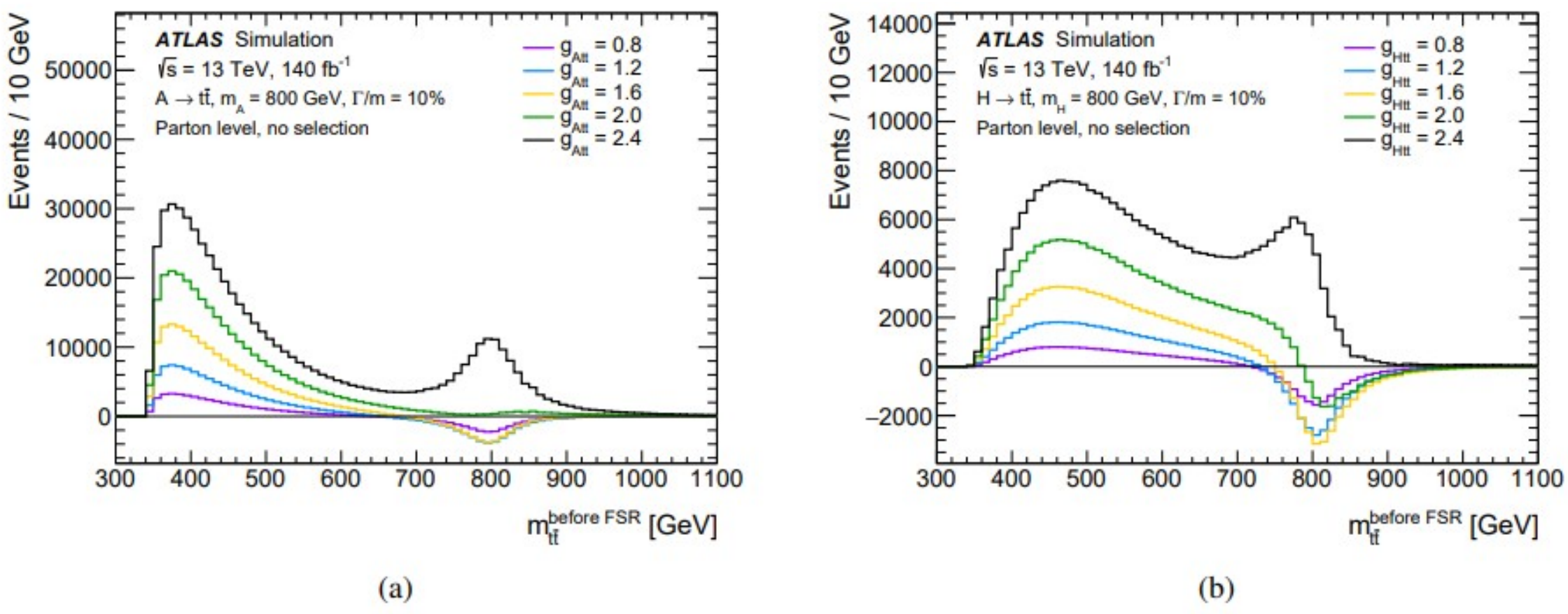

*Figure 15: Signal-plus-interference distributions in mtt at parton level (a) for a single pseudo-scalar A with mass mA=800 GeV and (b) for a single scalar H with mass mH=800 GeV, both with relative total width of 10%. The distributions are shown for different values of the coupling gA/Htt.*

Contrarily to the toponium, A490 has a relative width Γtt/M of order 5%. This ratio is larger for Γtot/M with the contribution of the reported extra modes.

The coupling of T700 to top pairs is unknown since it cannot be separated from H650. CMS concludes [2] that, once taken into account the toponium, the rest of the spectrum shows no feature indicating the presence of other resonances, CP-even or CP-odd. We will challenge this statement confronting the limits given by the two experiments on gHtt around 700 GeV and on gAtt around 490 GeV.

## II.1 ATLAS results

Figure 16 shows no clear evidence for resonances. Figures 17 and 18 clearly suggest the **presence of a H resonance around 700 GeV** with a pattern presumably due to the ggF contribution loop effect.

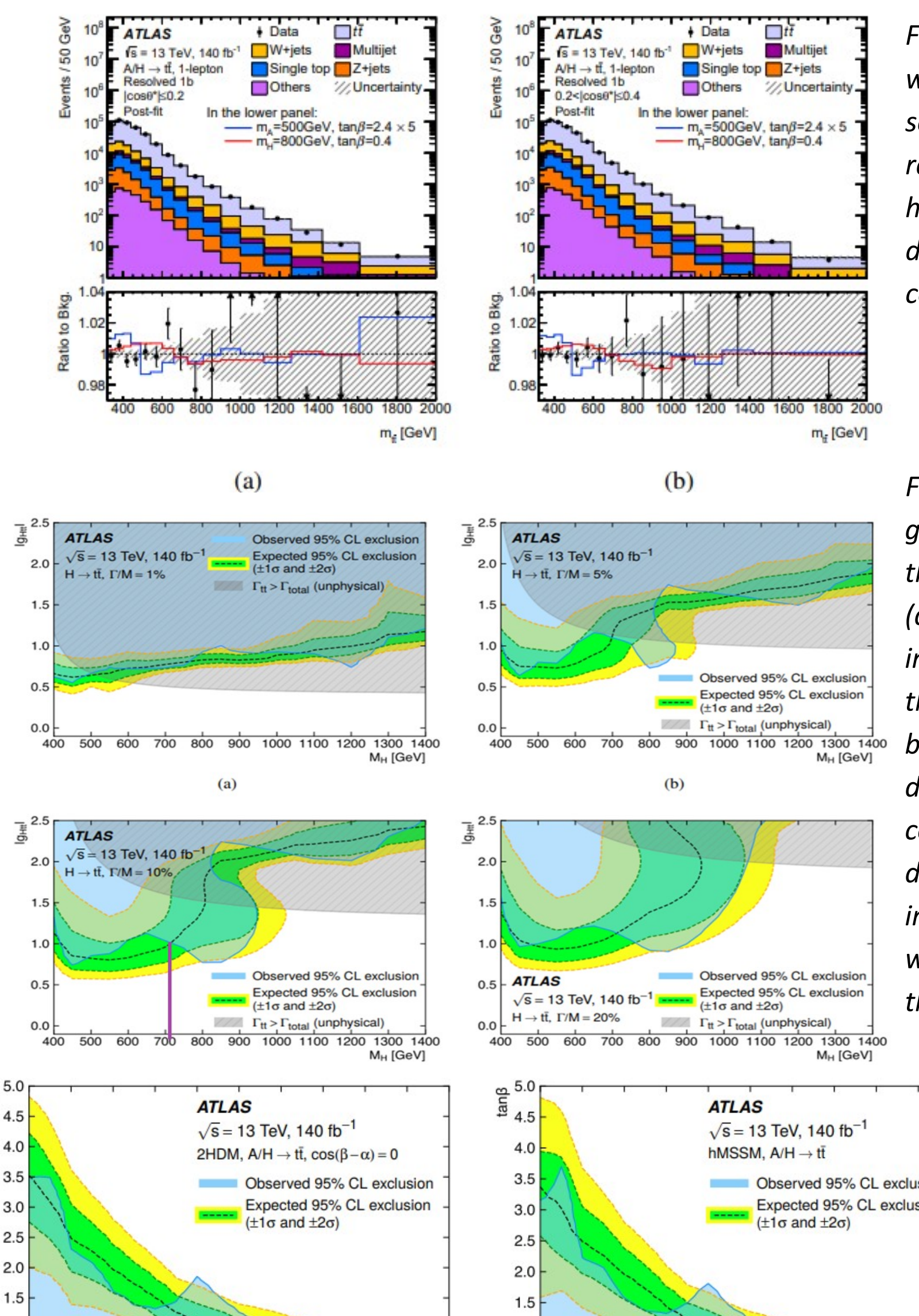

*Figure 16: Mass distribution of the t t̄ events with one lepton and 1 b jet for two angular selections (a) and (b). Figure 16b shows the ratio of the data to the prediction. The blue histogram correspond to the predicted mass distribution for mA=500 GeV, the red one corresponding to mH=800 GeV.*

*Figure 17: Constraints on the coupling strength gHtt as a function of mH for different values of the relative width of H: (a) 1%, (b) 5%, (c) 10%, (d) 20%. The observed exclusion region are indicated by the blue contour. The boundary of the expected exclusion region under the background-only hypothesis is marked by the dashed line. The surrounding shaded bands correspond to the ±1 and ±2 standard deviation uncertainty. The hatched grey area indicates the unphysical region of phase space were the partial width into top pairs is larger than the total width of H.*

*Figure 18: Observed and expected exclusion contours in the mA/H – tan$\beta$ plane for (a) a type-II 2HDM in the alignment limit cos($\alpha-\beta$)=0 with mass degenerate pseudo-scalar and scalar state, mA=mH and (b) the hMSSM. The observed exclusion region are indicated by the shaded area. The boundary of the expected exclusion region under the background-only hypothesis is marked by a dashed line. The surrounding shaded bands correspond to a ±1 and ±2 standard deviation uncertainty.*

## II.2 CMS result

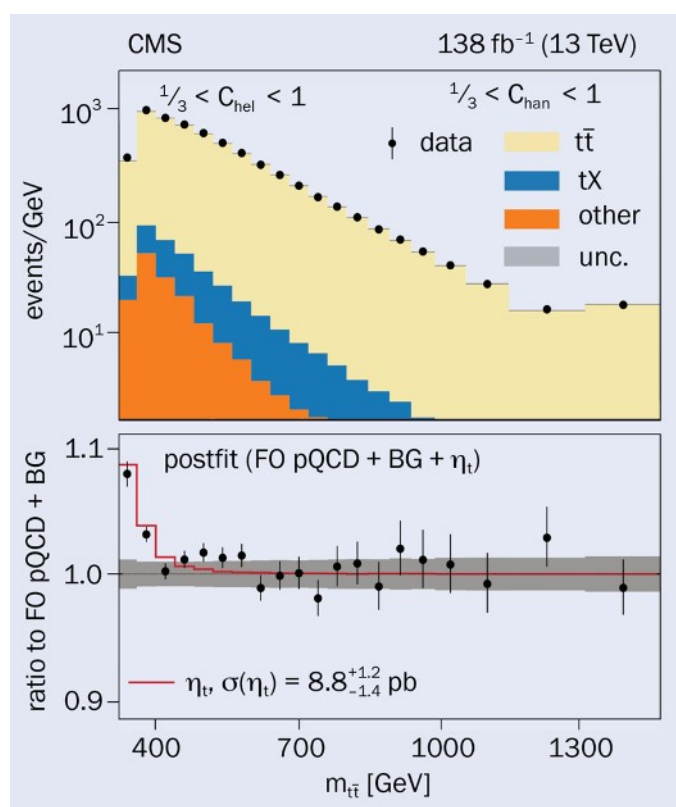

*Figure 19: The ratio of the $\ell^+\ell^-$ data to the background and the expected toponium signal is shown with appropriate angular selections. Adjusting these data one determines the cross section reported on this plot.*

At the EPS conference, ATLAS [7] has fully confirmed the toponium resonance seen by CMS (figure 19), measuring a consistent cross section, **σ=9±1.3 pb**, with an expected oponium cross section of **6.4 pb**, with large uncertainties as pointed out in [76]. There is room for **an extra contribution indistinguishable from the toponium** which, in our interpretation, is unlikely to come from the radion (see Appendix V). Could it come from T380? This seems unlikely given that one expects **σ**tt(380)~**σ**ZZ(380) and that figure 6 indicates that this cross section is of order 100 fb, in spite of the large uncertainties due to interference effects. Reference [42] suggests that there could be and additional **CP-odd resonance**. With more data, these scenarios could be distinguished by using a **VBF selection,** which is not yet available.

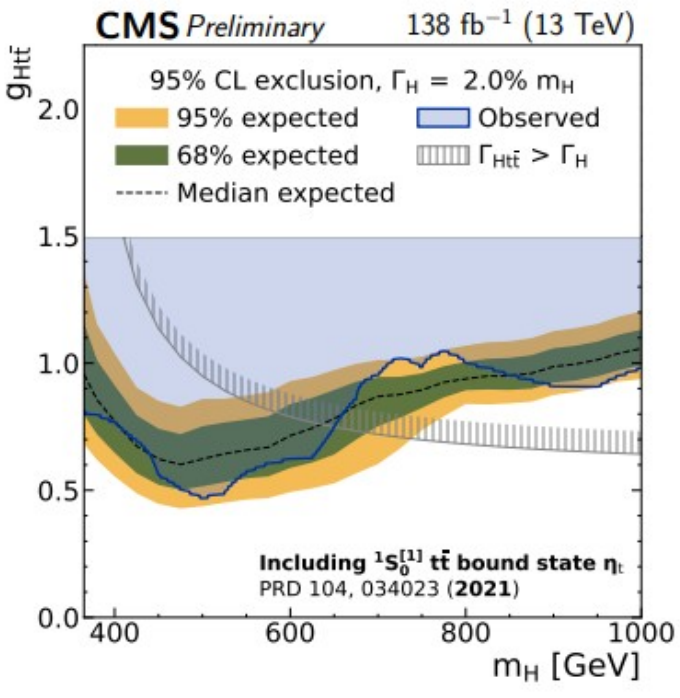

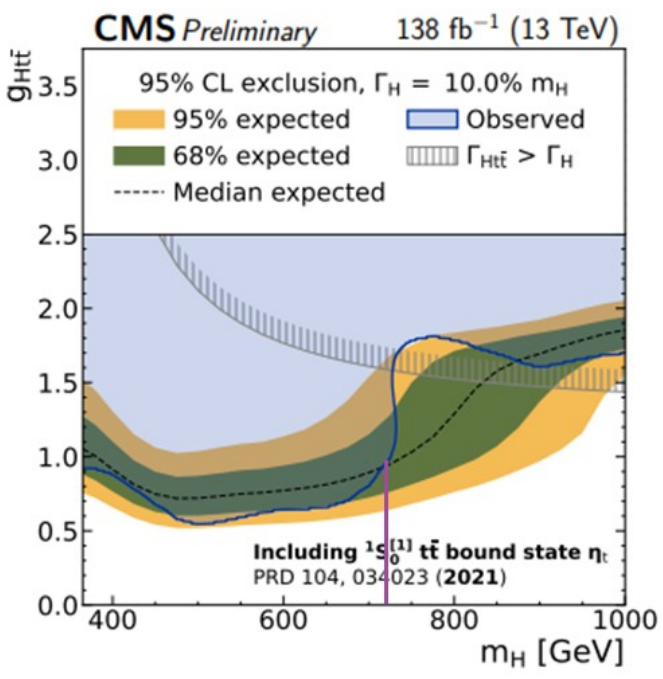

*Figure 20: Model-independent constraints on gHtt as a function of the H mass, for widths of 2% and* 10%.

For the H analysis, figure 20 shows, as expected, **a strong departure** from the predicted limit in the region 700-850 GeV which could have a similar origin to what is found by ATLAS. Indeed, for Γ/MH=10%, the predicted and the effective mass limits diverge in that mass region, crossing slightly above 700 GeV as indicated by the **magenta lines in figure 20**. The slopes of the predicted limits are however opposite which could be due to the fact that CMS only takes positive values for gHtt.

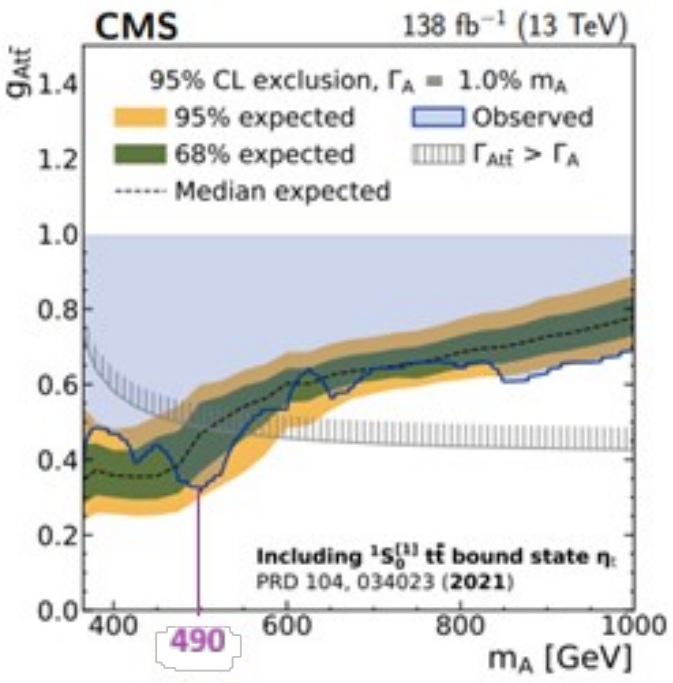

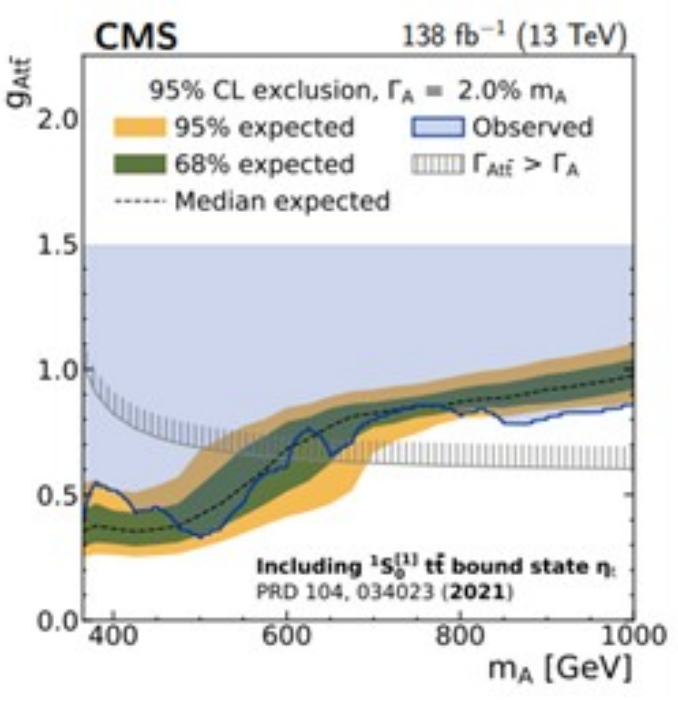

*Figure 21: Model-independent constraints on gAtt as a function of the A mass, for relative widths 1% and 2% . Same as for figure 20.*

For what concerns the A analysis, where the effect of the toponium is taken into account, figure 21 shows a 'deficit' in the mass region near 490 GeV. In this mass region the loop effect is already severe and interference with the QCD background does take place. Again this result may seem marginal but it coincides with ATLAS and with the other LHC indications (see Appendix I).

In conclusion, as expected, the presence of A490 and H650 decaying into top pairs cannot be established using conventional methods but seems well suggested by both ATLAS and CMS.

## II.3 The four top channel and the radion

The four top channel has been measured by ATLAS and CMS. It could come from the following diagram:

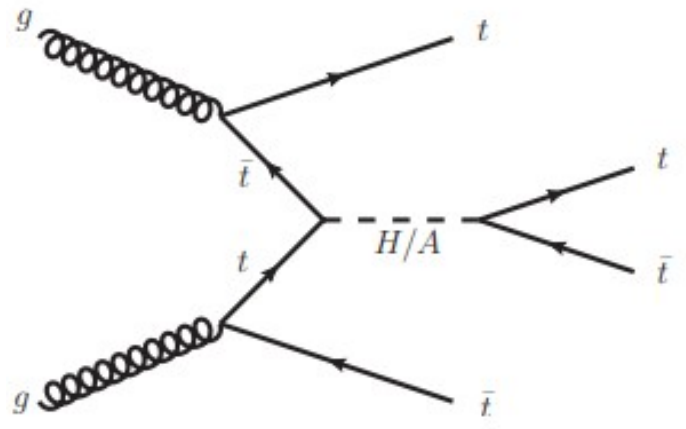

So far we have found that there are 2 pseudo-scalar, one scalar coupled to top pairs and a tensor: Toponium, A490, H650, T380. Even adding these various contributions does not allow to interpret the factor two excess observed by ATLAS and CMS with respect to the SM: 2 s.d. for ATLAS [9] and 1 s.d. [10] for CMS. With more statistics, this effect could become significant.
Promising seems a prediction given in [11], where one assumes a Randall Sundrum type model where the Kaluza Klein gluon $g_{KK}$ has a coupling to right handed top quarks five times larger than the QCD coupling, resulting in a wide resonance observable in the mode $gg \to g_{KK} \to t\bar{t}$ where the gKK is emitted by a **right handed top**:

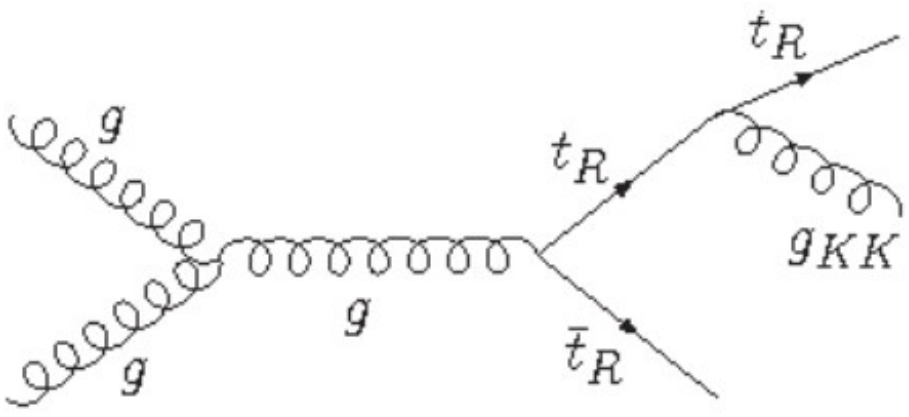

However this reference assumes a 4 TeV resonance, leading to a negligible excess in the four-top channel. CMS has searched[13] for $g_{KK}$ into top pairs [12] and reached a mass limit of 4.55 TeV. This limit can be avoided in an extended version of the RS model [13] where, in the extra dimension, top quarks are 'sequestered' from KK gauge bosons. In this case $g_{KK}$ could be much lighter than 4 TeV, but its coupling to top will be reduced, therefore unlikely to provide an adequate rate for the four-top observed excess.
In the RS model, the KK graviton $G_{KK}$ could go into a **pair of radions**. If these particles decay into top pairs, this would provide an alternative interpretation:

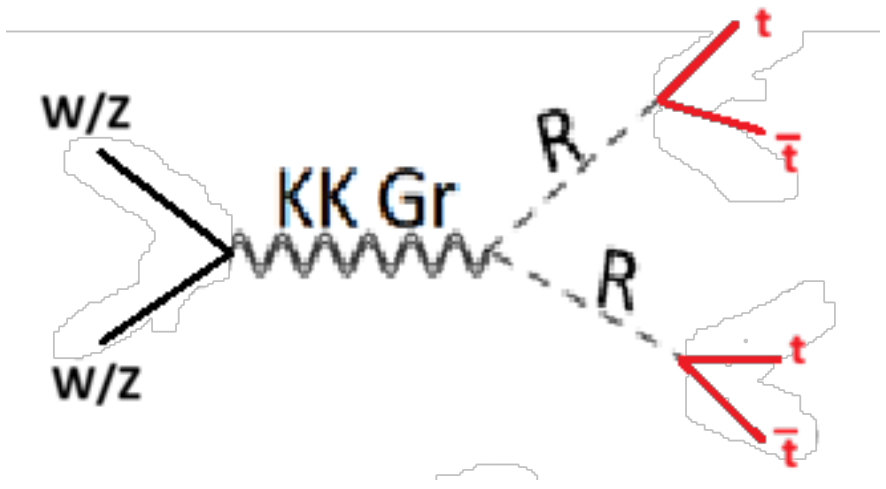

Recall also that RS predicts a sequence of such gravitons (see section I).

13 Many thanks to Yangfan Zhang for pointing out this result

The four top excess shows various distributions which agree with the SM. It seems therefore that in an interpretation by the process $G_{KK}$→rr, the mass of the **radion r** is close to the top pair threshold[14]. If not, one would predict measurable deviations in the variable $H_T$. Also this choice allows to have large contributions from the low rank tensors $G_{KK}$(1000), $G_{KK}$(1300) and $G_{KK}$(1620). Lets us therefore assume mr~400 GeV. One expects that the four top mass will peak just above threshold, say around a TeV.

We need a BSM contribution to the four top channel of about 10 fb. $G_{KK}$(1300) should contribute to about 1/3 of this cross section, that is 3±1.5 fb. This resonance should have a cross section below 180 fb and a total width of 130 GeV (see section II). Reference [13] predicts a width of $G_{KK}$ into rr of 4.7 GeV, therefore a cross section below 6.5 fb, which matches quite well the requirement, provided that r decays with a BR into top pairs of about 60%.

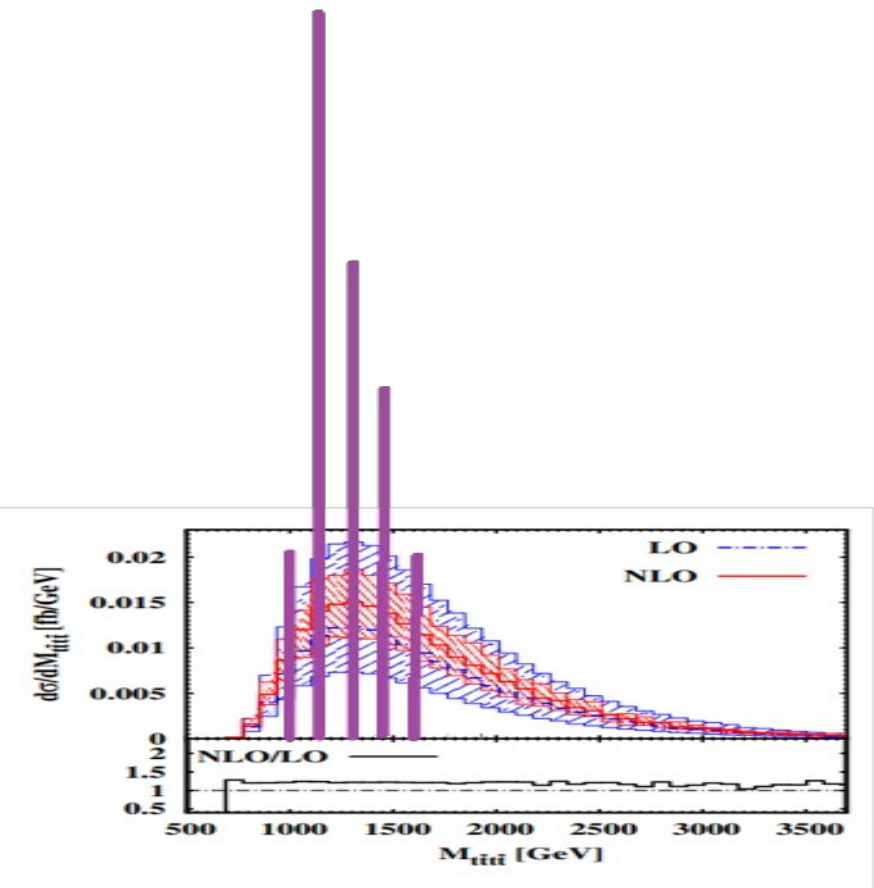

*Figure 22: Expected behaviour of the four top mass distribution in a KK scenario where there is a sequence of graviton excitations which decay into the sequence* $G_{KK}$*→rr→tttt.*

Mass resolution of the four top final state with leptonic selections and combinatorial effects will probably forbid to have direct evidence for the mass pattern displayed in figure 22. The impact of this component will therefore be very similar to the SM component contribution. With increased statistics it could become possible to perform a spin-parity analysis of the four top events to look for an evidence of an excess or tR in the LHC sample as predicted in these models.

## II.5 Consequences for MSSM and NMSSM models

Various observations suggest that there is one extra Higgs doublet with tan$\beta$ close to 1. In Appendix II we develop this aspect and draw the consequences for two SUSY inspired models. From the SM Higgs, MSSM predicts very heavy SUSY particles, NMSSM still allows the possibility for **SUSY discoveries at LHC**. Based on indications for three light scalars candidate, h95 and H/A152 [14] and H+130 [15], we tend to favour a model with 3 doublets, but it is fair to say that some of these evidences, if not all, could still vanish, allowing for an NMSSM interpretation of LHC data for what concerns the scalar sector.

## II.6 SUMMARY on top searches

There are 5 resonances which can be coupled to top pairs: the CP odd toponium, the radion, T700 both CP even and coupled to WW/ZZ, A490 CP odd, H(650) a priori decoupled from WW/ZZ. In the ggF mode they

14 Meaning that it could give an addition to the toponium cross section do not appear as bumps, with the exception of the toponium, but as deviations, positive or negative with respect to the SM background. This is manifest for A490 and H700.
The VBF mode, to our knowledge not yet studied, could bring some surprises.

# III. Consequences for future colliders

Searches from LHC indicate the presence of a **low tanβ solution** for MSSM/NMSSM. In the MSSM framework, this implies that SUSY particles are beyond LHC reach (see Appendix II). Also this framework cannot accommodate part of the scalar spectrum indicated by the data. NMSSM offers better prospects for SUSY discoveries and could also be compatible with part of the spectrum suggested by LHC indications.
e+e- colliders offer a cleaner environment, free from the $t\bar{t}$ QCD background present at LHC, but exploring the heavy doublet will require a centre of mass energy reaching at least 1.5 TeV for:

- Producing e+e- → $H^+$700+$H^-$700
- Producing e+e- → H650+A490

At a centre of mass energy of 1.5 TeV the cross sections for $H^+$700+$H^-$700 and H650+A490 are, respectively, 0.5 and 2.4 fb. Extrapolating ILC at 1.5 GeV one could presumably collect 12000 fb-1 which would allow to measure precisely these channels. CLIC provides a promising scenario with higher energy. /
Such colliders would also allow observing $T^{++}T^{--}$ and $T^+T^-$.
ILC plans to run at 550 GeV to measure Zhh and tth, which allows produce $T^+$375 through $ZW^+$ fusion and T380 through WW/ZZ fusion.
One could run such LC in an e-e- mode which would allow to produce $T^{--}$450 GeV at a similar energy.

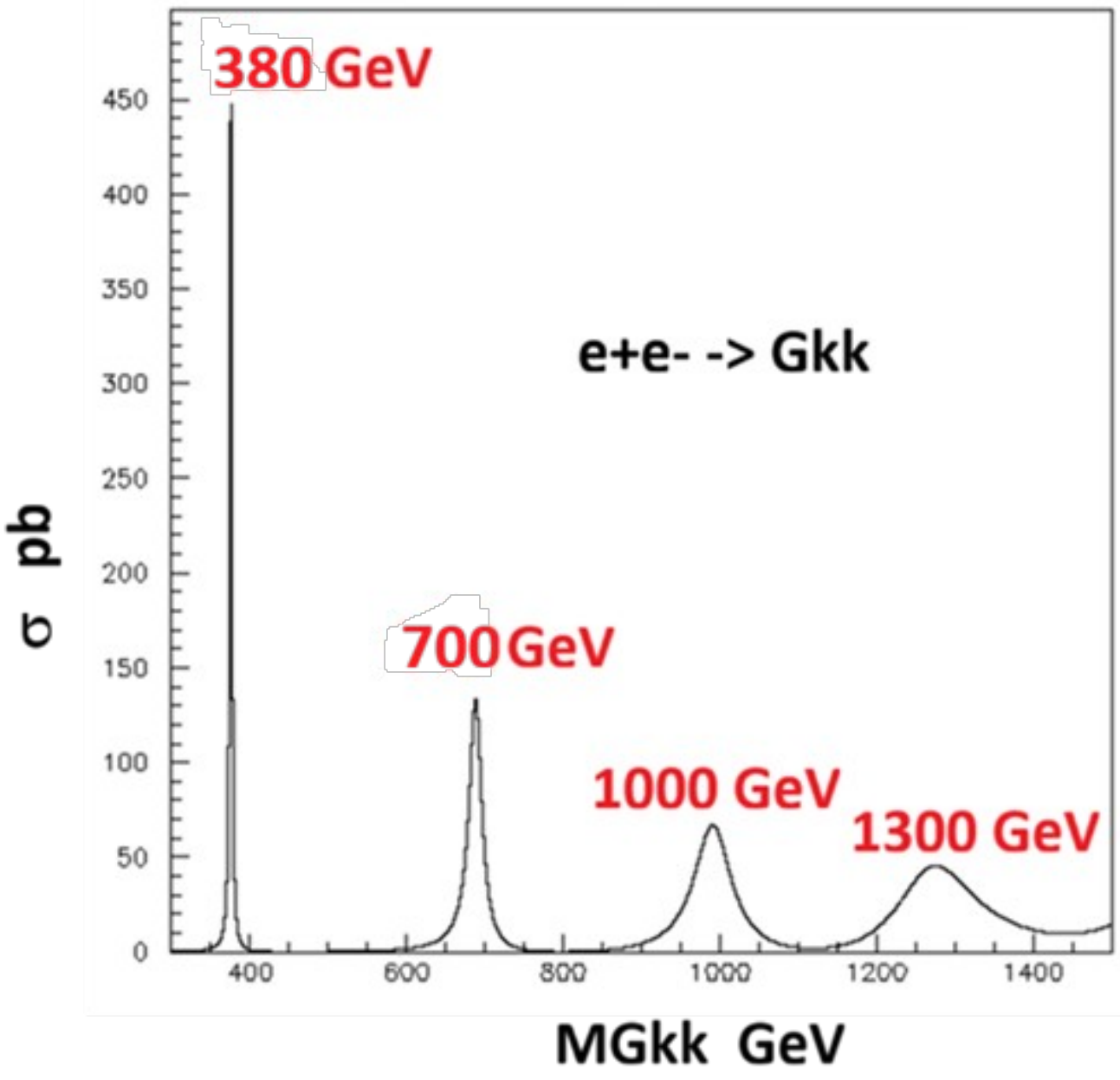

*Figure 23: e+e- cross sections for Kaluza Klein graviton resonances.*

An easier access to BSM physics at e+e- colliders, already suggested in [1], is to produce the **KK graviton resonances** which seem coupled to e+e-. If one assumes BR(e+e-)=0.2%, compatible with present indications, this gives a **cross section of 150 pb** for a centre of mass energy of 700 GeV. This means that a LC collecting 5000 fb-1 at 700 GeV would produce **$10^9$ events!**

A similar scenario could occur for **T380.** So far there is no sign at LHC of a coupling of this resonance to e+e-, which can be understood given the lower coupling and a higher level of background at such masses. σ(e+e-→T380) at resonance could reach **450 pb**.

Similarly, **T1000** is accessible with a TeV LC, providing a third KK graviton factory and an overwhelming proof of the RS model. This resonance is potentially very interesting since it would allow to produce the charged states through the processes **T1000→**$T^+T^-$ **and** $T^{++}T^{--}$.

**T1300** offers an ideal environment to observe the cascade **T1300→radion+radion → tttt**.

These cross sections are about **three orders of magnitude higher than the VBF cross sections at LHC**, with a negligible background and an ideal mass resolution.

Figure 23 gives the expected pattern at the Born term level. Interference effects between resonances are clearly displayed above 1 TeV. For the sake of simplicity, we assume that the various resonances share the same final states, which may not be the case if new channels open up for larger masses.

Observing this type of pattern would clearly constitute a full proof of the graviton interpretation, in particular for what concerns the very narrow state at 380 GeV.

Measuring angular distributions of the fermionic and bosonic final states would give additional clues. In such a scenario, toponium physics could offer new prospects. One would have a situation similar to what was studied in [24] assuming that the top quark mass, unknown in 1985, was below mZ/2, leading to an exploration of toponium physics [24].

# IV. Possible consequences for DM searches

Our scenario assumes that both SUSY and extra dimensions are present. SUSY provides a valid interpretation of dark matter but present mass limits from LHC suggest that **SUSY particles are heavier than expected.** Recently the Fermi-LAT experiment has provided some evidence for photon excesses which can be attributed to annihilations of **neutralinos with masses circa 500 GeV.** Reference [52] has questioned the compatibility of these excesses coming from the halo of our galaxy with the limits given by dwarf galaxies. To reconcile these observations, one would need a **resonance at about 1 TeV** which couples to these neutralinos. It turns out that one of the consequences of our work is that it is not unlikely to find a KK graviton resonance at the right mass which could do the job.

The process: $\chi500\chi500\rightarrow G_{KK}(1000)\rightarrow ZZ/WW/hh$

could explain the origin of the excess of photons around 20 GeV observed by Fermi-LAT given that fragmentation of the hadronic decays of ZZ/WW/hh reproduces the observed energy distributions. Recall also that this resonance could decay into two photons, giving a mass peak around 500 GeV which, however, has not been observed [54]. This is expected given the low branching ratio observed for $G_{KK}(700)\rightarrow\gamma\gamma$, at the % level as shown in figure 1 from [55].

**At LHC** the Fermi-LAT DM candidate could originate from **KK gravitons with masses above 1 TeV**, noting that these particles are produced with a **cross section of 140 fb**. Reference [83] describes possible detection methods which select VBF events with missing transverse energy.

One expects that a **TeV e+e- collider** producing this resonance would be an ideal set up to observe these neutralinos. This contribution could be measured by operating above the T1300 resonance, selecting radiative return to **tag events with invisible energy**. One could also observe the associated charginos, which would be pair produced with a substantial cross section. This would require a **LC operating up to 1.5 TeV.**

Even in a compressed type scenario where the chargino and neutralino masses are very close (see for instance [53]), an e+e- machine will be able to measure this channel. It is not excluded that HL-LHC could observe it. Direct measurement methods do not seem to provide an adequate sensitivity[15].

# Conclusion

A major result of LHC searches is a compelling evidence for a spin 2 resonance with a mass of 700 GeV, which can be interpreted as a **KK graviton**. It seems reminiscent of the initial indication of a 750 GeV resonance seen in two photons by ATLAS. In contrast with the standard RS predictions, we conclude that **this particle does not couple directly to gluons nor photons**. The most spectacular consequence of this interpretation is a dramatic reduction of the mass reach of LHC which is illustrated by figure 34 of Appendix V. While the genuine RS model predicts a RUN2 sensitivity reaching 4.5 TeV for the two photon mode and 1.8 TeV for the ZZ mode into four leptons, we see that, with the present integrated luminosity, LHC experiments barely observe the two photon mode at 700 GeV with 3 s.d.

For the ZZ mode, unfortunate angular selections which assume a scalar resonance prevent CMS from confirming the first evidence obtained by ATLAS with genuine cuts.

The WW mode is observed by CMS, indicating that this resonance is primarily produced through the VBF mechanism, a confirmation of the absence of the ggF mechanism.

The two photon mode has a BR of 0.3% which needs to be understood in terms of loop couplings involving the KK fermions and bosons. A calculation of this type has been already performed for the SM Higgs [59] but is not available for KK gravitons. A quantitative agreement would constitute a strong confirmation of the KK interpretation.

A significant success of this KK hypothesis comes from its prediction of **two neighbouring KK resonances of $G_{KK}$(700): $G_{KK}$(380) and $G_{KK}$(1000),** which are indicated by LHC data. The higher part of this sequence is indicated **up to 3 TeV** by analyses performed by ATLAS and CMS by selecting VV final states in hadronic modes (see figure 24 and 25 of Appendix I). Confirming such sequences should bring **the ultimate proof in favour of the RS graviton hypothesis**.

**$G_{KK}$(700)** is observed in **6 other channels** which allows to construct a global evidence reaching **6 s.d.**

The **e+e- channel**, observed both by ATLAS and CMS offers a very promising future for e+e- colliders reaching 1 TeV which could become **giga graviton factories** producing the 3 RS resonances at 380 GeV, 700 GeV and 1000 GeV centre of mass energies.

There is indirect evidence for **Z'/$Z_{KK}$ particles** from SLC/LEP precision measurements which predict a mass scale of order 2.5TeV/sinθ', where sinθ' is a mixing parameter >0.1. This interpretation relies on the BR into e+e-, of order 0.25%, indicated by ATLAS and CMS for the graviton resonance.

Evidence for **charged partners of $G_{KK}$(380)** forming an I=2 multiplet indicates a mechanism allowing to increase the domain of validity of **perturbative unitarity for the WW/ZZ->WW/ZZ process**.

15 ArXiv paper in preparation

In this note we have in addition presented various issues concerning the searches for BSM resonances at LHC. The main message of our work is that there could be a **very rich BSM spectroscopy at LHC** which requires taking care of several subtle effects (see also [56], [61] and [69]) which have been progressively identified, in particular:

- The impact of top loop corrections in the **ggF process** which require a reinterpretations of the various limits achieved so far, in particular in the **top pair channel** where interferences with the QCD contribution for high masses becomes imaginary and prevents observing resonances with conventional criteria
- **VBF,** which is still missing for top pair final states, partially escapes to these problems except for what concerns angular selections for **spin 2 resonances** which have marked differences with respect to the scalar resonances as shown in Appendix III

The **most solid result** of the $t\bar{t}$ searches is clearly the evidence for a **toponium type resonance**. This resonance escapes to the loop effect in ggF which is absent at threshold. It is however true that, given the mass resolution, one cannot prove that this resonance is **narrow**. At the moment the two experiments put different emphasis on $t\bar{t}$. CMS being mainly focussed on the identification of a **toponium** state while ATLAS deals with a search for H/A, the heavy scalars predicted by the 2HD model. Both ATLAS and CMS observe a departure between the predicted and the observed limits on the coupling **gH*tt*** for masses above 700 GeV, which agrees with our observations [1]. It is however fair to say that, in detail, these mass limits do not coincide. These examples demonstrate the **complexity of the $t\bar{t}$ channel**, where various effects need to be controlled to achieve meaningful results. The lesson of this exercise is quite clear: due to the loop effect in ggF, combined with interference with the background, **naïve bump hunting** for heavy resonances, which has been the current practice, is clearly not appropriate (see also [27] and [28]).

Our present conclusion is that the **$t\bar{t}$ channel presently provides 4 indications**: the **toponium** resonance, **A490, H650**, without forgetting the **radion** indicated in the four top channel.

For the scalar sector, various interpretations are possible. An NMSSM interpretation is not excluded even if the mass hierarchy of H650 and A490 comes as a surprise. NMSSM predicts a SUSY mass scale detectable at LHC. The **Scherk-Schwarz SUSY breaking mechanism** could be at work as suggested by the KK graviton tensors.
RUN2+RUN3 have already delivered 500 fb-1, meaning that with adequate selections it should be possible to reach a significance on ZZ in four leptons above 5 s.d. per experiment. **We very much hope that the ZZ analysis will be done assuming a J=2 resonance and isolating VBF.**

Using the various decay channels of the KK graviton resonances, ZZ/WW into leptons, hh sopphisticated selections, ZZ/WW in semi-leptonics and ZZ/WW hadronic modes, LHC should bee able to cover a very wide spectrum and give the ultimate confirmation of the presence of heavy gravitons.

**Acknowledgements:** *Useful and pleasant exchanges with* ***Abdelhak Djouadi, Ulrich Ellwanger, Adam Falkowski*** *and* ***Gilbert Moultaka*** *are gratefully acknowledged. Thanks to* ***Severin Lüst*** *for suggesting an essential new mechanism occurring in the Randall Sundrum model. More recent exchanges with* ***Veronica Sanz*** *are gratefully acknowledged. This work has greatly benefitted from discussions at Higgs Hunting 2025, in particular with* ***Mathieu Pellen, Benjamin Fuks, Gautier Hamel de Monchenault****,* ***Yangfan Zhang*** *and* ***Wanda Isnard.*** *Exchange of ideas about BSM searches at LHC with* ***Andreas Höcker, Guillaume Unal*** *and* ***Louis Fayard*** *are gratefully acknowledged. Thanks to* ***Roman Poschl*** *for active reading of our work and useful inputs on the four top sector and on T650→* $\tau\tau\gamma\gamma$.
*Thanks to* ***Akira Miyazaki*** *for drawing our attention about the paper of Hitoshi Murayama about DM and to* ***François Couchot*** *for useful discussions on this topic. Thanks to* ***Daniel Treille*** *for a critical reading of our work.*
*We are grateful to* ***Yves Sirois*** *for providing useful infos on the* **LHC BSM Working Group.**

**References:**

*[1]*New resonances at LHC *Alain Le Yaouanc(IJCLab, Orsay), François Richard(IJCLab, Orsay*
*e-Print: 2506.094*
*Recent oral presentations:*

- *3rd ECFA workshop on e+e- Higgs, Electroweak and Top Factories  e-Print: 2408.12178*
- *LCWS 2025 Valencia https://agenda.linearcollider.org/event/10594/sessions/5644/#20251021*
- IRN Terascale workshop in Orsay, April 2026
  https://indico.in2p3.fr/event/38920/contributions/171801/

*[2]Observation of a pseudoscalar excess at the top quark pair production threshold*
*CMS Collaboration Aram Hayrapetyan(Yerevan Phys. Inst.)et al.(Mar 28, 2025) e-Print: 2503.22382*
*see also: Search for heavy neutral Higgs bosons decaying into a top quark pair in 140 $fb^{-1}$ of proton-proton collision data at CMS    CMS PAS HIGG-22-013*
*[3]Bound-state effects on kinematical distributions of top quarks at hadron colliders*
*Y. Sumino and H. Yokoya, , JHEP 09 (2010) 034, doi:10.1007/JHEP09(2010)034,*
*arXiv:1007.0075. [Erratum: doi:10.1007/JHEP06(2016)037]*
*[4]Toponium physics at the Large Hadron Collider*
*Benjamin Fuks(Paris, LPTHE)(May 6, 2025)*
*Contribution to: Moriond EW 2025 e-Print: 2505.03869*
*[5]Contrasting pseudoscalar Higgs and toponium states at the LHC and beyond*
*Abdelhak Djouadi(Annecy, LAPP and Granada U., Theor. Phys. Astrophys.), John Ellis(King's Coll. London and CERN), Jeremie Quevillon(Annecy, LAPP)(Dec 19, 2024)*
*Published in: Phys.Lett.B 866 (2025) 139583 e-Print: 2412.1513*
*[6] Search for heavy neutral Higgs bosons decaying into a top quark pair in 140 fb-1 of proton-proton collision data at sqrt(s)= 13 TeV with the ATLAS detector*
*ATLAS Collaboration Georges Aad(Marseille, CPPM)et al. (Apr 29, 2024)*
*Published in: JHEP 08 (2024) 013   e-Print: 2404.18986*
*[7]Search for heavy pseudoscalar and scalar bosons decaying to a top quark pair in proton-proton collisions at sqrt(s)=13 TeV*
*CMS Collaboration Aram Hayrapetyan(Yerevan Phys. Inst.) et al.(Jul 7, 2025)  e-print:2507.05119*
*Observation of a cross-section enhancement near the $t\bar{t}$ production threshold in sqrt(s)=13 TeV pp collisions with the ATLAS detector*
*ATLAS Collaboration Georges Aad(Marseille, CPPM) et al. (Jan 16, 2026)  e-Print: 2601.11780*
*[8]The Anatomy of electro-weak symmetry breaking. I: The Higgs boson in the standard model*
*Abdelhak Djouadi(Montpellier U.and Orsay, LPT)(Mar, 2005)*
*Published in: Phys.Rept. 457 (2008) 1-216 e-Print: hep-ph/0503172*
*The Anatomy of electro-weak symmetry breaking. II. The Higgs bosons in the minimal supersymmetric model*
*(Montpellier U. and Orsay, LPT) (Mar, 2005)*
*Published in: Phys.Rept. 459 (2008) 1-241 e-Print: hep-ph/0503173*
*[9]Observation of four top quark production in proton-proton collisions at s=13TeV*
*CMS Collaboration Aram Hayrapetyan(Yerevan Phys. Inst.)et al. (May 22, 2023)*
*Published in: Phys.Lett.B 847 (2023) 138290   e-Print: 2305.13439*
*[10]Observation of four-top-quark production in the multilepton final state with the ATLAS detector*
*ATLAS Collaboration Georges Aad(Marseille, CPPM) et al.(Mar 27, 2023)*
*Published in: Eur.Phys.J.C 83 (2023) 6, 496, Eur.Phys.J.C 84 (2024) 156 (erratum)*
*e-Print: 2303.15061*
*[11]Boosted four-top production at the LHC: A window to Randall-Sundrum or extended colour symmetry*
*Debajyoti Choudhury(Delhi U.), Kuldeep Deka(Delhi U. and New York U., Abu Dhabi), Lalit Kumar Saini(Delhi U. and SGTB Khalsa Coll. and HBNI, Mumbai and Bhubaneswar, Inst. Phys.) (Apr 5, 2024)*
*Published in: Phys.Rev.D 110 (2024) 7, 075020 e-Print: 2404.04409*
*[12]Search for resonant tt production in proton-proton collisions at sqrt(s)=13 TeV*
*CMS Collaboration Albert M Sirunyan(Yerevan Phys. Inst.) et al. (Oct 13, 2018)*
*Published in: JHEP 04 (2019) 031 e-Print: 1810.05905  4.55 TeV !*
*[13]LHC Signals for KK Graviton from an Extended Warped Extra Dimension*
*Kaustubh Agashe(Maryland U.), Majid Ekhterachian(Maryland U.), Doojin Kim(Texas A-M), Deepak Sathyan(Maryland U.) (Aug 14, 2020)*
*Published in: JHEP 11 (2020) 109 e-Print:  2008.06480*

*[14]Combined explanation of LHC multilepton, diphoton, and top-quark excesses*
*Guglielmo Coloretti (Zurich U. ,PSI Villigen), Andreas Crivellin(Zurich U. and PSI, Villigen),*
*Bruce Mellado(U. Witwatersrand, Johannesburg, Sch. Phys. and iThemba LABS)Dec 28, 2023*
*Phys.Rev.D 110 (2024) 7, 073001 e-Print: 2312.17314*
*[15]Search for a light charged Higgs boson in t→H±b, with H±→cb, in the lepton+jets final state in proton-proton collisions at sqrt(s)=13 TeV with the ATLAS detector*
*ATLAS Collaboration Georges Aad(Marseille,CPPM) et al. (Feb 22, 2023)*
*Published in: JHEP 09 (2023) 004 e-Print: 2302.11739*
*[16] Warped Gravitons at the LHC and Beyond*
*K. Agashe (Syracuse), H. Davoudiasl (Brookhaven), G. Perez (Stony Brook), A. Soni (Brookhaven)(July, 2007)*
*Published in: Phys.Rev. D76:036006,2007, e-print: hep-ph/0701186*
*[17]Search for Higgs boson pair production in association with a vector boson in pp collisions at sqrt(s)=13TeV with the ATLAS detector*
*ATLAS Collaboration Georges Aad(Marseille, CPPM) et al. (Oct 11, 2022)*
*Published in: Eur.Phys.J.C 83 (2023) 6, 519, Eur.Phys.J.C 83 (2023) 519 e-Print: 2210.05415*
*[18]Combination of Searches for Resonant Higgs Boson Pair Production Using pp Collisions at sqrt(s)=13 TeV with the ATLAS Detector*
*ATLAS Collaboration Georg Aad(Marselle, CCPM)et al (Nov 27, 2023)*
*Published in: Phys.Rev.Lett. 132 (2024) 23, 231801 e-Print: 2311.15956*
*[19]Combination of searches for singly and doubly charged Higgs bosons produced via vector-boson fusion in proton–proton collisions at sqrt(s)=13 TeV with the ATLAS detector*
*ATLAS Collaboration Georges Aad(Marseille, CPPM)et al. (Jul 15, 2024)*
*Published in: Phys.Lett.B 860 (2025) 139137e-Print: 2407.10798*
*[20] Search for charged Higgs bosons produced in vector boson fusion processes and decaying into vector boson pairs in proton–proton collisions at sqrt(s)=13 TeV*
*CMS Collaboration Albert M Sirunyan(Yerevan Phys. Inst.) et al. (Apr 10, 2021)*
*Published in: Eur.Phys.J.C 81 (2021) 8, 723 e-Print: 2104.04762*
*[21]Combination of searches for heavy spin 1 resonances using 139 fb−1 of proton-proton collision data at sqrt(s)= 13 TeV with the ATLAS detector*
*ATLAS Collaboration Georges Aad(Marseille, CPPM)et al. (Feb 16, 2024)*
*Published in: JHEP 04 (2024) 118 e-Print: 2402.10607*
*[22]It is a Graviton! or maybe notRicky Fok (York U., Canada), Carol Guimaraes(York U., Canada), Randy Lewis (York U., Canada), Veronica Sanz(York U., Canada and CERN) (Mar, 2012)*
*Published in: JHEP 12 (2012) 062 e-Print: 1203.2917*
*[23]Enhanced Higgs Mass in Compact Supersymmetry*
*Kohsaku Tobioka(Tel Aviv U. and Weizmann Inst. and KEK, Tsukuba), Ryuichiro Kitano(Tokyo U., IPMU and KEK, Tsukuba and Sokendai, Tsukuba), Hitoshi Murayama(LBL, Berkeley and Tokyo U., IPMU and UC, Berkeley)(Nov 12, 2015)*
*Published in: JHEP 04 (2016) 025 e-Print: 1511.04081*
*[24]TOPONIUM PHYSICS AT LEP*
*W. Buchmuller(CERN), Andre Martin(CERN), Johann H. Kuhn(Munich, Max Planck Inst.), F. Richard(Orsay, LAL), P. Roudeau(Orsay, LAL)et al.(Dec, 1985) Contribution to: LEP Jamboree*
*https://lib-extopc.kek.jp/preprints/PDF/1986/8604/8604083.pdf*
*[25]Toponia at the HL-LHC, CEPC, and FCC-ee*
*Yang Bai(Wisconsin U., Madison and Argonne), Ting-Kuo Chen(Wisconsin U., Madison), Yiming Yang(Wisconsin U., Madison)(Jun 17, 2025) e-Print: 2506.14552*
*[26]Complementarity between ILC250 and ILC-GigaZ*
*A. Irles (Orsay, LAL), R. Pöschl (Orsay, LAL), F. Richard (Orsay, LAL), H. Yamamoto(Tohoku U.)*
*(May 1, 2019) e-Print:1905.00220*
*[27]Higgs pair production in the 2HDM: impact of loop corrections to the trilinear Higgs couplings and interference effects on experimental limits*
*S. Heinemeyer (Madrid, IFT), M. Mühlleitner (KIT, Karlsruhe, TP), K. Radchenko (DESY), G. Weiglein (Desy and Hambirg U., Inst. Theor. Phys. II) (Mars 21, 2024)*
*Published in: Eur.Phys.J.C 85 (2025) 4, 437 e-Print: 2403.14776*
*[28]Deconstructing resonant Higgs pair production at the LHC: effects of coloured and neutral scalars in the NMSSM test case*
*Stefano Moretti(Southampton U. and Uppsala U.), Luca Panizzi(Calabria U.), (Stockholm U.)Jörgen Sjölin, Harri Waltari(Uppsala U.) (Jun 10, 2025) e-Print: 2506.09006*

[29] Search for electroweak-scale dijet resonances using trigger-level analysis with the ATLAS detector in 132 fb−1 of pp collisions at 13V TeV
ATLAS Collaboration Georges Aad(Marseille, CPPM) et al. (Sep 1, 2025)
e-Print: 2509.01219
[30]Search for the nonresonant and resonant production of a Higgs boson in association with an additional scalar boson in the $\gamma\gamma\tau\tau$ final state in proton-proton collisions at sqrt(s)= 13 TeV
CMS Collaboration Aram Hayrapetyan(Yerevan Phys. Inst.)
et al. (Jun 28, 2025) e-Print: 2506.23012
[31]Search for heavy resonances decaying into a W or Z boson and a Higgs boson in final states with leptons and b-jets in 36 fb−1 at sqrt(s)=13 TeV in pp collisions with the ATLAS detector
ATLAS Collaboration Morad Aaboud (Oujda U.)et al.(Dec 18, 2017)
Published in: JHEP 03 (2018) 174, JHEP 11 (2018) 051 (erratum) e-Print:1712.06518
Search for a CP-odd Higgs boson decaying to Zh in pp collisions at sqrt(s) = 13 TeV with the ATLAS detector
ATLAS collaboration Mar 23, 2016 34 pages
Report number: ATLAS-CONF-2016-015
[32]Search for a heavy scalar boson decaying into a Higgs boson and a new scalar particle in the four b-quarks final state using proton-proton collisions at sqrt(s)=13 TeV CMS-PAS-HIG-20-012
see also : Additional scalar bosons - CMS Orateur: Efe Yazgan (National Taiwan University)
https://indico.ijclab.in2p3.fr/event/10259/timetable/#20240923.detailed
[33]MSSM Higgs benchmark scenarios for Run 2 and beyond: the low $\tan\beta$ region
Henning Bahl(DESY), Stefan Liebler(KIT, Karlsruhe,TP), Tim Stefaniak(DESY)(Jan 17, 2019)
Published in: Eur.Phys.J.C 79 (2019) 3, 279 e-Print: 1901.0
[34]Extended Higgs Sector: 2HDM, MSSM and NMSSM
E. Boos(SINP, Moscow), R.Nevzorov (Moscow, ITEP )(2016) Contribution to: LHCP 2015, 81-86
[35]Additional Higgs Bosons near 95 and 650 GeV in the NMSSM
Ulrich Ellwanger (IJCLab, Orsay), Cyril Hugonie (U. Montpellier 2, LUPM) (Sep 14, 2023)
Published in: Eur.Phys.J.C 83 (2023) 12, 1138 e-Print: 2309.07838
[36]Quantum gravity and extra dimensions at high-energy colliders
Gian F. Giudice(CERN), Riccardo Rattazzi(CERN), James Wells(CERN)(Nov, 1998) Published in: Nucl.Phys.B 544 (1999) 338 e-Print:hep-ph/9811291
[37]Resonant Diphoton Phenomenology Simplified
Giuliano Panico(Barcelona, IFAE), Luca Vecchi(SISSA, Trieste and Padua U. and INFN, Padua), Andrea Wulzer(INFN, Padua and Padua U.)(Mar 14, 2016)
Published in:JHEP 06 (2016) 184 e-Print:1603.04248
[38]LHC constraints on hidden gravitons
J.A. R. Cembranos(Madrid U.), R.L. Delgado(Madrid, Polytechnic U. and Munich, Tech. U. and INFN, Florence),
H. Villarrubia-Rojo(Zurich, ETH)(Aug 2, 2021)
Published in: JHEP 01 (2022) 129 e-Print: 2108.00930
[39]Tau polarization and Randall-Sundrum scenario at e+ e- colliders
Namit Mahajan(Delhi U.)(Jul, 2002)
Published in: J.Phys.G 29 (2003) 2677-2684 e-Print: hep-ph/0207065
[40]KK gravitons and unitarity violation in the Randall-Sundrum model
Bohdan Grzadkowski(Warsaw U.), John F. Gunion(UC, Davis)(Oct, 2006)
Published in: Phys.Lett.B 653 (2007) 307-319 e-Print: hep-ph/0610105
[41]Higgs boson: The scalar/tensor case
olfgang Kilian(Siegen U.), Thorsten Ohl(Wurzburg U.), Jürgen Reuter(DESY), Marco Sekulla(Siegen U. and KEK, Tsukuba and KIT, Karlsruhe)(Oct 30, 2015)
Published in: Phys.Rev.D 93 (2016) 3, 036004 e-Print: 1511.00022
[42]New physics in toponium's shadow?
Thomas Flacke(Korea Inst. Advanced Study, Seoul), Benjamin Fuks(Paris, LPTHE), Dongchan Kim(Korea Inst. Advanced Study, Seoul and Korea U.), Jinheung Kim Seung J. Lee(
Korea Inst. Advanced Study, Seoul) et al.(Dec 2, 2025) e-Print: 2512.03220
[43]Slight excess at 130 GeV in search for a charged Higgs boson decaying to a charm quark and a bottom quark at the Large Hadron Collider
[A.G. Akeroyd(Southampton U.), Stefano Moretti(Southampton U. and Uppsala U.), Muyuan Song(Southampton U. and Peking U., CHEP)(Feb 7, 2022)
Published in: J.Phys.G 49 (2022) 8, 085004 e-Print: 2202.03522

[44] Search for dijet resonances in events with an isolated charged lepton using sqrt(s)=13 TeV proton-
proton collision data collected by the ATLAS detector
ATLAS Collaboration (Georges Aad (Marseille, CPPM) et al.). Feb 26, 2020. 41 pp.
CERN-EP-2019-276 e-Print: arXiv:2002.11325
[45]Emerging Narrow Resonance at 152 GeV
Srimoy Bhattacharya, Mukesh Kumar, Rachid Mazini, Bruce Mellado
(Nov 6, 2025)
Contribution to: DIS2025 e-Print: 2511.04335
[46]S. Weinberg, Gauge Theory of CP violation, Phys. Rev. Lett, 37 (1976) 657
[47]More Effective RS Field Theory
Severin Lüst(Paris, LPTHE), Michael Nee(Harvard U., Phys. Dept.), Lisa Randall(Harvard U., Phys. Dept.)(Oct 13, 2025)
e-Print: 2510.1177
[48]Zwirner, Fabio (CH-1211 Geneva 23) Publication 2004 In: Pramana - J. Phys. 62 (2004) 497-511
DOI 10.1007/BF02705104 https://www.ias.ac.in/article/fulltext/pram/062/02/0497-0511
[49]CMS collaboration, Search for heavy scalar resonances decaying to a pair of Z bosons in the
4-lepton final state at 13 TeV, CMS Physics Analysis Summary CMS-PAS-HIG-24-002, CERN, Geneva (2024)
[50]Search for dijet resonances in events with an isolated charged lepton using sqrt(s)=13 TeV
proton-proton collision data collected by the ATLAS detector
ATLAS Collaboration Georges Aad(Marseille, CPPM) et al. (Feb 26, 2020)
Published in JHEP 06 (2020) 151 e-Print: 2002.11325
[51]Sum rules for Higgs bosons
J.F. Gunion(Santa Barbara, KITP and UC, Davis), H.E. Haber(Santa Barbara, KITP and UC, Santa Cruz),
J. Wudka(UC, Davis) (Jul, 1990)
Published in: Phys.Rev.D 43 (1991) 904-912
[52]Resonant Annihilation of WIMP Dark Matter for Halo Gamma Ray Signal Hitoshi Murayama(UC, Berkeley and
Tokyo U., IPMU and LBNL, Berkeley) (Dec 1, 2025) e-Print: 2512.01404
[53]Non-universal SUSY models, (g−2)μ, mH and dark matter
John Ellis(King's Coll. London and CERN), Keith A. Olive(Minnesota U.), Vassilis C. Spanos(Minnesota U. and Athens U.)
(Jul 11, 2024)
Published in: Eur.Phys.J.C 84 (2024) 10, 1121 e-Print: 2407.08679
[54]20 GeV halo-like excess of the Galactic diffuse emission and implications for dark matter annihilationTomonori
Totani(Tokyo U., ICEPP and Tokyo U., RESCEU)(Jul 9, 2025) Published in: JCAP 11 (2025) 080 e-Print: 2507.07209
[55]Search for resonances decaying into photon pairs in 139 fb−1 of pp collisions at sqrt(s)=13 TeV with the ATLAS
detector
ATLAS Collaboration Georges Aad(Marseille, CPPM) et al. (Feb 26, 2021)
Published in: Phys.Lett.B 822 (2021) 136651 e-Print: 2102.13405
Search for new physics in high-mass diphoton events from proton-proton collisions at ss = 13 TeV
CMS Collaboration Aram Hayrapetyan(Yerevan Phys. Inst.)et al. (May 15, 2024)
Published in: JHEP 08 (2024) 215 e-Print: 2405.09320
[56]Additional evidence of a new 700 GeV scalar resonance
M. Consoli(INFN, Catania), L. Cosmai (INFN, Bari), F. Fabbri (INFN, Bologna), G. Rupp (IST, Lisbon (main))
(Jan 7, 2025) e-Print: 2501.03708
[57]A Large mass hierarchy from a small extra dimension
Lisa Randall(Princeton U. and MIT) and Raman Sundrum(Boston U.)(May 4, 1999)
Published in: Phys.Rev.Lett. 83 (1999) 3370-3373 e-Print: hep-ph/9905221
[58] Probes of localized gravity: On and off the wall
H. Davoudiasl(SLAC), J.L. Hewett(SLAC), T.G. Rizzo(SLAC)(Jun, 2000)
Published in: Phys.Rev.D 63 (2001) 075004 e-Print: hep-ph/0006041
[59]The Custodial Randall-Sundrum Model: From Precision Tests to Higgs Physics
Sandro Casagrande(Munich, Tech. U., Universe), Florian Goertz(Mainz U., Inst. Phys.), Uli Haisch(Mainz U., Inst. Phys.),
Matthias Neubert(Mainz U., Inst. Phys. and Heidelberg U.), Torsten Pfoh(Mainz U., Inst. Phys.)(May, 2010)
Published in: JHEP 09 (2010) 014 e-Print: 1005.4315
[60]Warped Compactifications in Particle Physics, Cosmology and Quantum Gravity
Prateek Agrawal(Oxford U., Theor. Phys.), Cari Cesarotti(Harvard U.), Andreas Karch(Texas U.), Rashmish K.
Mishra(Harvard U.), Lisa Randall(Harvard U.)et al.(Mar 14, 2022)
Contribution to: Snowmass 2021 e-Print: 2203.0753

[61]Interference effects in new physics searches
Tania Robens (Boskovic Inst., Zagreb)(Jan 30, 2026)e-Print:26 02.00256
[62]Search for High Mass Resonances Decaying into W+W− in the Dileptonic Final State with 138 fb−1fb−1 of Proton-Proton Collisions at sqrt(s)=13 TeV
CMS Collaboration Sadhana Verma(Indian Inst. Tech., Madras)for the collaboration. (Jul 17, 2024)
Published in: Springer Proc.Phys. 304 (2024) 1168-1170 CMS-PAS-HIG-20-016
Contribution to: 25th DAE-BRNS High Energy Physics Symposium, 1168-1170
[63]Collider constraints on massive gravitons coupling to photons
David d'Enterria(CERN), Malak Ait Tamlihat(Rabat U.), Laurent Schoeffel(IRFU, Saclay), Hua-Sheng Shao(Paris, LPTHE), Yahya Tayalati(Rabat.U and Unlisted, MA)(Jun 27, 2023)
Published in: Phys.Lett.B 846 (2023) 138237 e-Print: 2306.15558
[64]Kaluza-Klein gravitons at LHC2Barry M. Dillon(Sussex U.), Veronica Sanz(Sussex U.)(Mar 31, 2016)
Published in: Phys.Rev.D 96 (2017) 3, 035008 e-Print: 1603.09550
[65]Gravity-mediated (or Composite) Dark MatterHyun Min Lee(Chung-Ang U.), Myeonghun Park(CERN), Veronica Sanz(York U., Canada and Sussex U., Astron. Ctr.) (Jun 18, 2013)
Published in: Eur.Phys.J.C 74 (2014) 2715 e-Print: 1306.4107
[66]Phenomenology of a stabilized modulus
Walter D. Goldberger(Caltech), Mark B. Wise(Caltech) (Nov, 1999)
Published in: Phys.Lett.B 475 (2000) 275-279 e-Print: hep-ph/9911457
Modulus stabilization with bulk fields
Walter D. Goldberger(Caltech), Mark B. Wise(Caltech)(Jul, 1999)
Published in: Phys.Rev.Lett. 83 (1999) 4922-4925 e-Print: hep-ph/9907447
[67]TASI lectures on extra dimensions and branes
Csaba Csaki(Cornell U., LEPP)(Apr, 2004)
Contribution to: Theoretical Advanced Study Institute in Elementary Particle Physics (TASI 2002): Particle Physics and Cosmology: The Quest for Physics Beyond the Standard Model(s), 605-698
e-Print: ep-ph/0404096
[68]Production and decay of a heavy radion in Rundall-Sundrum model at the LHC
Yoshiko toponium Ohno(Ochanomizu U., Inst. Human. Sci.), Gi-Chol Cho(Ochanomizu U.)(Jan, 2013)
Published in: EPJ Web Conf. 49 (2013) 18003 Contribution to: HCP 2012 e-Print: 1301.7514
[69]Complementarity of di-top and four-top searches in interpreting possible signals of new physics
Henning Bahl(U. Heidelberg, ITP), Philipp Gadow(Hamburg U.), Romal Kumar(DESY), Krisztian Peters(DESY), Panagiotis Stylianou(Cyprus U.)et al.(Feb 16, 2026) e-Print: 2602.15
[70]Search for heavy resonances decaying into a pair of Z bosons in the ℓ+ℓ−ℓ′+ℓ′− and ℓ+ℓ−νν¯ final states using 139 fb−1 of proton–proton collisions at sqrt(s)=13 TeV with the ATLAS detector
ATLAS Collaboration Georges Aad(Marseille, CPPM)et al.(Sep 30, 2020)
Published in: Eur.Phys.J.C 81 (2021) 4, 332 e-Print: 2009.14791
Search for heavy ZZ resonances in the ℓ+ℓ−ℓ+ℓ and ℓ+ℓ−νν¯ final states using proton–proton collisions at sqrt(s)=13 TeV with the ATLAS detector
ATLAS Collaboration M. Aaboud et al. (Dec 18, 2017)
Published in: Eur.Phys.J.C 78 (2018) 4, 293 e-Print: 1712.06386
[71]Diphoton portal to warped gravity
Adam Falkowski(Orsay, LPT), Jernej F. Kamenik(Ljubljana U. and Stefan Inst., Ljubljana)(Mar 22, 2016)
Published in: Phys.Rev.D 94 (2016) 1, 015008 and e-Print: 1603.06980
[72]Implications of the 750 GeV gamma-gamma Resonance as a Case Study for the International Linear Collider
LCC Physics Working Group Keisuke Fujii (KEK, Tsukuba)et al. (Jul 13, 2016)
e-Print: 1607.03829
[73]Two-Higgs-doublet models in light of current experiments: a brief review
Lei Wang(Yantai U.), Jin Min Yang(Beijing, Inst. Theor. Phys. and Beijing, GUCAS), Yang Zhang(Zhengzhou U.)(Mar 14, 2022)
Published in: Commun.Theor.Phys. 74 (2022) 9, 097202 e-Print: 2203.07244
[74] Forward-backward asymmetries of the bottom and top quarks in warped extra-dimensional models: LHC predictions from the LEP and Tevatron anomalies
Abdelhak Djouadi(Orsay, LPT and CERN), Gregory Moreau(Orsay, LPT), Francois Richard(Orsay, LAL) (May, 2011)
Published in: Phys.Lett.B 701 (2011) 458-464 e-Print: 1105.3158

*[75]Resolving the A(FB)**b puzzle in an extra dimensional model with an extended gauge structure*
*Abdelhak Djouadi(Orsay, LPT), Gregory Moreau(Orsay, LPT), Francois Richard(Orsay, LAL)(Oct, 2006)*
*Published in: Nucl.Phys.B 773 (2007) 43-64 e-Print: hep-ph/0610173*
*[76]Top quark pairs in the threshold at the LHC*
*Maria Vittoria Garzelli(U. Hamburg (main) and CERN), Giovanni Limatola(U. Hamburg (main)), Sven-Olaf Moch(U. Hamburg (main)), Matthias Steinhauser(KIT, Karlsruhe), Oleksandr Zenaiev(U. Hamburg (main)) (Apr 3, 2026)*
*Published in: J.Subatomic Part.Cosmol. 5 (2026) 100368 e-Print: 2604.09485*
*[77]Signals of W′ and Z′ bosons at the LHC in the SU(3)×SO(5)×U(1) gauge-Higgs unification*
*Shuichiro Funatsu(CCNU, Wuhan, Inst. Part. Phys. and Hua-Zhong Normal U., LQLP), Hisaki Hatanaka(Unlisted, JP), Yutaka Hosotani(Osaka U.), Yuta Orikasa(Prague, Tech. U.), Naoki Yamatsu(Kyushu U.) (Nov 10, 2021)*
*Published in: Phys.Rev.D 105 (2022) 5, 055015 e-Print: 2111.05624*
*[78]Composite Leptons and Quarks from Hexad Preons*
*Shun-Zhi Wang(Shanghai U. and Peking U., CHEP)(Dec, 2011)*
*e-Print: 1112.0181*
*[79]Search for new heavy resonances decaying to WW, WZ, ZZ, WH, or ZH boson pairs in the all-jets final state in proton-proton collisions at s=13TeV*
*CMS Collaboration Armen Tumasyan(Yerevan Phys. Inst.)et al.(Sep 30, 2022)*
*Published in: Phys.Lett.B 844 (2023) 137813 e-Print:2210.00043*
*[80]Probing gauge-Higgs unification models at the ILC with quark–antiquark forward–backward asymmetry at center-of-mass energies above the Z mass*
*A. Irles(Valencia U., IFIC), J.P. Márquez(Valencia U., IFIC), R. Pöschl(IJCLab, Orsay), F. Richard(IJCLab, Orsay), A. Saibel(Valencia U., IFIC)et al.(Mar 14, 2024)*
*Published in Eur.Phys. J.C 84 (2024) 5, 537 e-Print: 2403.09144*
*[81]Physics Case for the International Linear Collider*
*Keisuke Fujii(KEK, Tsukuba), Christophe Grojean(DESY and ICREA, Barcelona), Michael E. Peskin(SLAC), Tim Barklow(SLAC), Yuanning Gao(Tsinghua U., Beijing, CHEP)et al.(Jun 19, 2015)*
*e-Print: 1506.05992*
*[82]Search for heavy diboson resonances in semileptonic final states in pp collisions at s=13s=13 TeV with the ATLAS detector*
*ATLAS Collaboration Georges Aad(Marseille, CPPM)et al. (Apr 30, 2020)*
*Published in: Eur.Phys.J.C 80 (2020) 12, 1165 e-Print: 2004.14636*
*[83]Probing Freeze-In Dark Matter via a Spin-2 Portal at the LHC with Vector Boson Fusion and Machine Learning*
*Junzhe Liu(Vanderbilt U.), Alfredo Gurrola(Vanderbilt U.)(Apr 3, 2026)*
*e-Print: 2604.02604*
*[84]Search for resonant and nonresonant production of pairs of dijet resonances in proton-proton collisions at ss = 13 TeV*
*CMS Collaboration Armen Tumasyan(Yerevan Phys. Inst. and Yerevan State U.)et al. (Jun 20, 2022)*
*Published in: JHEP 07 (2023) 161, JHEP 05 (2025) 113 (erratum) e-Print: 2206.09997*
*Searches for New Physics at High Object Masses with CMS*
*Andrea Malara (Apr 16, 2026)*
*e-Print: 2604.15049*
*[85]Ultraheavy diquark decaying into vectorlike quarks at the LHC*
*Ioana Duminica(Bucharest, IFIN-HH and Bucharest U.), Calin Alexa(Bucharest, IFIN-HH), Ioan M. Dinu(Bucharest, IFIN-HH), Bogdan A. Dobrescu(Fermilab), Matei-Stefan Filip(Bucharest, IFIN-HH and Bucharest U.)(Mar 21, 2025)*
*Published in: Phys.Rev.D 111 (2025) 11, 115025 e-Print: 2503.17031*
*[86]Baryon number in warped GUTs: Model building and (dark matter related) phenomenology*
*Kaustubh Agashe(Johns Hopkins U.), Geraldine Servant(Argonne and Chicago U., EFI and Saclay, SPhT)(Nov, 2004)Published in: JCAP 02 (2005) 002 e-Print: hep-ph/0411254*

# APPENDICES

## I. SUMMARY OF BSM INDICATIONS AT LHC

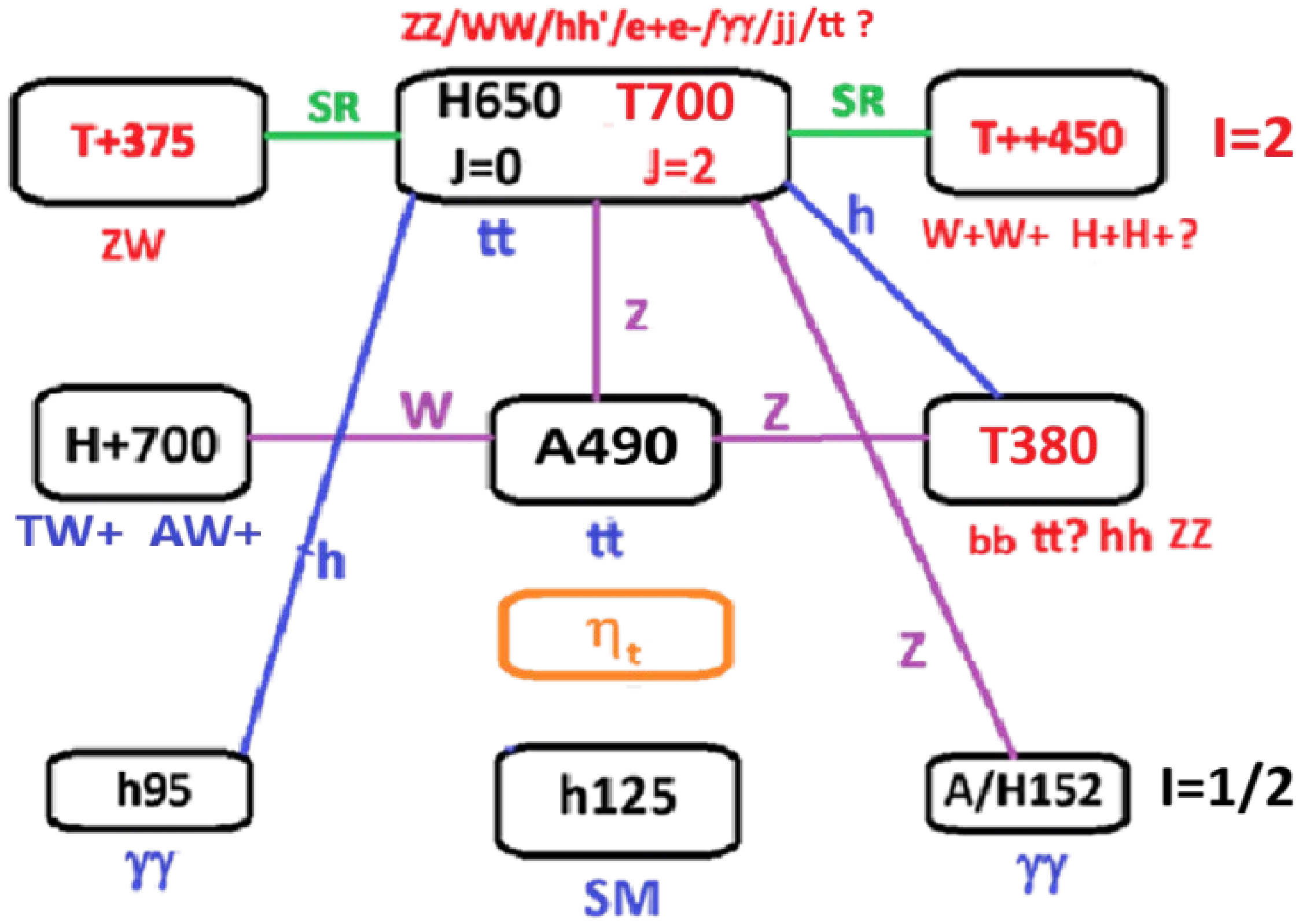

Above diagram, edited from [1], summarises our present findings. There are about twenty indications which cross the 3 s.d. CL local evidence. They comprise 3 types of isospin:

- the conventional J=0 Higgs-like I=1/2 (in black)
- the J=2 candidates, comprising a I=2 double charged state seen in W+W+ (in red)
- The toponium candidate I=0 (orange)

RUN3 should confirm the true ones. The heart of this complex mixture between scalars and tensors are two nearby resonances: T700 and H650.

**T700** has a mass known to **±10 GeV** and a width **Γ=20±5 GeV**, precisely deduced from **$ZZ\to\ell^+\ell^-+\ell'^+\ell'^-$, $\gamma\gamma$ and e+e- final states.** These final states are compatible with a KK graviton resonance. The spin of this resonance explains the disappearance of the ZZ mode when applying a scalar selection, both in ATLAS and CMS. This resonance is also observed into WW but with poor mass resolution. The coupling of T700 to ZZ and WW could imply, through the unitary sum rules **SR (**see [51] and Appendix IV), the existence of the charged resonances $T^+$ and $T^{++}$ which are indicated by LHC data.

**H650**, the J=0 candidate, is observed into A490+Z→$t\bar{t}+\ell^+\ell^-$, with a significance of 2.85 s.d. This mode does not allow a precise determination of the mass and width of this resonance. We also believe that it receives substantial contributions from T380h125 and top pairs. The top pair signal suffers from the top loop effect (see section II).

Recently, ATLAS has performed a search for **electroweak-scale dijet resonances** using a trigger-level analysis efficient for resonances with masses above 600 GeV [29]. This is of interest for H650 and T700 with dominant hadronic decays from ZZ, WW, hh', tt, bb, AZ, and T380h. It turns out that this method shows a local **3.5 s.d. excess** around 650 GeV which would correspond to a σBR(jj) of order 500 fb, which roughly matches with our expectations.

Below are summarised the significances of these various channels for H650 and T700 with the **associated** references.

| Channel | Local significance s.d. | Reference |
| --- | --- | --- |
| T700→ZZ→$\ell^+\ell^-+\ell'^+\ell'^-$ | 3.5 ggF 2.7 VBF | ATLAS2009.14791 |
| T700→WW→$\ell^+\ell^-+\nu\nu$ | 3.8 VBF | CMS-PAS-HIG-20-016 |
| T700→h125h95 $b\bar{b}/\tau\tau+\gamma\gamma$ | 3.8 | CMS 2310.01642, 2506.23012 |
| H650→T380h125→ $b\bar{b}b\bar{b}$ | 4.1 | CMS-PAS-HIG-20-012 |
| T700 → $\gamma\gamma$ | 3.5 | ATLAS 2102.13405 |
| H700+T700→Jet-Jet | 3.4 | ATLAS 2509.01219 |
| H650→A490Z→$t\bar{t}\,\ell^+\ell^-$ | 2.85 | ATLAS 2311.04033 |
| T700→e+e- | 3 | ATLAS1903.06248CMS2103.02708 |
| H650→$t\bar{t}$ | 3 | ATLAS 2404.18986CMS 2507.05119 |

The plot below displays the significance of the various channels.

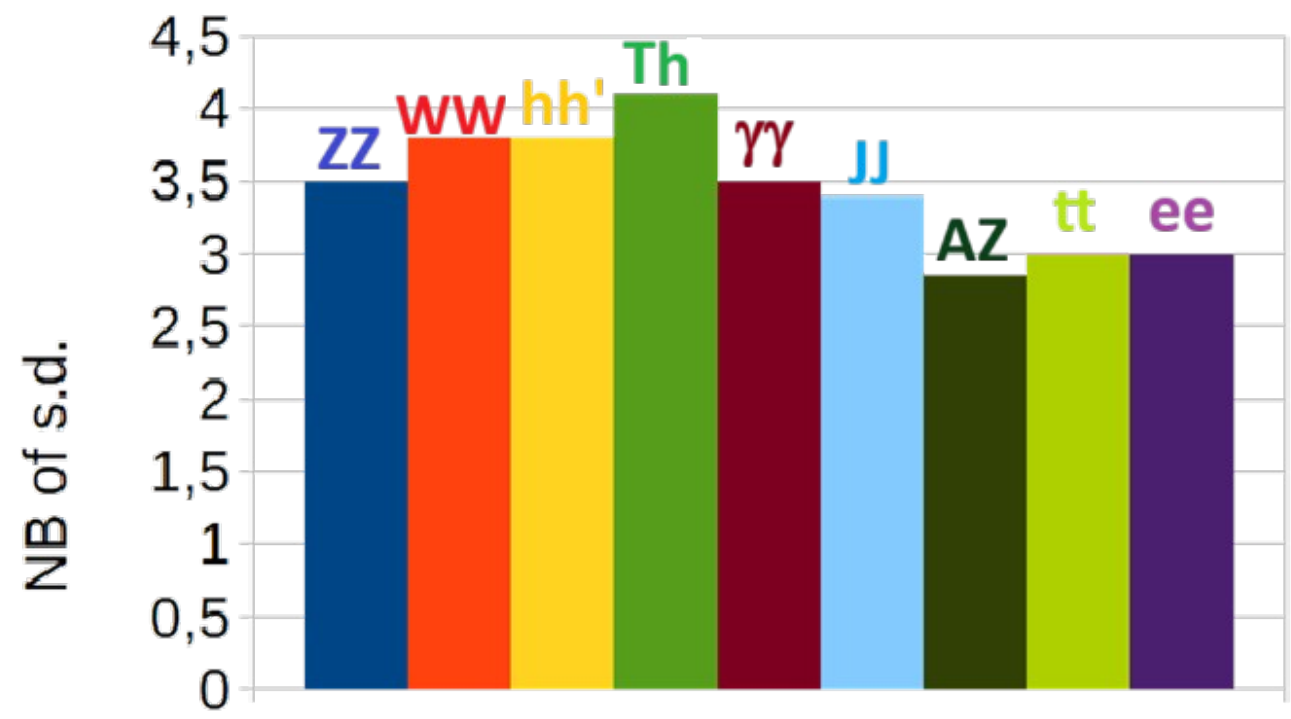

CMS has performed a search for the **ZZ+WW modes in hadronic final states** [79]. This method applies for searches of very heavy recurrences of KK gravitons where the SM background dies out. It benefits from **a detection efficiency increase by two orders of magnitude** with respect to the 4 lepton and two photon modes. Figure 24 shows some evidence for excesses above 2 TeV. These resonances are very wide as shown in the Table 1 of section I and one expects that they will join and form a **wide structure**, as indicated by this figure. With an almost **vanishing background**, the 3.6 s.d. excess around 3 TeV correspond to $A\sigma\varepsilon \sim 0.1$ fb cross section, that is Aε~20%, assuming a 50% BR into hadrons (see Table 1 from section I).

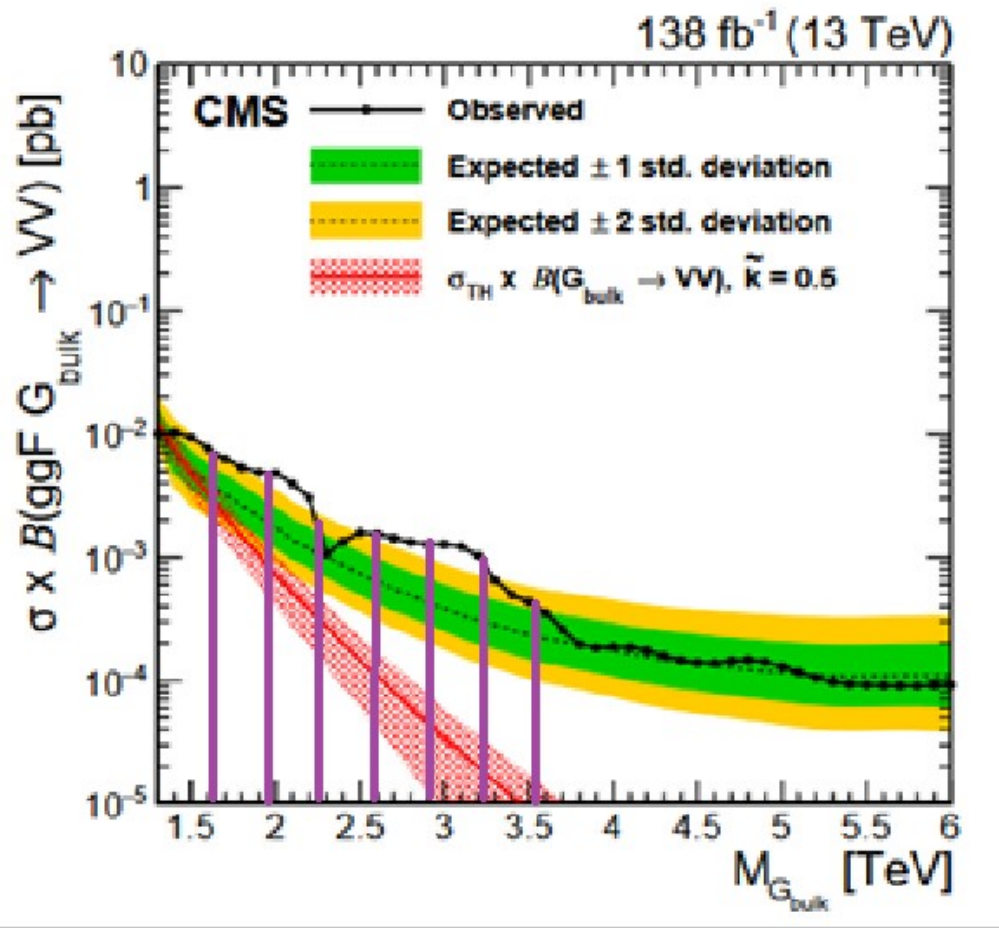

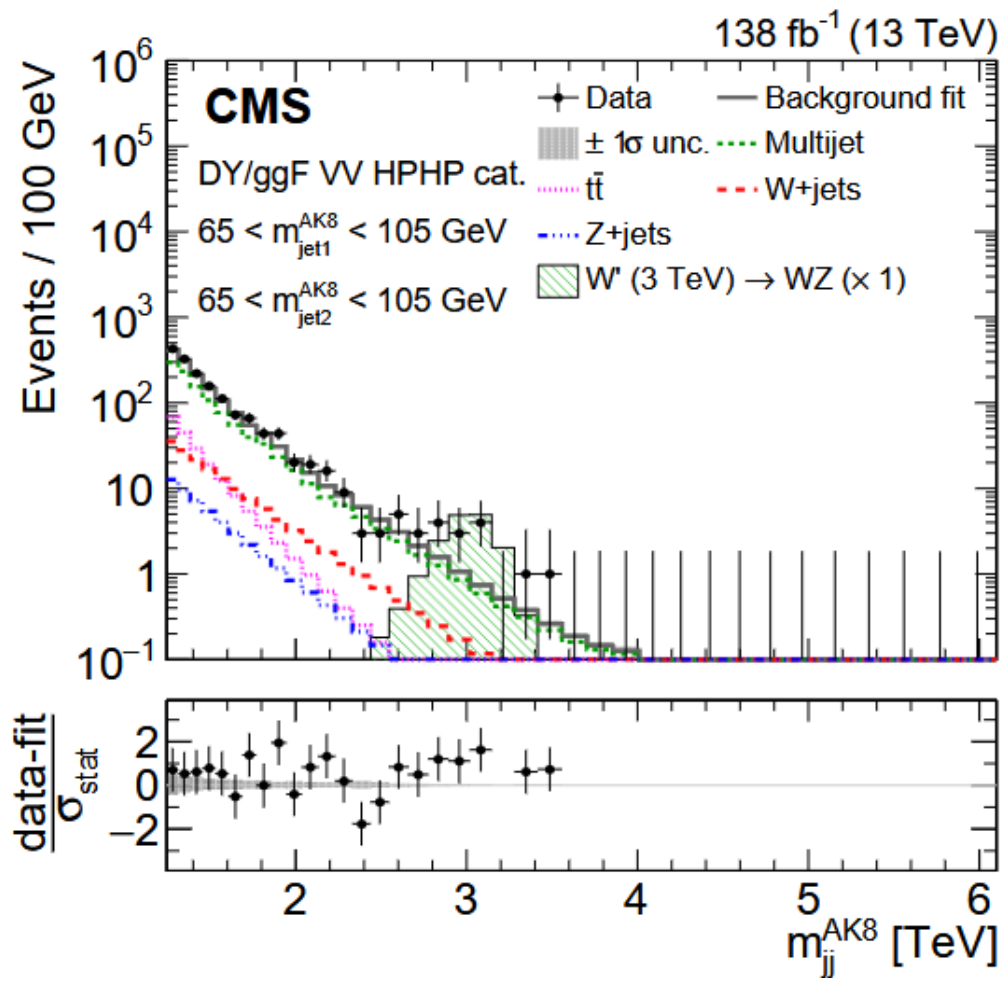

*24a: Mass limits for $G_{KK}$→WW+ZZ in jet final states. The magenta lines indicate the expected positions of the KK graviton resonances. A ~5 fb VBF cross section for masses above 2 TeV is predicted. These wide (~TeV width) resonances VBF are expected to overlap and produce wide structure as suggested by the figure. Fig 24b: Mass distribution of the VV candidates with a visible excess around 3 TeV.*

ATLAS [82] has performed a search on the **ZZ channel.** Using semi-leptonic combinations. They use the **VBF process**, which is optimal according to our findings, and also observe an excess in the tail of the reconstructed mass distribution, figure 25. One can interpret the observed excess at the end of the distribution as coming from the KK gravitons with masses 1322 and 1650 GeV. Table I predicts a cross section ~5.5 fb for the ZZ channels. The observed effect excess corresponds to **Aε ~2.5%**.

This low value probably reflects the fact that there are various effects (interference, reduced luminosity for the high mass tails) which reduce the effective cross section.

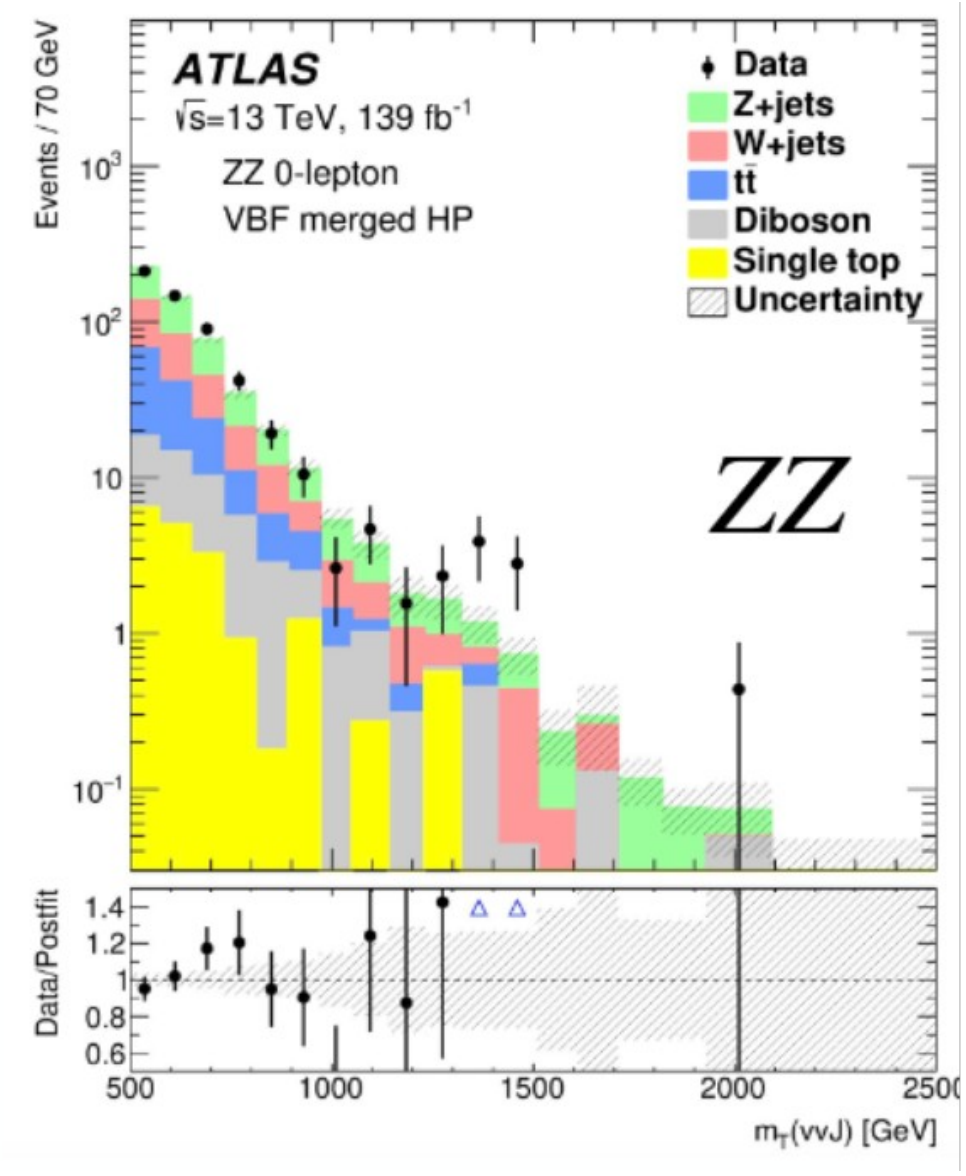

*Figure 25: VBF-> ZZ mass mass distribution using hadronic modes with a HP (high purity) selection.*

The $t\,\bar{t}$ channel excesses do not allow a simple estimate of the statistical significance but are evidently present in both experiments (section II.1 and II.2). We estimate a combined significance above 3 s.d.

Using the $\chi^2$ deduced from these nine channels which have compatible masses, within an interval 600 to 750 GeV, one reaches a local significance of 6.3 s.d. If, conservatively, one takes into account that the number of mass intervals from ZZ threshold up to the end of populated masses, **the global significance is ~6 s.d.** Again, this significance presumably combines the effect of **two resonances.**

**A490** is the second pivot of this structure since it is indicated as decaying from H650 and from $H^+$700 and into T380+Z, with T380→h+h. Its decay into $t\,\bar{t}$ is weakly but coincidentally indicated by ATLAS and CMS. It was first observed into Zh by ATLAS with 36 fb-1 [31], which requires confirmation.

The heavy charged Higgs candidate labelled **$H^+$700** is observed in two instances:

- An inclusive search [50] requiring a lepton with PT>60 GeV shows an excess which can be interpreted as H+→A490+W
- In H+→T380+W→hh+W→$b\bar{b}b\bar{b}$ +W (see figure 9a)

Its mass is not precisely determined but according to the 3HD model from [43], one expects that it should be above 700 GeV.

In above diagram, we have represented in **red** what we believe are the contributions of the various tensors. **T380** is indicated (see II.1) in 4 modes.

Recently [32], an additional candidate has emerged in 4b: Y700→ X400+h125→$b\bar{b}b\bar{b}$ which could be **T700→T380+h125** or H650→T380+h125, assuming that T380 decays into $b\bar{b}$.

Finally, a scenario with resonances decaying into **hh** and 'polluting' the measurement of the **triple SM Higgs coupling** is quite likely to occur and was already discussed in [1]. One expects contributions from the various KK graviton resonances $G_{KK}$→h+h, as already indicated by LHC data. This implies that the measured hh cross section will **significantly exceed the SM prediction**. This measurement will therefore require a full understanding of the various KK resonances, which represents a **formidable challenge** for LHC.

# II. MSSM/NMSSM at low tanβ

MSSM does not allow to accommodate the '650 GeV' resonance seen into ZZ/WW since it predicts decoupling of the heavy Higgs into ZZ/WW. A less obvious contradiction with our findings is the mass difference between A490 and H650. The doxa of MSSM, seems to be MA>MH which is true at the tree level, ignoring radiative corrections. One has:

$M_h^2=M_Z^2\cos^2 2\beta$ obviously badly violated for low $\tan\beta$ and $M_A^2= M_h^2+M^2_Z\sin^2 2\beta$.

For tanβ close to 1, EW radiative corrections need to be very large to reproduce Mh as shown in figure 26. The 'proof' of MH>MA comes from the 3 reactions:

- $A490 \rightarrow h125+h125+Z \rightarrow b\bar{b}b\bar{b}+\ell^+\ell^-$
- $H650 \rightarrow A490+Z \rightarrow t\,\bar{t}+\ell^+\ell^-$
- $H^+700 \rightarrow A490+W^+ \rightarrow t\,\bar{t}+\ell^+\nu$

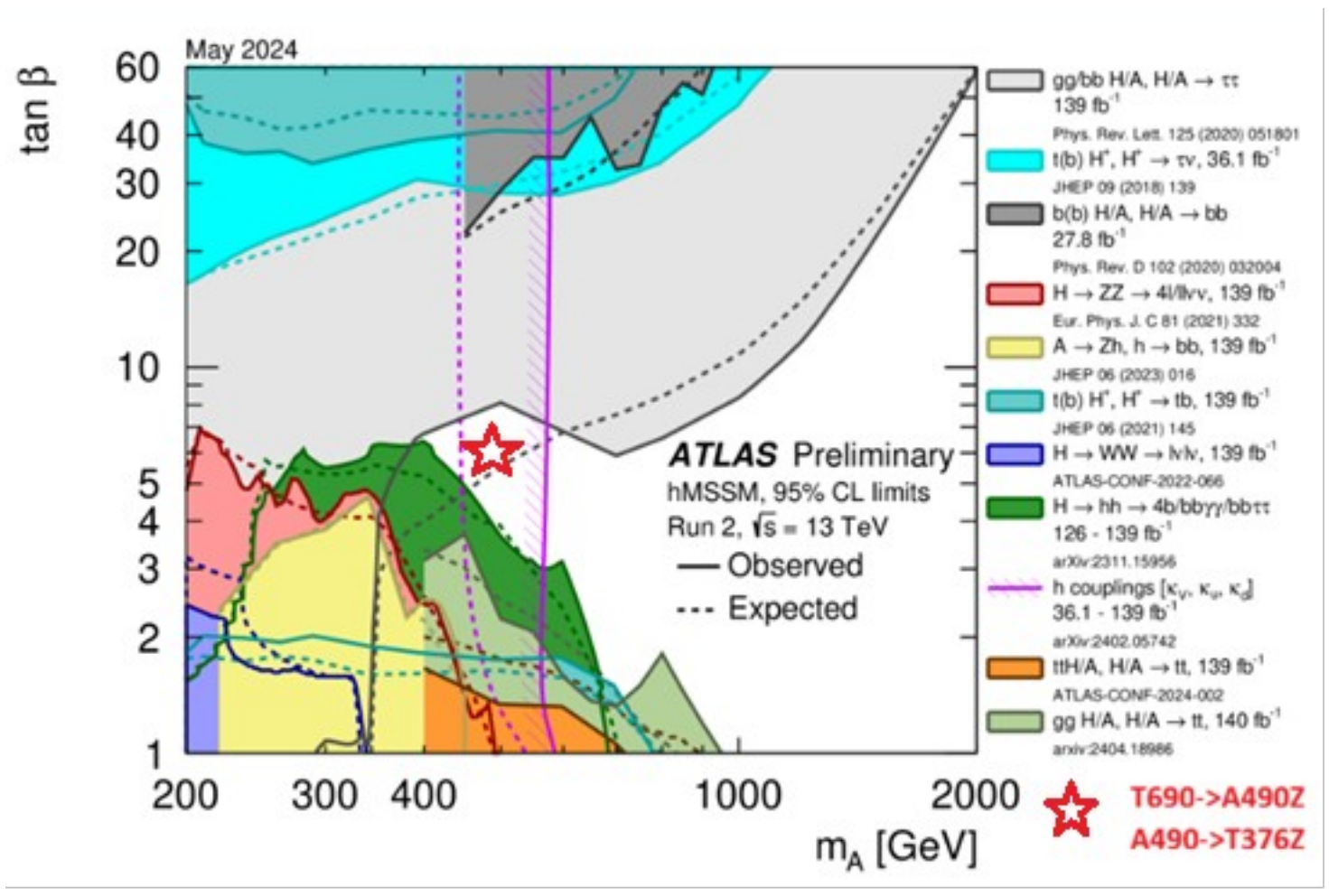

*Figure 26: Combined MSSM exclusions from ATLAS[16] . The red star solution comes from direct searches indications discussed in Appendix I. Within MSSM this solution is barely excluded my h coupling measurements but not by direct searches.*

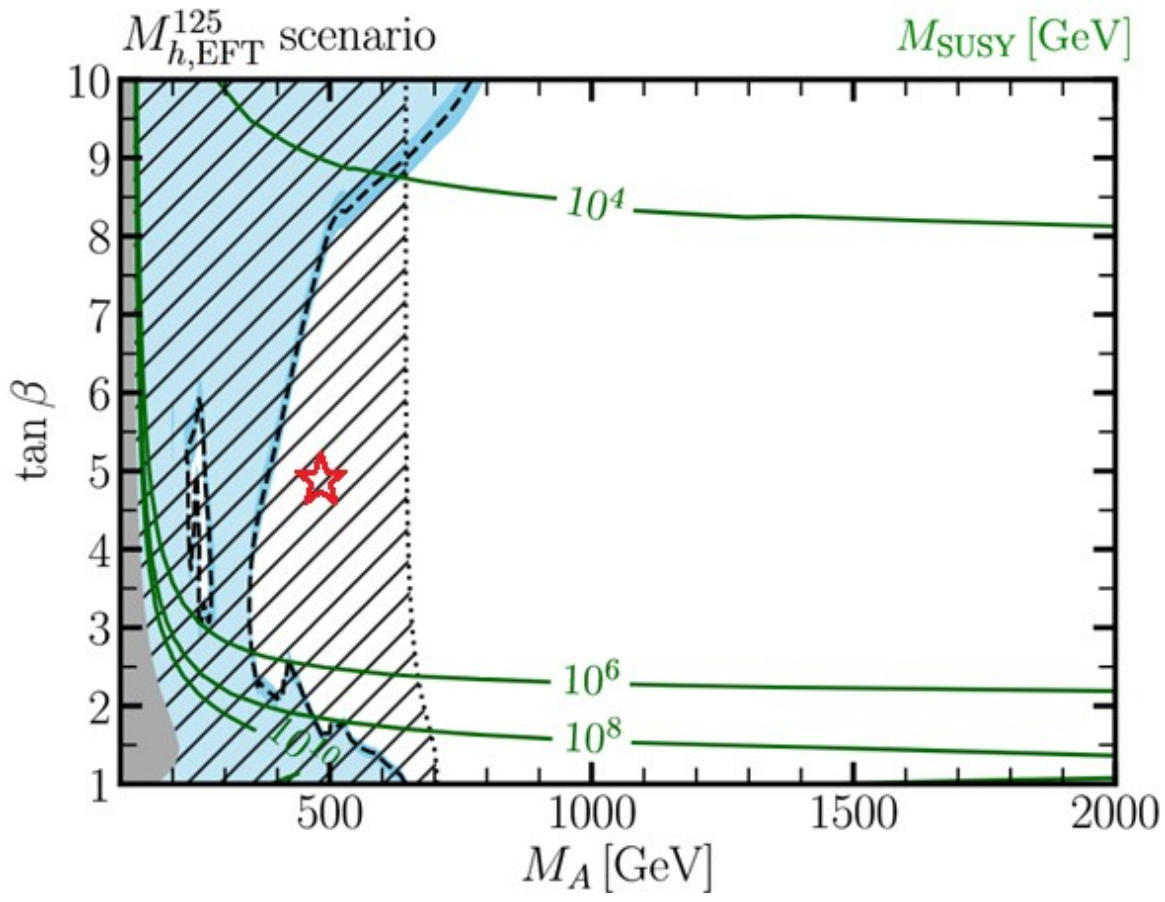

*Figure 27: The green solid lines indicate the required SUSY scale in GeV. The red star indicates our solution which is disfavoured by Higgs measurements. The blue region is excluded at the 95% CL by LHC searches for additional Higgs bosons.*

Reference [72] provides the allowed mass regions which fulfill the following basic criteria:

- Perturbative unitarity
- SM Higgs couplings
- Oblique parameters

Figure 27 shows the allowed regions for type II coupling and we see that our solution is perfectly allowed. The agreement between the SM and h125 measurements implies that, within MSSM, $\alpha \rightarrow \beta-\pi/2$ hence, both in type I and II, gHtt and gAtt are close to 1/tanβ. For tanβ close to 1, as indicated by the analysis, figure 27 [33] predicts that the SUSY scale should reach 1000 TeV to satisfy this solution. The contours of

16 https://cds.cern.ch/record/2898861/files/ATL-PHYS-PUB-2024-008.pdf

figure 28 indicate that this corresponds to a solution where H and A dominantly decay into t $\bar{t}$, with contribution from H→h+h and A→h+Z at the level of 10%, therefore within the sensitivity of future searches. The channel H650→ A+Z, which is usually kinematically closed, has already been indicated by ATLAS. At low tanβ, extended versions, like **NMSSM**, are able to accommodate **a TeV SUSY scale**, observable at LHC (see for instance [34]). This scenario would also provide an extended isoscalar sector which could accommodate h95, A152 and H650 candidates [35]. It seems however that such a scenario usually predicts mass degeneracy between $H^{+}$, A and H.

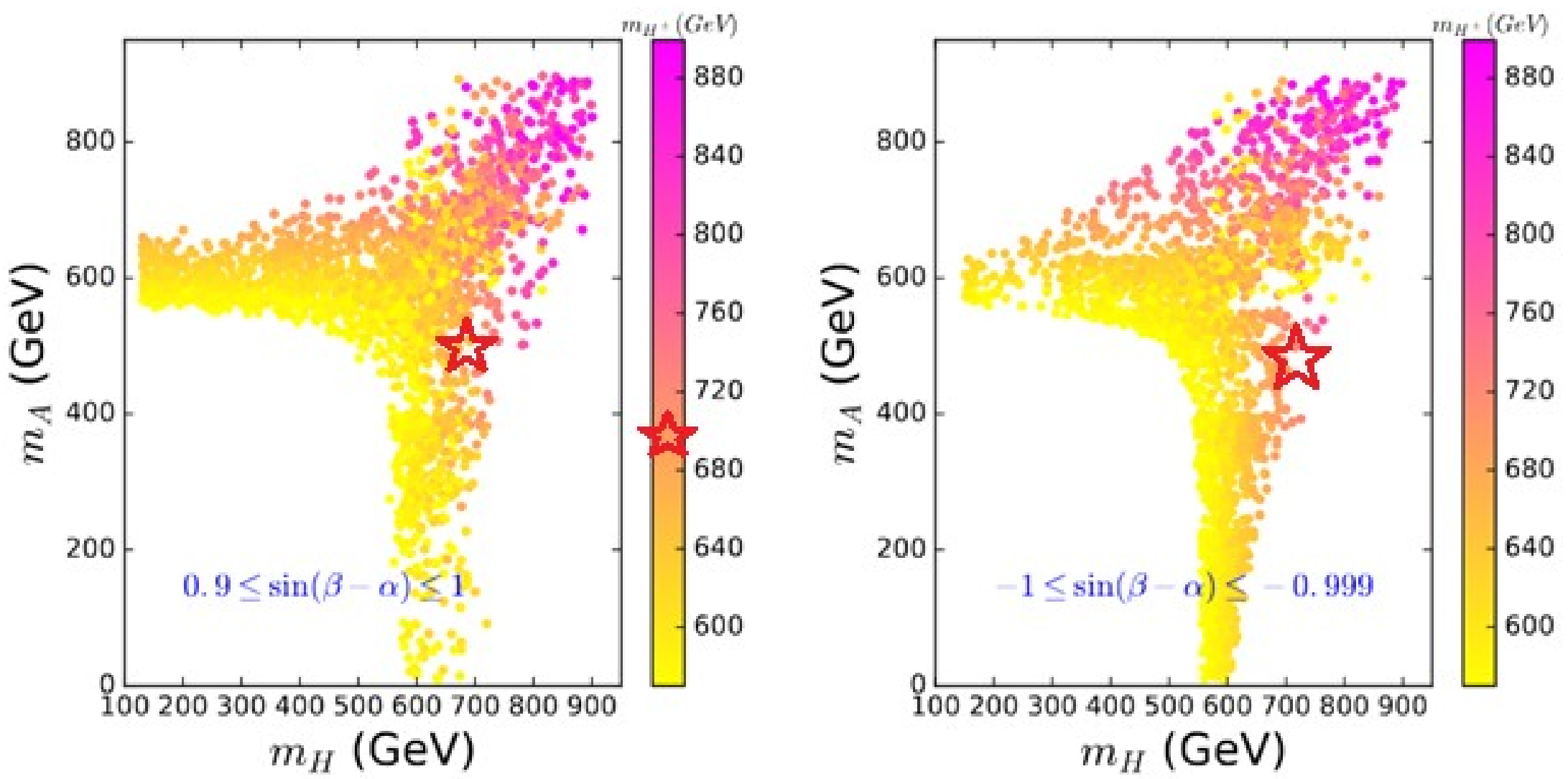

*Figure 28: scatter plot mA and mH satisfying vaccuum stability, perturbativity, and the oblique parameters for 570 GeV < mH± < 900 GeV and the SM Higgs measurements. The red star indicate our solution.*

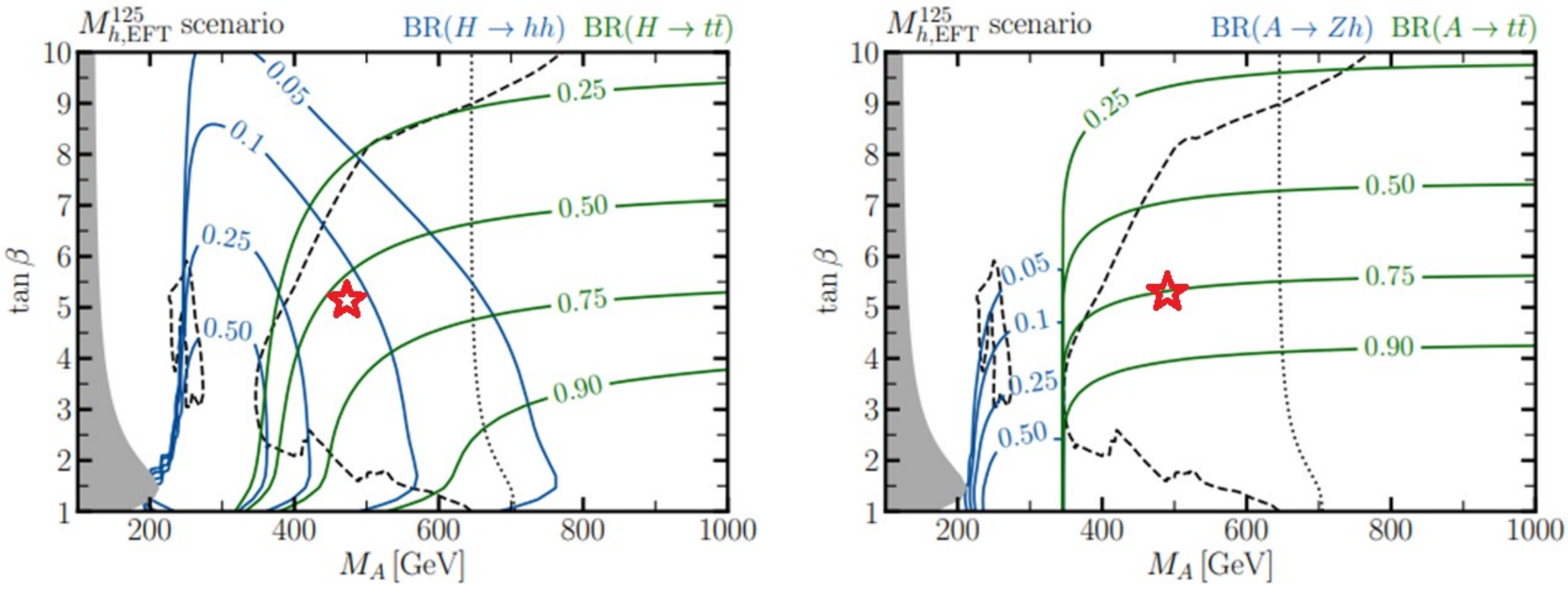

*Figure 29 left: Branching ratios of H into hh (blue) and into top quarks (green). In each plot the dashed line shows the region excluded by direct searches, dotted by measurements. 29 right: BR of A Into top pairs should be dominant.*

Recall that one predicts the following couplings[17]:

| | type-I | type-II |
|---|---|---|
| $\xi_h^u$ | $\sin(\beta-\alpha)+\cos(\beta-\alpha)/\tan\beta$ | $\sin(\beta-\alpha)+\cos(\beta-\alpha)/\tan\beta$ |
| $\xi_h^d$ | $\sin(\beta-\alpha)+\cos(\beta-\alpha)/\tan\beta$ | $\sin(\beta-\alpha)-\cos(\beta-\alpha)\cdot\tan\beta$ |
| $\xi_h^l$ | $\sin(\beta-\alpha)+\cos(\beta-\alpha)/\tan\beta$ | $\sin(\beta-\alpha)-\cos(\beta-\alpha)\cdot\tan\beta$ |
| $\xi_H^u$ | $\cos(\beta-\alpha)-\sin(\beta-\alpha)/\tan\beta$ | $\cos(\beta-\alpha)-\sin(\beta-\alpha)/\tan\beta$ |
| $\xi_H^d$ | $\cos(\beta-\alpha)-\sin(\beta-\alpha)/\tan\beta$ | $\cos(\beta-\alpha)+\sin(\beta-\alpha)\cdot\tan\beta$ |
| $\xi_H^l$ | $\cos(\beta-\alpha)-\sin(\beta-\alpha)/\tan\beta$ | $\cos(\beta-\alpha)+\sin(\beta-\alpha)\cdot\tan\beta$ |
| $\xi_A^u$ | $1/\tan\beta$ | $1/\tan\beta$ |
| $\xi_A^d$ | $-1/\tan\beta$ | $\tan\beta$ |
| $\xi_A^l$ | $-1/\tan\beta$ | $\tan\beta$ |

17 see for instance ATLAS-PHYS-PUB-2013-016

# III. Angular distributions

| Process<br>Jz: initial=m final=m' | Shape z=cos$\theta$ |
| --- | --- |
| $V_LV_L \to G \to W_LW_L/Z_LZ_L$ [18]<br>m=0 m'=0 | $(3z^2-1)^2$ |
| $gg \to G \to W_LW_L/Z_LZ_L$<br>m=2 m'=0 | $(z^2-1)^2$ |
| $W_LW_L/Z_LZ_L \to G \to \gamma\gamma$<br>m=0 m'=2 | $(z^2-1)^2$ |
| $gg \to G \to \gamma\gamma$<br>m=2 m'=2 + m'=-2 | $1+6z^2+z^4$ |
| $e^-_{R/L}e^+_{R/L} \to V/G$ | 0 |
| $e^-_Re^+_L \to V \to \ell^-_R\ell^+_L$<br>$e^-_Re^+_L \to V \to \ell^-_L\ell^+_R$<br>Average SM | $(1+z)^2$<br>$(1-z)^2$<br>$1+z^2$ |
| $e+e/q\overline{q} \to G \to \ell^-\ell^+$<br>m=1 m'=1 + m'=-1 | $1-3z^2+4z^4$ |
| $e+e-/\ q\overline{q} \to G \to \gamma\gamma$<br>m=1 m'=2 + m'=-2 | $1-z^4$ |
| $e+e-/q\overline{q} \to G \to W_LW_L/Z_LZ_L$ | $z^2(1-z^2)$ |

In this Appendix we intend to show how to characterise KK graviton resonances using angular distributions.

An important result of is that $gg \to G \to ZZ$ radically differs from $V_LV_L \to G \to ZZ$ meaning that the disappearance of the signal the ggF+VBF with the scalar selections practised by ATLAS and CMS prove the dominance of the VBF process implying that **ggF/VBF<0.5** which results on the very stringent upper limit on the coupling of to ZZ.

Note that $gg \to G \to \gamma\gamma$ differs from the scalar distribution with 50% of the events having $|z|>0.9$. For $V_LV_L \to G \to \gamma\gamma$, one has $(1-z^2)^2$ hence only 0.12% of the events have $|z|>0.9$.

It is therefore quite clear that KK gravitons can be easily separated from scalar resonance but not from spin 2 **composite resonances** [22]. This could however not be true for the **off-shell case** where different operators could set it and influence differently a composite resonance and an elementary graviton. In above table, we have collected a set of angular distributions [36,37], some of them being shown in figure 29.

In [36], additional channels have been computed:

- $q+\overline{q} \to g+G_{KK}$

18 Neglecting the t-channel exchange contribution see [39]

- q+g→q+$G_{KK}$
- g+g→g+$G_{KK}$

which could significantly contribute to produce gravitons. In [38], limits on graviton coupling to the energy-momentum tensor were derived from LHC data in the so-called ‘hidden graviton’ scenario.

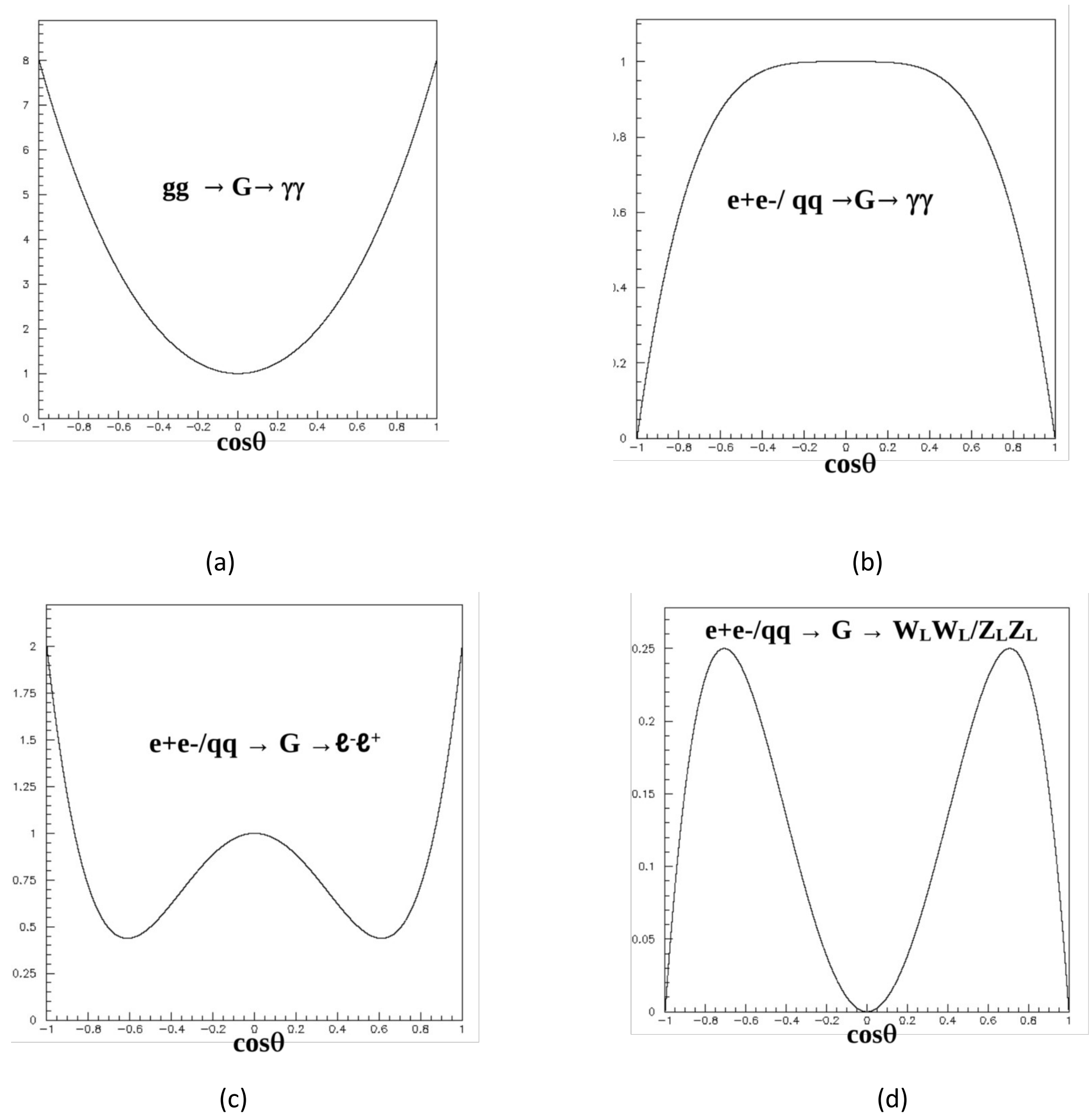

(a) (b)

(c) (d)

*Figure 30: RS predictions for (a) gg→ $\gamma\gamma$, (b) e+e-/$q\bar{q}$ → $\gamma\gamma$, (c) e+e-→μ+μ- and (d) e+e-→ WW/ZZ in arbitrary units.*

An important prediction from the RS model is that it only couples to spin aligned configurations, meaning that m=±2 for photons and gluons and m=±1 for leptons. Also, the graviton couples to WL/ZL meaning that for these states m=0. To understand the reason for this behaviour one can take the process e-(p1)e+(p2) → $G_{KK}$→ℓ⁻(k1)ℓ⁺(k2) for which the coupling of the graviton to the energy stress tensor gives the following matrix element [39]:

$$\mathcal{M}_G = \left(-\frac{\imath}{\Lambda_\pi^2}\right)\frac{1}{s-M_G^2}\Big[(k'\cdot p')\bar{u}(k_1)\gamma_\mu v(k_2)\;\bar{v}(p_2)\gamma^\mu u(p_1) + \bar{u}(k_1)\;\not{p}'v(k_2)\;\bar{v}(p_2)\;\not{k}'u(p_1)\Big]$$

From this expression one trivially understands why, as for the SM, only aligned spin configurations contribute.

Figure 30 shows the results obtained for (a) gg→γγ, (b) $q\bar{q}$→ γγ , (c) e+e-→μ+μ- and (d) e+e-→ WW.

(a) and (b) differ markedly. In addition one expects a significant contribution from the VBF process WW/ZZ→$G_{KK}$. The gg process dominates for the first two recurrences, due to the large gg luminosity.

For e+e-→ $G_{KK}$, the situation is simpler given that, on top of the huge graviton resonance, there will be a negligible contribution from SM processes. One expects that the final states will be dominated by WW/ZZ/hh with a significant contribution from γγ, gg and light fermions pairs. The contribution from top pairs should be larger. Measuring precisely these channels will help understanding the location of these particles in the extra-dimension, providing an essential test of the RS model.

# IV. Perturbative unitarity?

The reaction $W_LW_L$→$W_LW_L$ receives contributions from the three SM particles Z/photon/Higgs in the s and t channels. For large s values, these contributions compensate such that the resulting amplitude does not diverge, preserving unitarity at large s. This remains true adding extra doublets, given that heavy neutral scalars decouple from WW/ZZ and do not contribute at large s.

In the MSSM model, H650 decouples from ZZ/WW which therefore does not contribute to the unitarity sum rule.

In the Georgi Machacek (GM) model, one has compensation between the heavy neutral bosons exchanged in the s/t channel and the doubly charged scalars $H^{++}$ from the u channel, also preserving unitarity. For this mechanism to function one needs the following relation between the scalar coupling constants [51]:

$$g^2(4m_W^2 - 3m_Z^2 c_W^2) \overset{\rho \simeq 1}{\simeq} g^2 m_W^2 = \sum_k g^2_{W^+W^-H_k^0} - \sum_l g^2_{W^+W^+H_l^{--}}$$

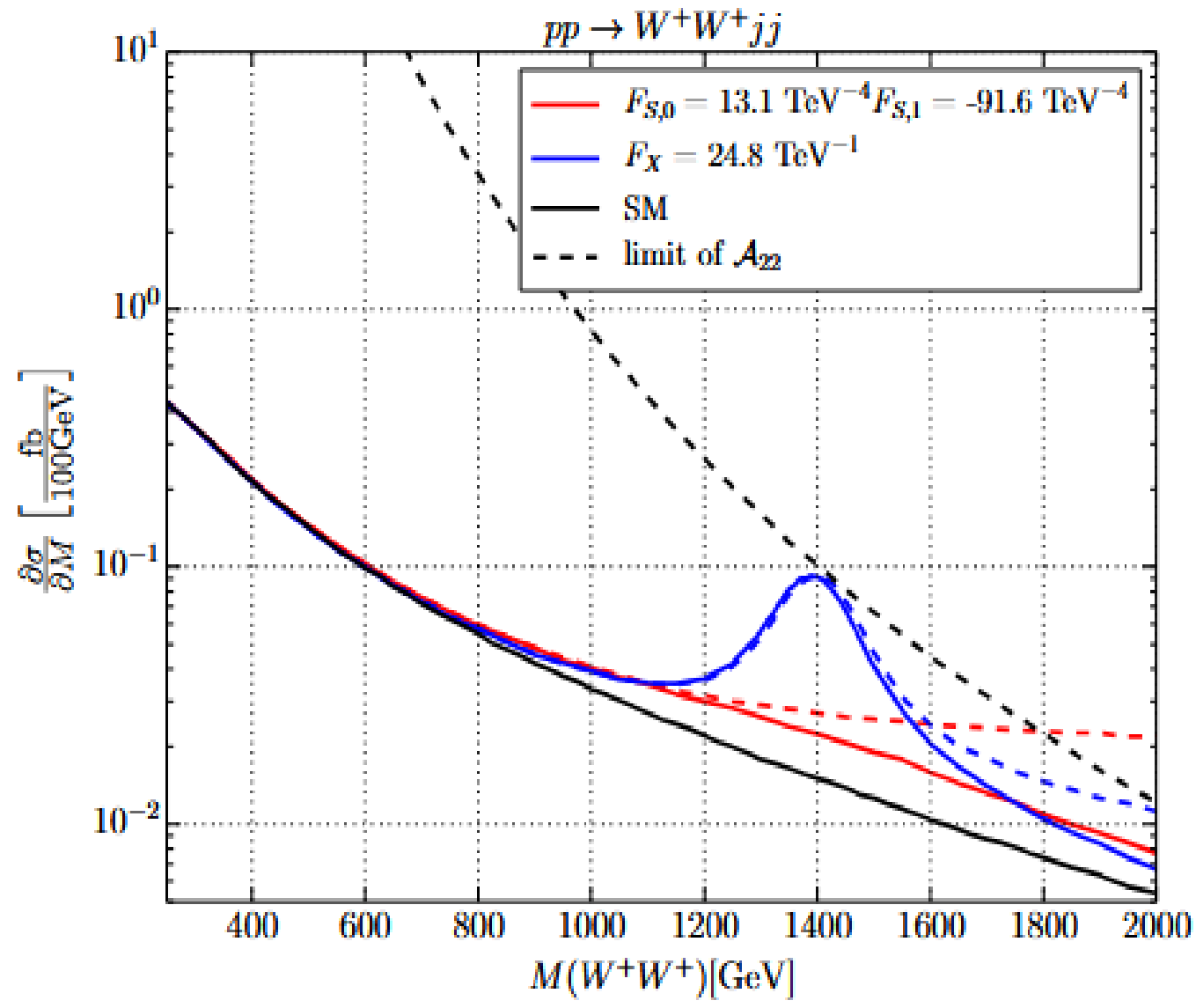

*Figure 31: Expected shape for an isotensor tensor distribution using the unitarisation procedure (blue curve) for a 1400 GeV resonance which has a width of 140 GeV.*

If one adds a heavy neutral graviton contribution, there will be no more compensation unless one also has **doubly charged tensors** exchanged in the **u channel**. This seems to be the case given that one observes a resonance $T^{++}$ almost degenerate in mass with T380.

Reference [40] computes explicitly the case with I=2 (isotensor) and J=2 (tensor), which is our hypothesis. It shows that for a graviton exchange there is imperfect compensation for the processes $WL^+WL^+ \to WL^+WL^+$ and $WL^+WL^- \to WL^+WL^-$ meaning that these amplitudes grow like s but with cancellations due to $T^{++}$.

Reference [40] claims that this effect can be taken care of by the so-called T-matrix unitarization procedure, resulting in a genuine Breit-Wigner near resonance.

These arguments repeat themselves for $Z_LZ_L \to W_LW_L$, with a cancellation due to the resonance $T^+375$ seen in ZW in the LHC data.

A valid criterion can be given by the value of $\sqrt{s}$ at which unitarity is violated. Reference [40] gives the following formula:

$$s \sim m_X^5/30\Gamma_X m^2_{whz}$$

where $m_{whz}$~100 GeV is the average mass between h125, Z and W and where $\Gamma_X$ is the width of the resonance. For the first resonance the unitarity limit is reached at 29 TeV and for the second 51 TeV. These high values are reassuring since no one can tell what happens at such scales.

For a model with an isoscalar tensor (I=0), these limits would be significantly reduced to 12 TeV and 21 TeV. It is therefore unclear whether the T700 resonance, and higher order KK resonances, also need this type of protection from $T^{++}$ resonances. If so, LHC could also observe $T^{++}700$ and $T^+700$.

A major difference between our views and the standard RS case is our assumption that T380 could belong to a I=2 isomultiplet. References [40] and [41] have anticipated this scenario invoking perturbative unitarity.

# V. Randall Sundrum models

In this section, we intend to describe various features of the RS models which are of relevance for the present work. Up to date references to this model can be found in [13].

## V.1 Bulk location effects

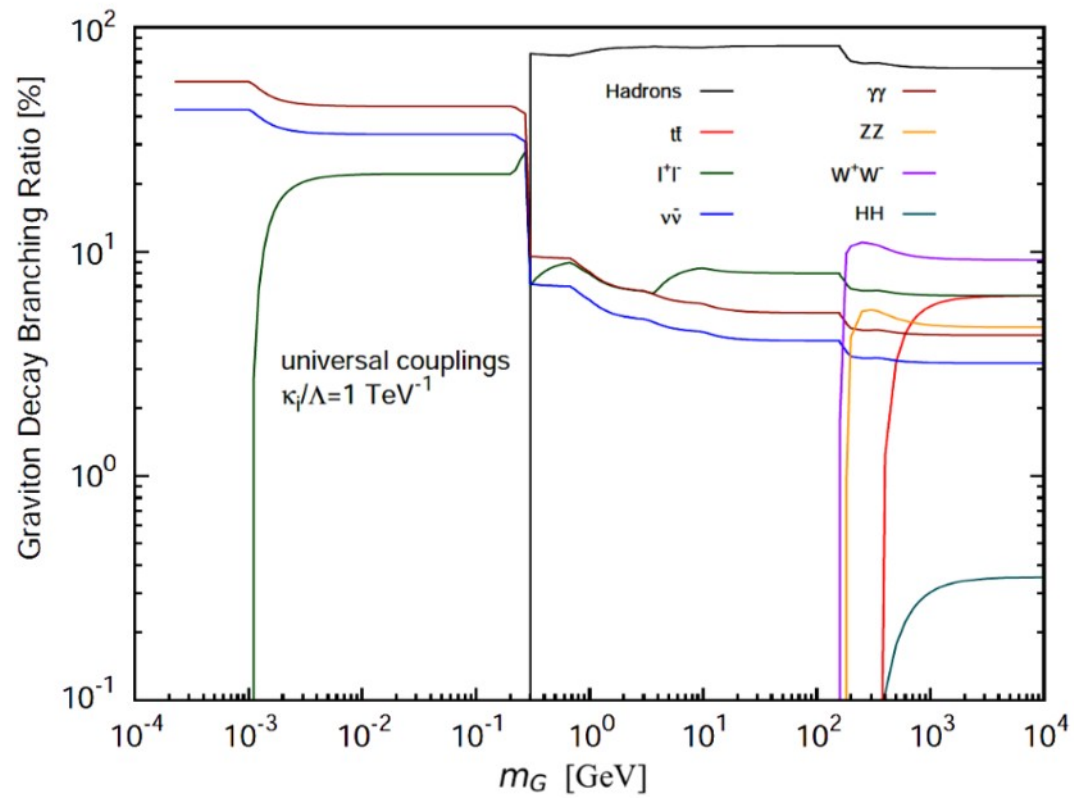

*Figure 32: from [63], graviton BR assuming universal coupling of the graviton to the stress tensor*

In its primitive version [57], this model assumes that there are two branes embedded in a D=4+1 space. One of them contains the SM particles, at the electroweak scale, the other being at the Planck scale. Warping insures an exponential transition between these two scales, solving the hierarchy problem. '*In this first version – the so-called **'RS1' version** – only gravitons propagate in between the two branes, therefore generating KK spin 2 resonances at a scale starting at about half a TeV.*'

Very soon [58] after this primitive version, it became clear that allowing SM particles to propagate between the two branes – the so-called **'bulk version'** – provided an interpretation of the hierarchy between the fermion masses, an other enigma which is not understood in most other models. This in turn automatically generated additional KK particles.

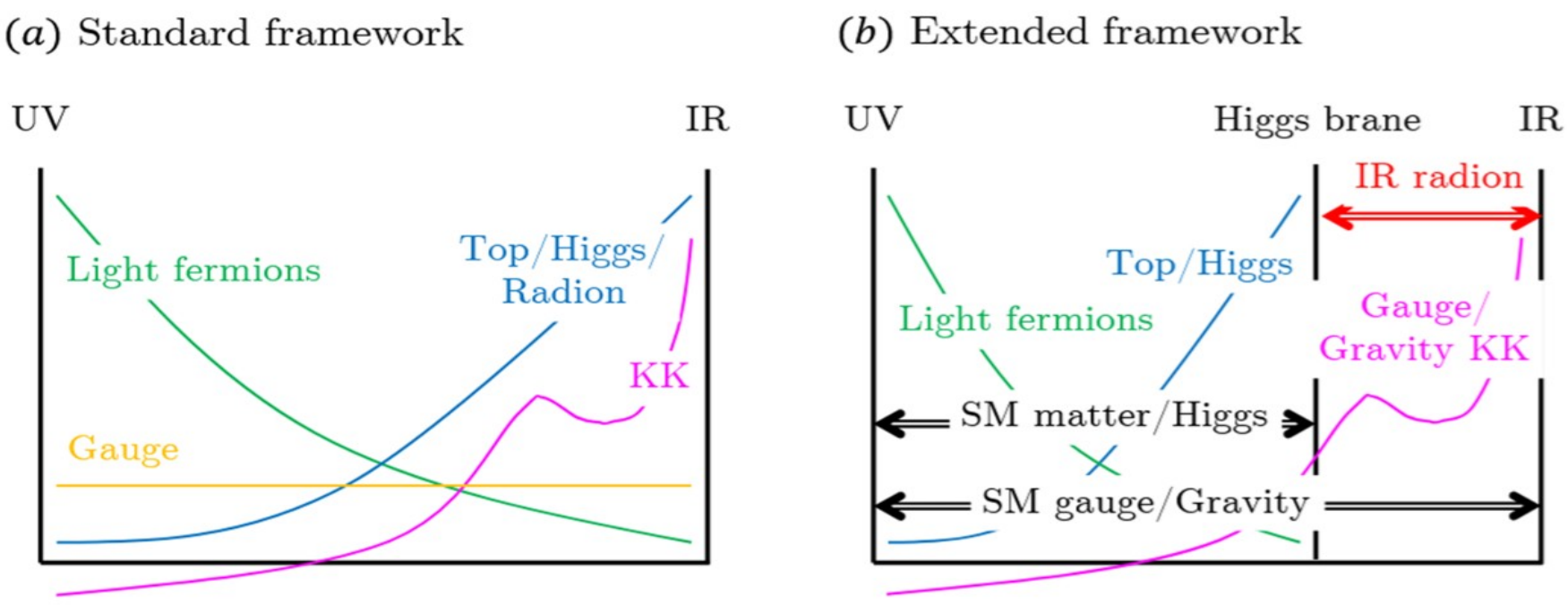

**Figure 33** Left: Warped extra-dimensional model with SM fields in the same bulk (standard framework). Schematic shapes of extra-dimensional wavefunctions for various particles (zero mode SM fermions and gauge particles, a radion, and a generic KK mode) are shown. Right: Warped extra-dimensional model with *all* SM gauge fields and gravity in the full bulk, but matter and Higgs fields in subspace (extended framework). Schematic shapes of extra-dimensional wavefunctions for various particles (zero mode SM fermions and gauge bosons, an IR radion, and gauge/gravity KK modes) are shown. The double-lined arrows indicate that the associated fields propagate within the marked range in the extra dimension.

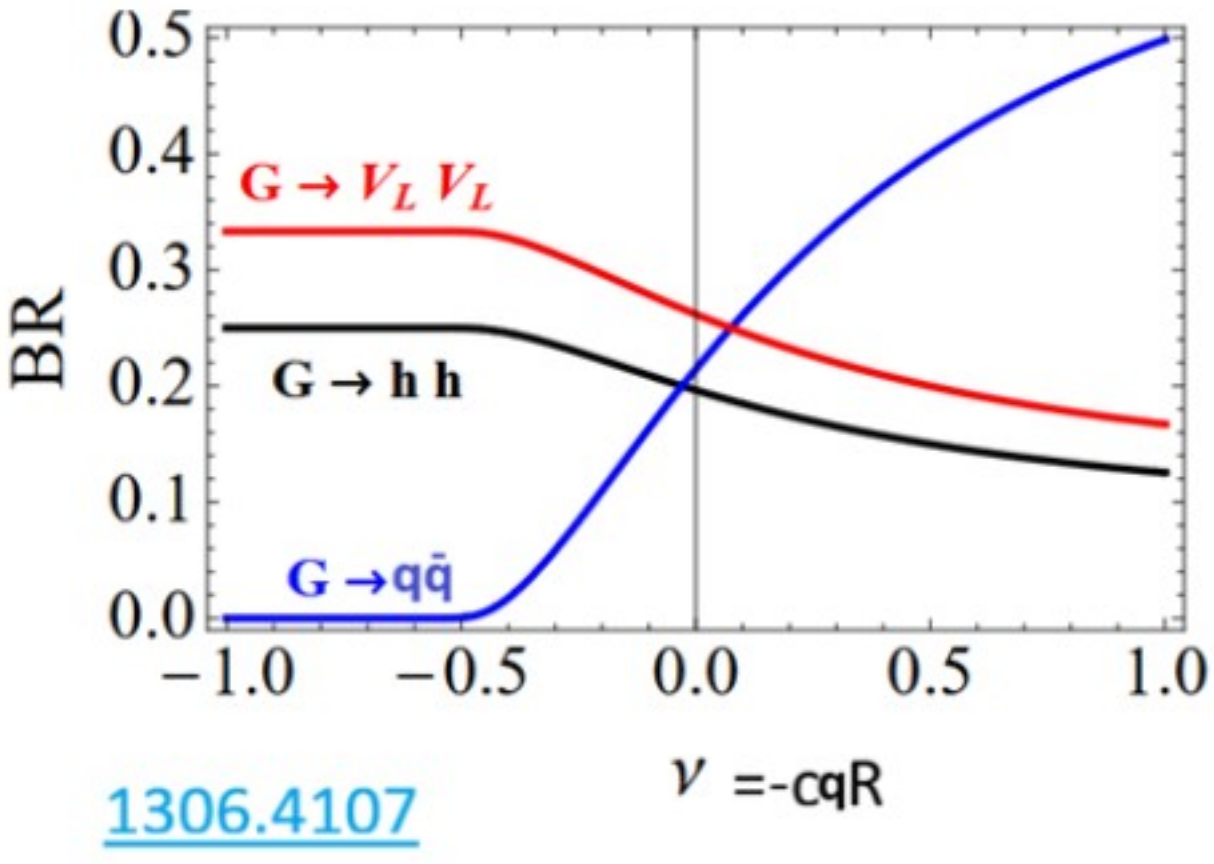

*Fig. 34: Branching ratio of graviton to the hh/VV and tt versus the bulk mass term.*

Allocating various positions in the bulk for SM particles automatically **breaks down universality of couplings** for the particles forming the stress tensor to the graviton, as represented in figure 31. This universality assumes a dominance into jet-jet decays from gluon and light quark contributions. It also assumes that heavy gravitons are almost universally coupled to W/Z/γγ/e+e-/νν/tt. In the new paradigm this is not so. It becomes false for electrons and neutrinos which are weakly coupled since they belong to the Planck brane, hence their low mass. This is also false for light quarks which no more contribute to hadronic decays, but remains true for gluons which travel freely between the two branes.

A similar example of flexibility of these models is shown in [65] where one assumes that fermions are separated from W/Z/h brane. Figure 34 shows that for $\nu$t<-0.5, fermions are decoupled from KK gravitons, which seems to be the case. For $\nu$eR=-0.45, using figure 33 one deduces that BR(ee)~0.25% compatible with LHC result ~0.25%. One still assumes a lower value for eL, as belonging to the same isodoublet as a neutrino, which allows to generate properly the electron mass.

## V.2 Gravitons of Higher rank

For a KK graviton with n=2, there is a **factor 25 decrease** of its coupling to **photons and gluons** with respect to n=1 ([16], [22]):

$$C_{00n}^{AAG} = e^{k\pi R}\frac{2\left[1-J_0\left(x_n^G\right)\right]}{k\pi R\left(x_n^G\right)^2\left|J_2\left(x_n^G\right)\right|}$$

with the consequence that T380 should indicate a much stronger **γγ signal** than T700, which is not the case, from which one concludes that these particles **do not couple directly to photons**.

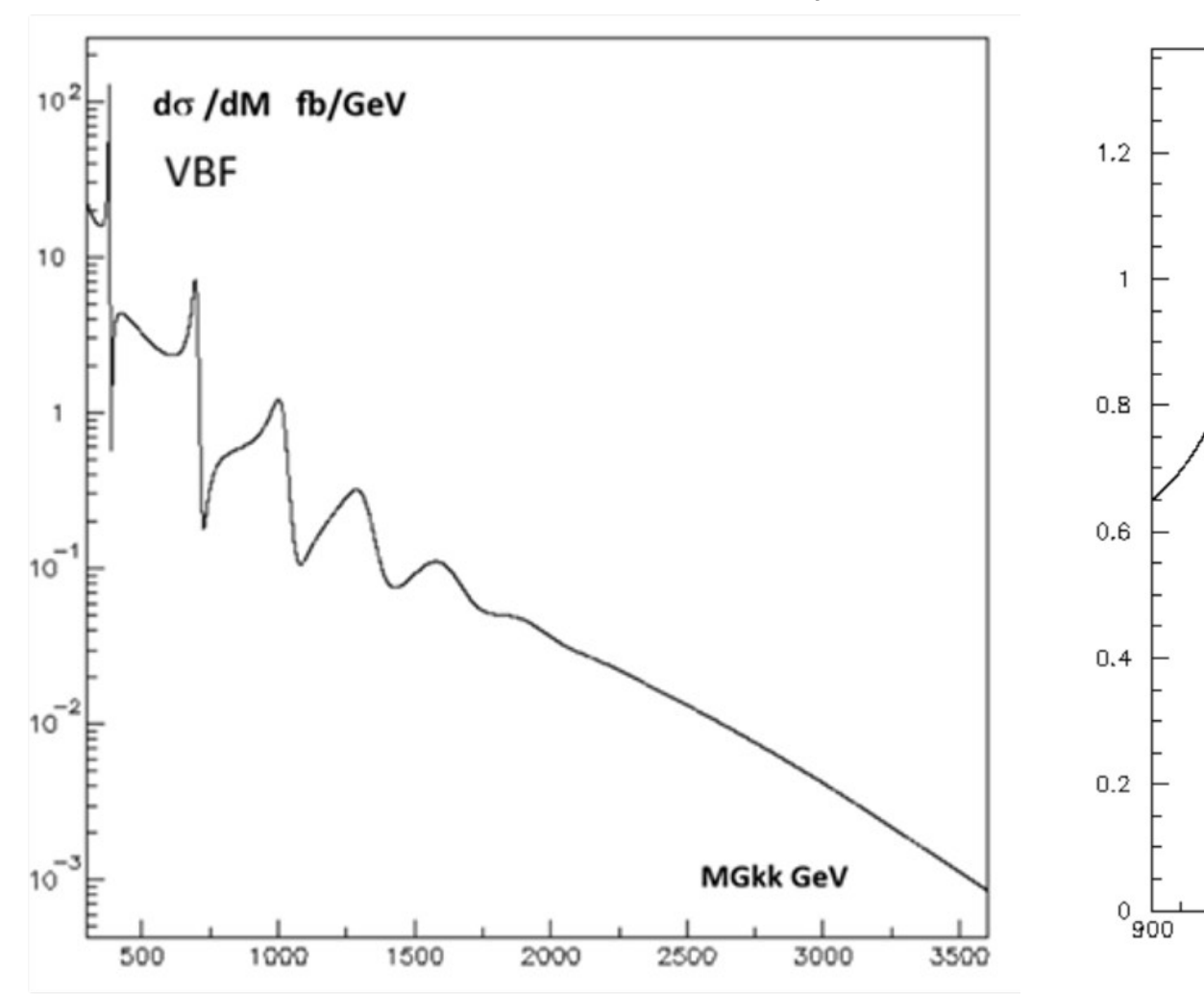

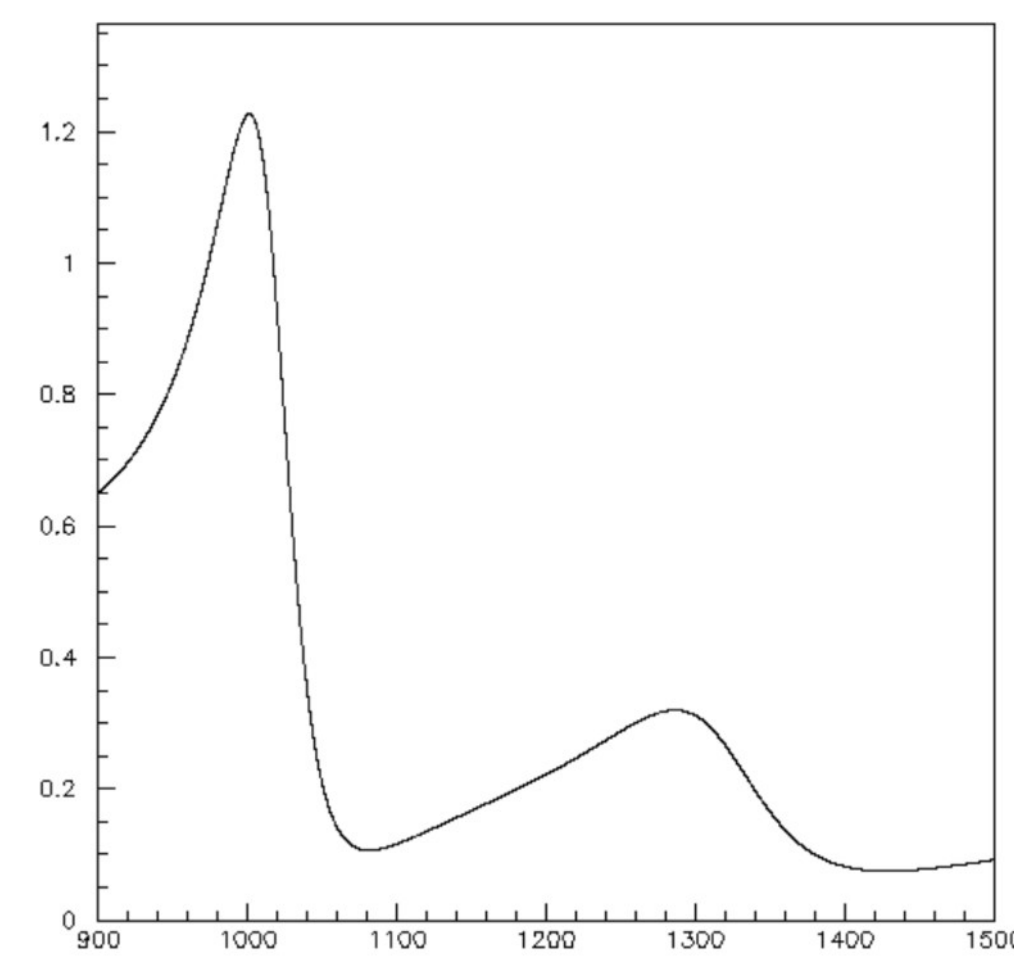

*Figure 35a: VBF differential cross section for KK gravitons. Figure 35b: VBF cross section for 3d and 4th recurrence.*

In practice gravitons of higher rank can still be produced by the VBF process. Using available VBF results for heavy Higgses, one can easily derive the KK graviton cross section. Results are displayed in figure 35. It allows to understand that the two first narrow resonances siting on very large SM background are only sensitive to two photon and purely leptonic decay modes, as seems the case. Figure 10 indicates that the third and fourth modes allow a clean detection for the mode 3 and 4. The following modes can only be detected in semi-leptonic and purely hadronic modes which seems to the case as indicated by figure 24 and 25.

| **GKK** | **4 leptons γγ** | | **hh** | **ZZ semi-lep VBF** | | | **ZZ+WW had** | | |
|---|---|---|---|---|---|---|---|---|---|
| m $G_{KK}$ GeV | 380 | **700±20** | 1000 | 1322 | 1640 | 1945 | 2260 | 2590 | 2905 |
| $\Gamma$ $G_{KK}$ GeV | 3 | **20±7** | 60 | 130 | 260 | 440 | 680 | 1042 | 1440 |
| VBF→$G_{KK}$ fb | 360 | **100±50** | 38 | 15 | 7 | 3.4 | 1.7 | 0.9 | 0.4 |

Figure 35b shows an asymmetry of the resonances which tend to drop sharply above the peak and slowly below, the former being due to an interference effect with the next resonance, the later being due to the strong mass dependence (power 7) of the VBF luminosity which rises on the negative side.

From the table one understands the following:

- Given the large background, the two first resonances being narrow are best observed in ZZ in the 4 lepton mode and in
- The hh mode provides the best visibility for the third resonance
- ZZ final states selected in semileptonic decays with a VBF selection allow to observe 4th and 5th resonances
- ZZ+WW in hadronic modes allow to observe the heaviest resonances

## V.3 Duality with a composite world

If one assumes that the KK gravitons couple directly to gluons and photons, one concludes that LHC is excluding these particles up to **4.5 TeV**, as shown in figure 36a from [55].

This assumption is of course wrong for the Higgs boson which couples indirectly to gluons and photons through **loops**.

For the process pp→$G_{KK}$→ZZ, if one assumes that $G_{KK}$ couples directly to gluons one would derive a mass limit of **1.8 TeV** as shown in figure 33b.

Given that T380 and T700 have much lower cross section than predicted by 33b, one concludes that the KK gravitons behaves like the Higgs boson with no direct coupling to photons and gluons.

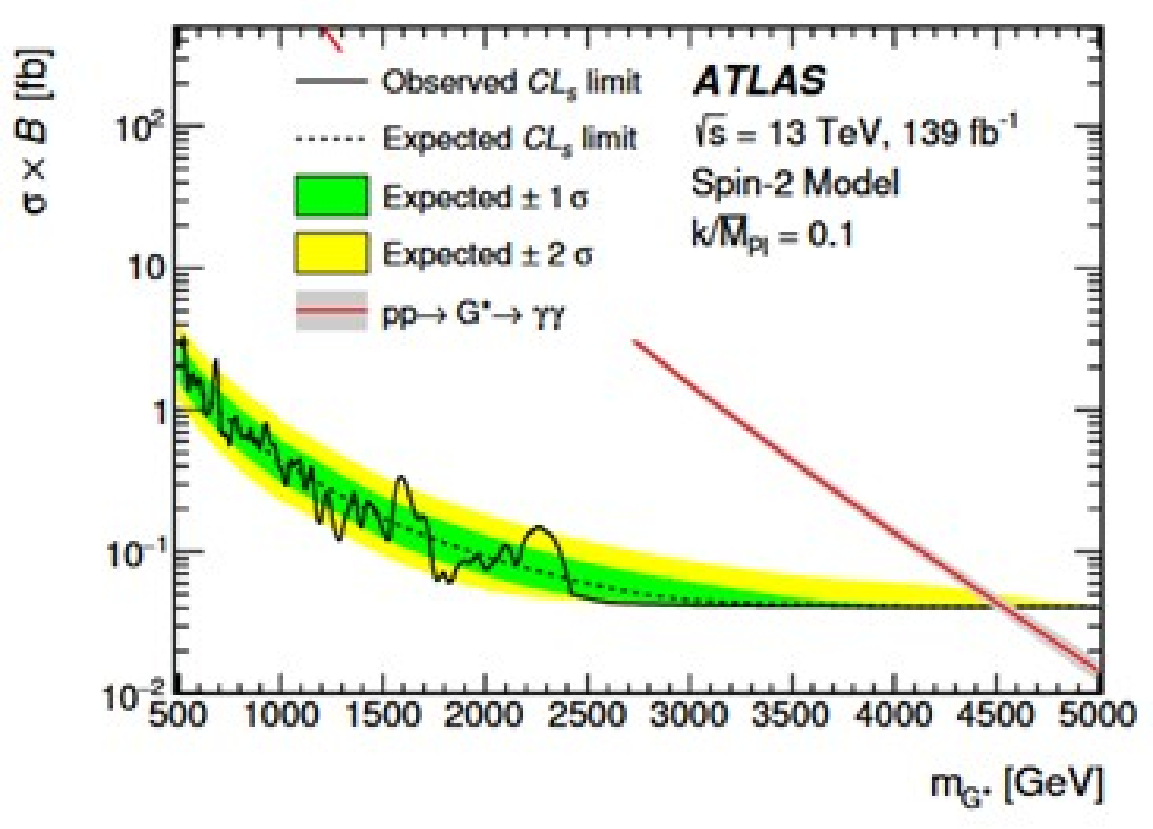

(a)

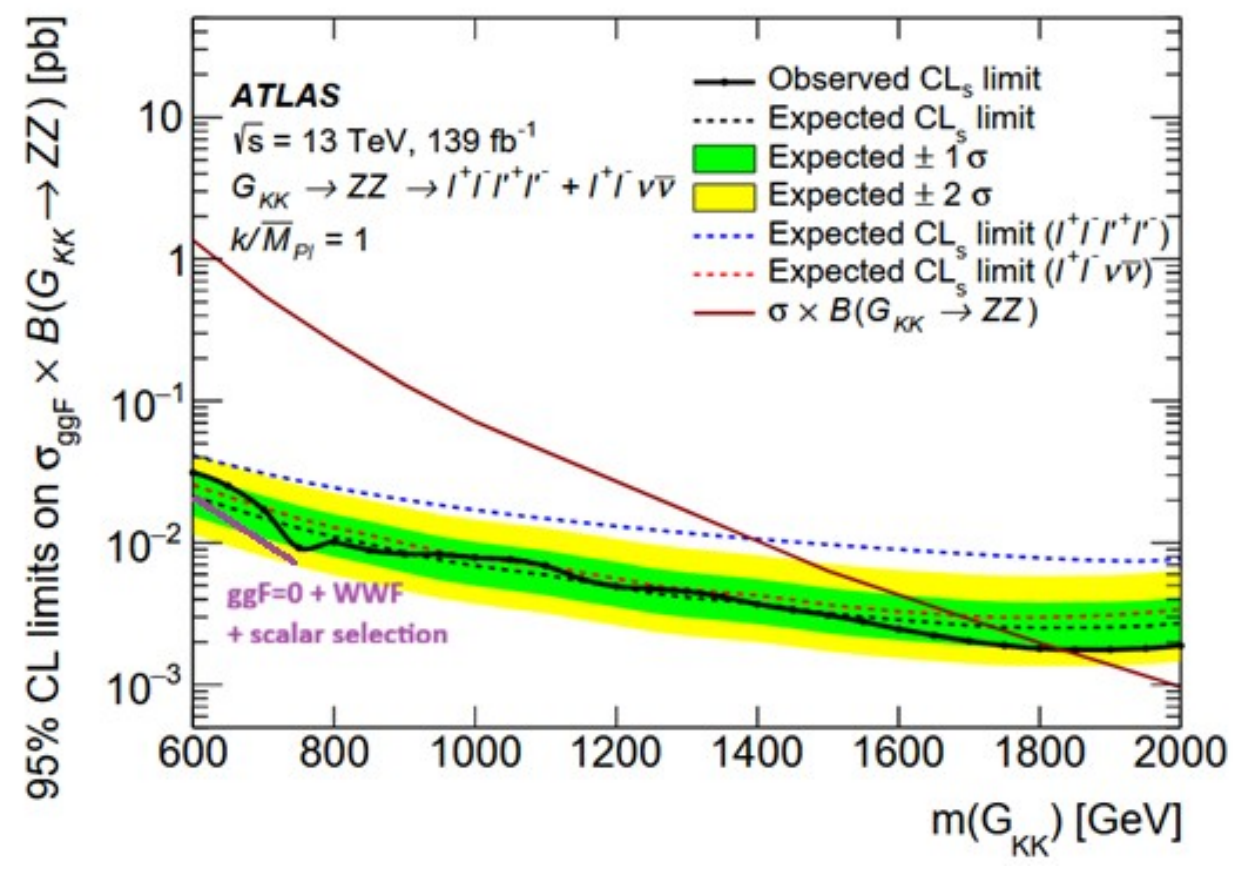

(b) *Figure 36a: The expected upper 95% CL limit on KK gravitons in the reaction pp→ $G_{KK}$->$\gamma\gamma$ [55]. Figure 36b: Expected upper limit on pp→$G_{KK}$-> ZZ achieved by ATLAS using 4 leptons and 2 leptons+2 neutrinos final states. The red curve is obtained including a ggF prediction for a n=1 graviton, while the magenta curve is our prediction with gluon decoupling, which falls below the observed limit. The ggF prediction for n=2 is simply obtained by dividing the red curve by 25.*

The RS description of KK gravitons in 5D space is assumed to be dual to a 4D description where particles are composite entities made of preons which are confined by strong forces. This assumption allows to understand an essential feature of our experimental findings which is the absence of direct coupling of gravitons to gluons and photons, reflecting the **absence of colour charge and electric charge of the preons** constituting the gravitons and the Higgs scalars. While the 5D picture is in essence geometric, assuming that interactions do not take place when particles do not overlap in the extra dimension, the fact that there is overlap does not necessarily mean that particles will interact.

As already stated this absence of direct coupling has **dramatic consequences** for searches of gravitons at LHC. As for the Higgs one is left with indirect couplings through **loops** with however a major difference: these loops will not only come from the top quark and the W but also by KK bosons and fermions. This type of exchange is also expected for the SM Higgs [59] but, with the present precision reached by LHC, has remained unnoticed.

Unfortunately there exist no calculation of this effect for LHC. Observations by ATLAS and CMS of the mode in two photons (figure 1) indicate that this contribution amounts to a width of 60±20 MeV.

## V.4 The 750 GeV episode

Although creating a great disillusion, this episode had the indisputable merit to trigger an intense work on graviton phenomenology. The first indication for the '750' came from an indication from ATLAS (see figure 1). One of the prominent interpretation was already in terms of KK gravitons. The table below gives the predicted branching ratios for a graviton, see [71] and [72]. It displays a high **dispersion of predictions** between the models, meaning that our observations are not so surprising. Note also that it assumed that '750' is the 1[st] KK.

| | IR | MIN | MED | MAX | GMAX |
|---|---|---|---|---|---|
| Br($X \to \gamma\gamma$) [%] | 4.3 | 8.5 | 7.0 | 0.5 | 2.3 |
| Br($X \to ZZ$) [%] | 4.8 | 7.9 | 7.8 | 2.9 | 12 |
| Br($X \to WW$) [%] | 9.5 | 16 | 15 | 5.6 | 21 |
| Br($X \to Z\gamma$) [%] | 0 | 0 | 0 | 0 | 1.1 |
| Br($X \to hh$) [%] | 0.3 | 0 | 0.4 | 1.4 | 6.9 |
| Br($X \to t\bar{t}$) [%] | 5.1 | 0 | 8.3 | 85 | 56 |
| Br($X \to b\bar{b}$) [%] | 6.4 | 0 | 5.2 | 0.4 | 0.04 |
| Br($X \to jj$) [%] | 66 | 68 | 61 | 4.5 | 0.5 |
| Br($X \to e^+e^-$) [%] | 2.1 | 0 | 0 | 0 | 0 |
| $\Gamma(X \to \gamma\gamma)$[MeV] | 0.25 | 0.15 | 0.18 | 2.5 | 25 |
| $\Gamma(X \to \mathrm{tot})$[MeV] | 5.7 | 1.8 | 2.6 | 500 | 1060 |
| PLC: $\sigma_{eff}(\gamma\gamma \to X)$ [fb] | 40 | 24 | 29 | 400 | 4000 |
| LC: $\sigma(e^+e^- \to X)$ [pb] | 0.4 | 0 | 0 | 0 | 0 |

A **significant coupling to e+e- is only present the IR case** when all SM particles belong to the so called infra-red brane IR. It allows **decoupling of the gluons and light quarks in scenarios MAX and GMAX** as seems the case in real life. In GMAX and MAX a **$t\bar{t}$ dominance is predicted,** against our findings.

The observed excess into 2 photons observed initially by ATLAS at a mass of 750 GeV corresponds to a cross section of order 5±1.5 fb while the result of figure 1 is about 1±0.0.3 fb. This seems to mean:

- That there was an initial positive **statistical fluctuation** which explains the apparent disappearance of the preliminary signal
- This signal was slightly displaced in mass, 750 GeV instead of 700 GeV, which can be attributed to a wrong calibration which has now been fixed in ATLAS
- The larger width of the reconstructed signal (see fig. 1) also suggests possible imperfect calibration

With the full RUN2 analysis ATLAS observes an indication at 700 GeV but with a **narrower width** and accompanied by 8 more indication which greatly comforts the evidence.

## V.6 Precision measurements

### V.6.1 Indirect effects

For this phenomenology, one defines a unique **mass scale parameter mKK**, not directly related to the mass of a given KK particle. T and S constraints imply [59] that **mKK>30 TeV**. With **custodial symmetry** one has **mKK>1.8 TeV**. KK gravitons could be lighter. Reference [71] suggests that this can be accomplished by assuming a breakdown of universality within the kinetic terms. The custodial symmetry implies an extension of the SM with additional symmetries, with the **presence of a Z'** with a mass similar the KK vectors.

### V.6.2 Input from EW precision measurements

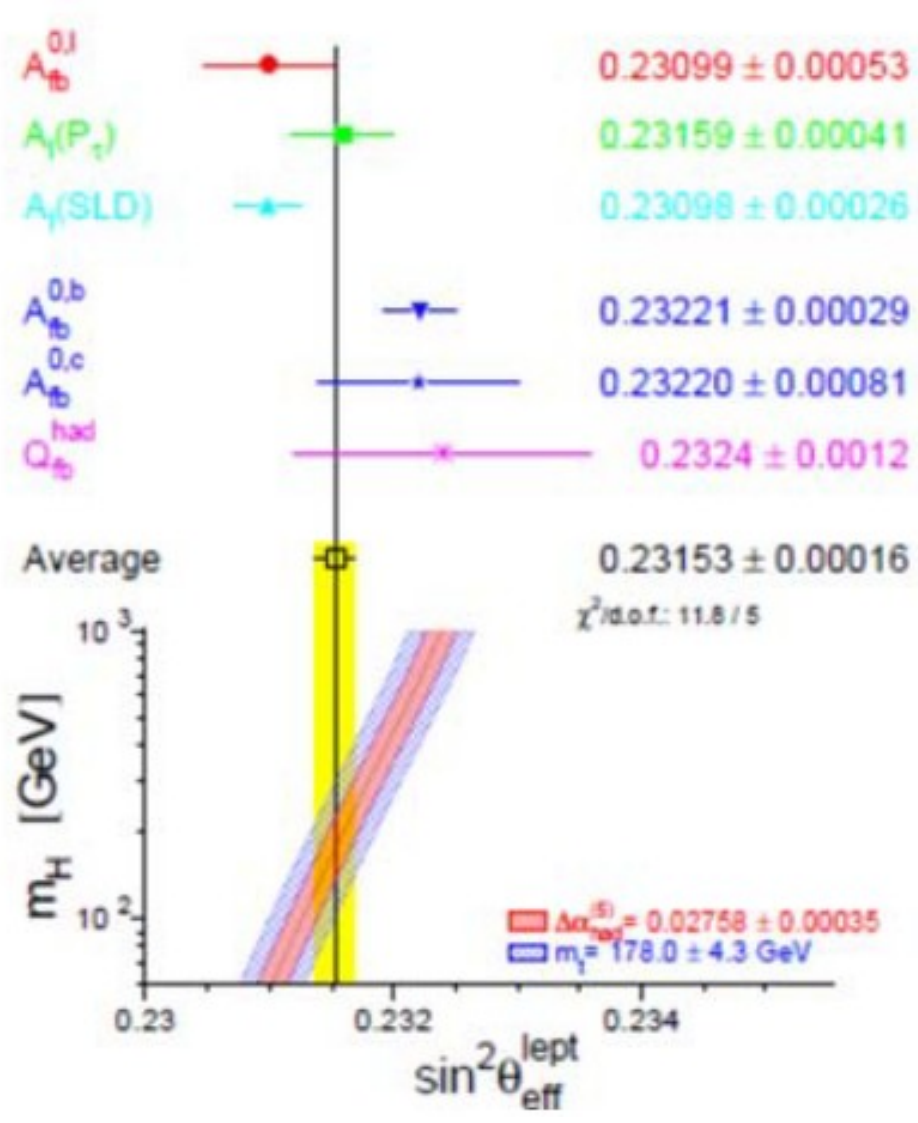

*Figure 37: LEP and SLD results*

There are several deviations suggesting that RS effects could be at work: AFBb at LEP, ALR at SLD and AFBt at TeVatron. Figure 37 shows that the two most precise determinations of $\sin^2\theta$lep are away by more than 3 s.d. Averaging between these two values, one obtains an **apparent 'conformity'** with SM predictions.

Interpretations in terms of RS have been provided in [74,75]. They can be reconciled with $Z'/Z_{KK}$ masses of order 3 TeV assuming $\sin\theta'$=0.1 for the mixing angle related to a Z'. Then one has ceR=0.56, meaning that the coupling of $G_{KK}$ to e+e- will be vanishingly small.

If one assumes a $Z'/Z_{KK}$ mass much heavier, ceR decreases and the coupling of $G_{KK}$(700) to e+e- can become substantial. The dependence of the coupling of fermions to $G_{KK}$ can be deduced from **figure 33** given that ceR=-$\nu$e. Satisfying the ALR result of figure 34 and the measured value of BR(ee), requires that ceR=0.45 and **$mZ_{KK}$=2.5TeV/$\sin\theta'$**. Since perturbativity requires that $\sin\theta'$>0.1 (meaning **$mZ_{KK}$<25 TeV)**, this particle should be observed either indirectly [26] at a LC and/or directly at LHC for larger values of $\sin\theta'$.

This interpretation implies that KK vectors are much heavier than the KK gravitons indicated by LHC. Note again that LEP PM have no direct implication on the $G_{KK}$ sector since one can assume that masses of KK vectors and tensors are independent.

Future Z factories, linear or circular, allow to improve the LEP/SLC accuracies by an order of magnitude, thus fully confirming/discarding the effect seen in figure 36. If confirmed, this prediction would constitute an impressive illustration of the **mass reach of precision measurements in e+e-,** far more spectacular than the prediction for the top quark and the Higgs boson.

### V.6.3 Effect on SM Higgs BR

It there are light $W_{KK}$ and/or light $q_{KK}$, they will contribute to the **loop corrections** driving **h125→gg/γγ** hence modifying LHC cross sections for these two channels. These effects become spectacular for KK masses ~2 TeV as shown in figure 38 [59], where one sees a **strong decrease for gg** and an **enhancement for γγ**. With a the present accuracy of ~10%, LHC is able to **exclude KK masses up to ~4 TeV**. Results from figure 36 have to be interpreted with a 'grain of salt'.

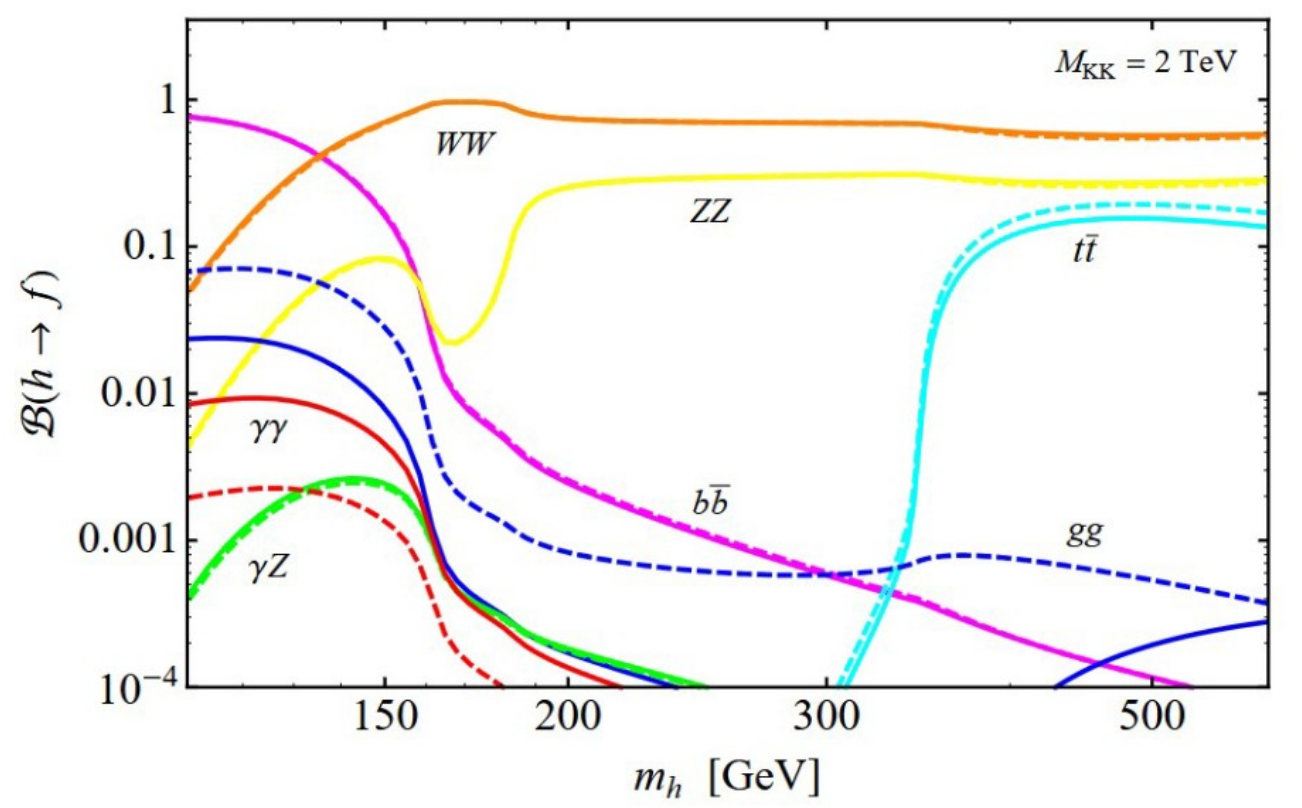

*Figure 38: Predicted BR for the SM Higgs boson [59] assuming the presence of light (~2 TeV) KK particles. The dashed curves correspond to the SM and show clear discrepancies with the full curves for gg and $\gamma\gamma$ final states.*

Again, what is called $\mathbf{m_{KK}}$ is a mass scale directly related to the KK quarks and bosons which contribute to loop corrections affecting the SM Higgs:

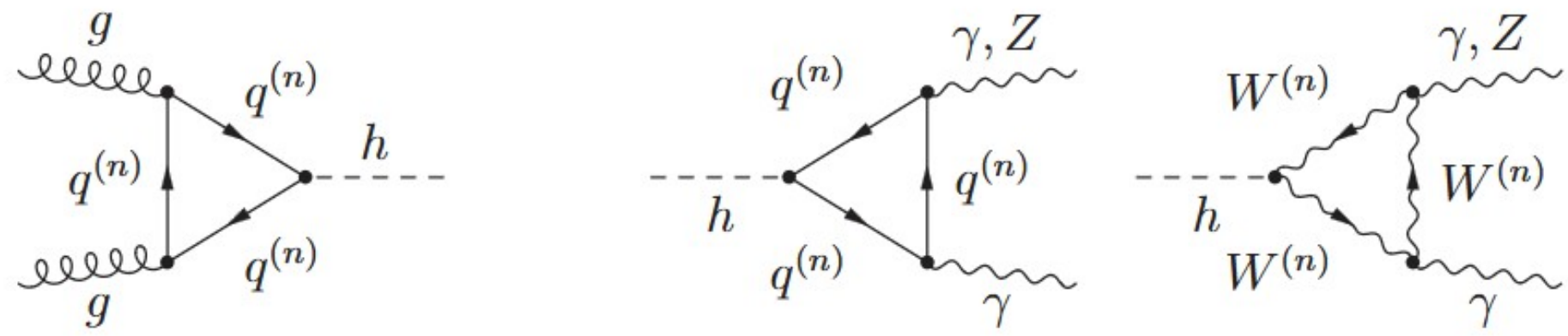

Similar diagram are expected for the KK graviton into two photons but they have not yet been calculated. These contributions will apply to the couplings of $G_{KK}$(380) and $G_{KK}$(700) and, even if these particles are not directly coupled to 2 photons, generate some non negligible couplings as suggested by figure 1.

Next, one would like to predict if Z'/$Z_{KK}$ can be detected at colliders.

### V.6.4 SUSY and Compositeness

Figure 39 from [81] shows the sensitivity of SM Higgs couplings to MSSM and compositeness contributions.

Note that if these two contributions were present, the most significant effects would tend to compensate, which illustrates the limits of these indirect measurements.

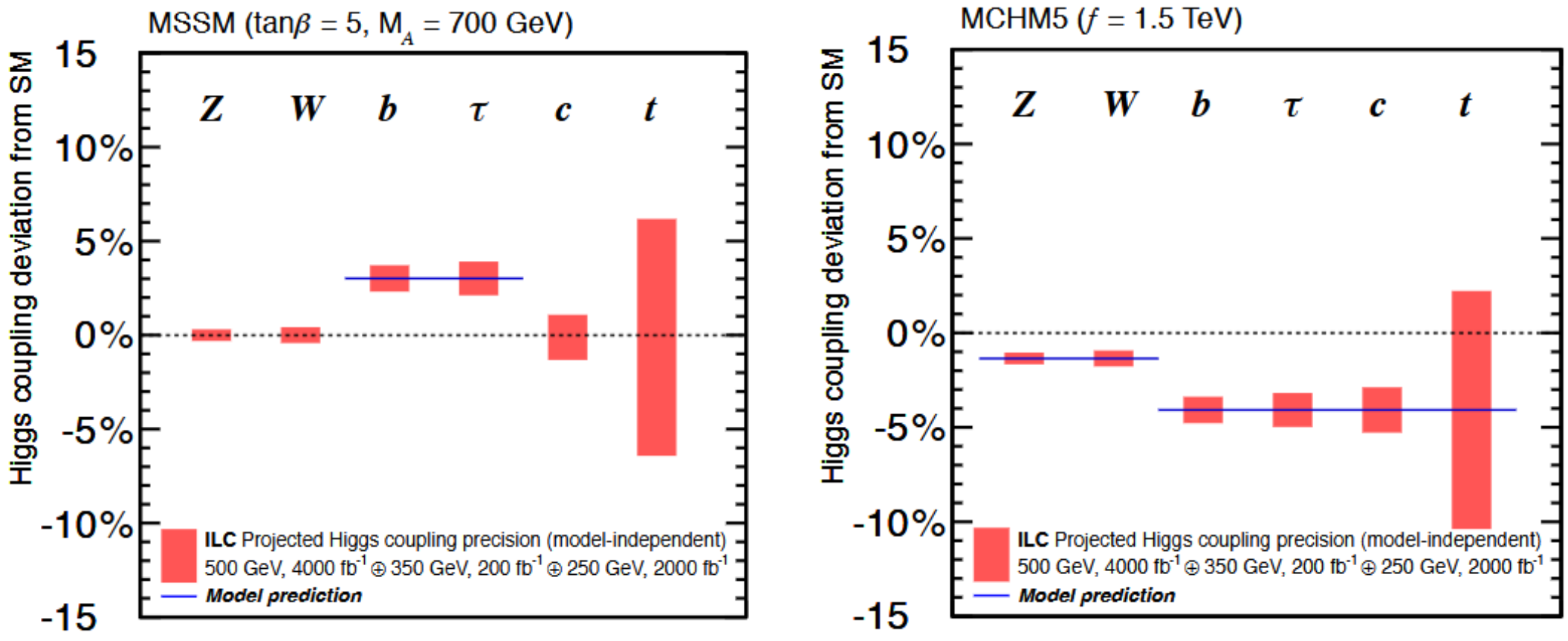

*Figure 39: Predicted deviation from SM of Higgs couplings in MSSM and in a composite model together with the predicted accuracies for a 500 GeV ILC.*

## V.7 $Z_{KK}$/Z' direct detection

In the RS scenario, Z'/$Z_{KK}$ is likely to be heavy, >5 TeV, as suggested by SM Higgs precision measurements discussed previously. At LHC it can be directly produced by annihilation of a valence quark and a **sea antiquark**. Note that the **$q\bar{q}$ parton luminosity** at 700 GeV is about an order of magnitude smaller than the gg luminosity. Even assuming that cqR has a value of order 0.4, one ends up with a **sensitivity well below 5 TeV** for $Z_{KK}$. Z' may have an enhanced coupling which would then allow to reach that mass region with HL-LHC.

Again the VBF process may lead to a more favourable situation but as is the case for the KK graviton, the production cross section may not allow to go well beyond a TeV.

In this scenario, indirect detection through precision **measurements in e+e- looks more promising.**

In the case of the **Hosotani scenario** [77,80], the situation is radically different since Z' has a copious coupling to light quarks already allowing to cover masses above 5 TeV with available LHC data.

## V.8 Evidence for a very heavy KK resonance

CMS (84) has shown some evidence for an 8 TeV resonance decaying into four jets forming a pair of 2 TeV resonances. Given the mass of this resonance, it seems that most production mechanism are 'out of steam' with the exception of valence quark annihilation, meaning that this particle is a diquark[85] DQ. At 8 TeV, the gg differential luminosity is smaller but comparable. One therefore cannot a priori exclude a KK scalar or tensor resonance produced through gg. A KK graviton seems excluded given its width at 8 TeV and its poor coupling to gg. Similarly a KK gluon could not do the job since it does not couple to gg.  A KK scalar from the Higgs boson is more likely but, again would have a prohibitive width.

While not present in the strict RS model, in the composite picture a KK diquark would give:

**qq→ DQ(8TeV)→ qKK(2TeV)qKK(2TeV)**

This process, if true, gives a good example of mass hierarchies between the various KK particles noting that reference [86] provides a prediction for the KK quark masses close to observation.

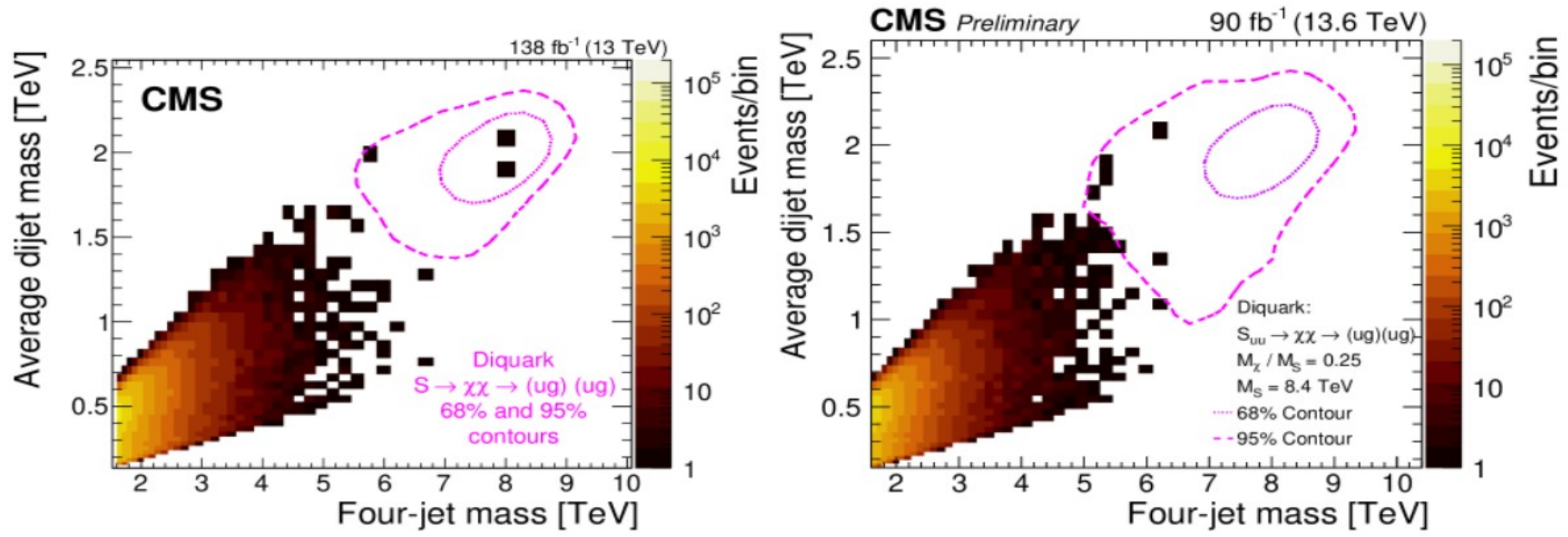

*Figure 50: Observed events in the m4j – m2j plane for Run2 with 138 fb-1 at 13 TeV (left) and the new run 3 with 90 fb-1 at 13.6 TeV (right). The dotted and dashed curves show the 68% and 95% probability contours, respectively, from a signal simulation of a diquark with a mass of 8.4 TeV, decaying into a pair of vector-like quarks, each with a mass of 2.1 TeV.*

The width of this resonance could be large although its extension to large masses is naturally limited by the parton luminosity cut-off. Figure 50 indicates that the amount of selected data varies significantly with the assumption on the selection contour: from 2 evts in figure 50 left and 0 events in 50 right, to, respectively, 7 and 5 selected events.

If one assumes that the CMS reconstruction has a good acceptance for such a process, the production cross section comes out to be of order 10 ab. The HL-LHC sensitivity should be able to provide an answer on this intriguing phenomenon.

Having said that, one has to accept that this effect was, by far, not predicted by the standard RS picture but this somewhat also applies to the KK graviton sector.

The heavy fermion hypothesis qKK(2 TeV) could, perhaps, be tested in pair production in ggF. The expected cross section is 6 orders lower that the top pair cross section that is of order 1 fb, which looks challenging.

## VI.9 The Radion

[67] suggests that the radion mass is lighter than the graviton mass. In our case this means lighter than 380 GeV or 700 GeV. It seems therefore that it could decay into top pairs and offer a solution to the excess observed in top pairs near threshold. The n=3 KK graviton recurrence could decay into two radions giving 4 tops, as well as the next KK states. For a **radion** near the top pair threshold one expects [68] WW/ZZ/tt=1/0.5/0.25. One can use cross section limits from the ZZ channel [70] to deduce cross section limits on top pairs. Near the top threshold, the VBF channel would give a limit of **50 fb** which is too weak to explain the discrepancy between the measured toponium cross section and the expected value. Note that the analysis deduced from [70] only applies to **scalars** and therefore cannot be applied to the tensor T380.

## V.10 The full picture ?

**Interaction between DM and KK gravitons,** could explain the **Fermi-LAT** observation as was already suggested in [64].

An emerging aspect of this model seems to be the **connections between SUSY and RS** which offers promising avenues to understand the symmetry breaking mechanism of SUSY (see [23] and [47]). These two approaches allow obvious complementarities:

- SUSY predicts the SM Higgs mass and offers a calculable theory
- RS gives a geometrical interpretation of the fermion masses and couplings
- RS is also a calculable theory
- SUSY offers a valid dark matter candidate, the neutralino
- RS connects particle physics to gravitation
- RS allows a connection to compositeness through the concept of duality

This list is certainly not exhaustive but illustrates the potential of combining these basic concepts.

We will not deal with the various and essential merits of the RS model with respect to **cosmology** [60].